\documentclass[11pt,a4paper]{article}

\usepackage{geometry}
\usepackage[utf8]{inputenc}
\usepackage{amssymb,amsfonts,amsmath,stmaryrd}
\usepackage[export]{adjustbox}
\usepackage{enumerate,float,indentfirst}
\usepackage{color}
\usepackage{tikz}
\usepackage{tikz-cd}
\usetikzlibrary{arrows,snakes,backgrounds}
\usepackage{url}
\definecolor{darkblue}{rgb}{0.1,0.1,.7}
\usepackage[colorlinks, linkcolor=darkblue, citecolor=darkblue, urlcolor=darkblue, linktocpage]{hyperref}
\usepackage{textcomp}
\usepackage{subfig}
\usepackage{amsthm}
\usepackage{bbold}
\usepackage{mathtools}
\usepackage{multirow}
\usepackage{braket}
\usepackage{leftindex}
\usepackage[normalem]{ulem}
\usepackage[square, comma, compress,numbers]{natbib}
\usepackage{stackrel}

\graphicspath{{figures/}}

\numberwithin{equation}{section}

\def\be{\begin{eqnarray}}
\def\ee{\end{eqnarray}}

\def\calL{\mathcal{L}}

\def\calF{\mathcal{F}}
\def\calO{\mathcal{O}}
\def\calR{\mathcal{R}}

\def\bz{\bar{z}}
\def\hb{\bar{h}}

\def\eps{\epsilon}
\def\vareps{\varepsilon}

\definecolor{red}{rgb}{1,0,0}
\definecolor{orange}{rgb}{1,0.5,0}
\definecolor{violet}{rgb}{0.7,0,1}

\newcommand{\abs}[1]{\left\lvert#1\right\rvert}

\newcommand{\ignore}[1]{}

\usepackage{listofitems}
\newcommand{\QCG}[1]{\readlist*\Fvar{#1}\begin{bmatrix} \Fvar[1] & \Fvar[3] & \Fvar[5]\\ \Fvar[2] & \Fvar[4] & \Fvar[6]\end{bmatrix}_{\Fvar[7]}}

  \newcommand{\sixj}[1]{\readlist*\Fvar{#1}\begin{Bmatrix} \Fvar[1] & \Fvar[2] & \Fvar[3]\\ \Fvar[4] & \Fvar[5] & \Fvar[6]\end{Bmatrix}}

\newtheorem{theorem}{Theorem}[section]

\theoremstyle{remark}
\newtheorem{remark}[theorem]{Remark}
\newtheorem{example}[theorem]{Example}

\begin{document}

\vspace*{-.6in} \thispagestyle{empty}
\begin{flushright}
\end{flushright}
\vspace{1cm} {\Large
\begin{center}
{\bf Quantum Groups as Global Symmetries}
\end{center}}
\vspace{1cm}
\begin{center}
{ Barak Gabai$^q$, Victor Gorbenko$^q$, Jiaxin Qiao$^q$, Bernardo Zan$^{\Tilde{q}}$ and Aleksandr Zhabin$^q$}\\[2cm] 

$^{q}$ \textit{Laboratory for Theoretical Fundamental Physics, Institute of Physics,\\ École Polytechnique Fédérale de Lausanne, Switzerland}\\
$^{\Tilde{q} }$ \textit{Department of Applied Mathematics and Theoretical Physics,\\
University of Cambridge, CB3 0WA, UK}
\vspace{1cm}

\vspace{1cm}\end{center}

\vspace{4mm}

\begin{abstract}
We study quantum field theories which have quantum groups as global internal symmetries.   We show that in such theories operators are generically non-local, and should be thought as living at the ends of topological lines. We describe the general constraints of the quantum group symmetry, given by Ward identities, that correlation functions of the theory should satisfy. We also show that generators of the symmetry can be represented by topological lines with some novel properties. We then discuss a particular example of $U_q(sl_2)$ symmetric CFT, which we solve using bootstrap techniques and relying on the symmetry. We finally show strong evidence that for a special value of $q$ a subsector of this theory reproduces the fermionic formulation of the Ising model. This suggests that a quantum group can act on local operators as well, however, it generically transforms them into non-local ones.

\end{abstract}
\vspace{.2in}
\vspace{.3in}
\hspace{0.7cm} October 2024

\newpage
\tableofcontents

\section{Introduction}
What is the most general set of symmetries that a Quantum Field Theory can possess?  Even if we restrict ourselves to symmetries that are global and internal, meaning that they act faithfully on the physical Hilbert space of the theory, and commute trivially with the Lorentz group, this question seems notoriously complex. The most standard case is that symmetry transformations form a Lie group or a discrete group and act locally on the fundamental local fields of the model. Recently, we witnessed very fruitful research developments that generalized the concept of symmetries in two ways. The first generalization is related to relaxing the group-like structure of a symmetry. It turns out that a discrete set of transformations commuting with a Hamiltonian does not have to correspond to a group, but instead can be generated by a set of topological defect lines that are generically non-invertible. In modern language these symmetries are called non-invertible, or categorical, see e.g. \cite{Chang:2018iay,Komargodski:2020mxz, shao2023s}. The second generalization is achieved by allowing the symmetry to act not only on local fields, but also on extended objects present in a theory. These transformations are called higher-form symmetries \cite{Gaiotto:2014kfa,Sharpe:2015mja, Cordova:2022ruw}. The two generalizations can be combined to produce non-invertible higher-form symmetries \cite{Kaidi:2021xfk,Choi:2021kmx,Schafer-Nameki:2023jdn,Brennan:2023mmt}. In this paper, we seek for a different direction to generalize the concept of symmetry. 

Part of our motivation comes from studying certain two-dimensional models of statistical physics. Consider, for example, the $O(n)$ loop model for a continuous value of $n$ \cite{Nienhuis:1982fx} . The operator spectrum of this model in the continuum is known \cite{diFrancesco:1987qf} and exhibits degeneracies that are not explained by the model's apparent global symmetry, that is $O(n)$. Certain versions of the model also have an $O(n)$-singlet operator that is nevertheless not generated by the RG flow, seemingly leading to fine-tuning \cite{Jacobsen:2002wu,Gorbenko:2020xya}. While protection of this particular operator was beautifully explained by Jacobsen and Saleur in a recent paper \cite{Jacobsen:2023isq} by discovering a non-invertible defect line present in the theory, the symmetry structure that explains the rest of the spectrum remains a mystery, and most certainly cannot be a regular symmetry group, nor the generalizations described above. This relatively simple example demonstrates that there exist symmetries that go beyond the presently existing classification and that  have a well-defined effect on physical observables. 

In this paper, instead of trying to directly answer the big question formulated above, we take a more practical approach. We take an existing algebraic structure, and seek for a field theory that has this structure as a symmetry.  Namely, we consider a particular type of deformation of Lie groups that bears the name Quantum Groups (QG). While we will give the technical definition of a QG below, let us give here an intuitive picture. Imagine an experimentalist who does a set of measurements on some Lorentz-invariant system and finds that it has a conserved $U(1)$ charge. In the next step she discovers that operators or states of the theory have a degenerate energy spectrum, and in particular there are several triplets of operators with charges +1, -1 and 0. At this point it is natural to conjecture that the symmetry is enlarged and is $SU(2)$ instead of just $U(1)$ and that operators are in its spin one representation. The smoking gun of $SU(2)$ would then be the vanishing of the three-point function of the three zero charge operators. It has to vanish because symmetry forces the correlator to be proportional to $\epsilon^{ijk}$. The experiment, however, shows that $\braket{ 0,0,0 }$ correlators are not zero, and instead they have the same functional form as the $\braket{ +,0,- }$ correlators, but differ by a certain multiplicative constant, the same for all triplets. This pattern of correlation functions is exactly what a quantum group symmetry would predict. QG, in our case $U_q (sl_2)$, is a certain continuous deformation of $su(2)$ labeled by a complex parameter $q$. The goal of this paper is to understand whether QFTs with QG symmetries exist, study the general consequences of such symmetries and provide an explicit example of a UV-complete (by a lattice theory) QFT of this type.

Needless to say, QGs appeared on numerous occasions in the studies of QFT, as well as quantum gravity, and we are not aiming here to give a full list of references. Let us nevertheless mention several benchmark developments. Quantum groups were first discovered as a tool to explain integrability of spin-chain Hamiltonians, as well as to construct new integrable systems \cite{Takhtajan:1979iv,Kulish:1981dli}, see \cite{Faddeev:1996iy,Franchini:2016cxs} for reviews. As a mathematical construction, they were developed in the works \cite{Drinfeld:466366,Drinfeld:1986in,Jimbo:1985zk,Faddeev:1987ih}. Soon after, QGs appeared in the study of Chern-Simons theories \cite{Witten:1988hf,Witten:1989rw,Guadagnini:1989tj,Reshetikhin:1991tc,Alekseev:1994pa,Fock:1998nu} and 3d topology \cite{Turaev:1992hq}. The connection exists because of the relation between QGs and representations of the braiding group, which appears naturally in theories with generalized statistics, see \cite{Fredenhagen:1988fj,Moore:1988qv}. The latter paper appears to be the first to introduce QG-related notions in the context of 2D CFTs, event though it does not mention QG explicitly. This was done in subsequent publications, including \cite{Moore:1989ni,Alekseev:1990vr,Felder:1991ab,Alvarez-Gaume:1988bek,Falcet:1991xt,Bazhanov:1994ft}. Through connection to non-abelian anyons, QGs also made their appearance in quantum Hall physics \cite{Slingerland:2001ea}. QG symmetries appeared in the context of AdS/CFT integrability \cite{Beisert:2008tw,Arutyunov:2013ega,Thompson:2019ipl}. More recently, QG were discussed in models of lower-dimensional holography, namely the double-scaled SYK model \cite{Berkooz:2018jqr,Lin:2023trc}. Possibly the relation is again due to Chern-Simons theories \cite{Mertens:2022ujr,Narovlansky:2023lfz,Gaiotto:2024kze}. 
Despite the abundance of physics research related to QGs, not much emphasis was made on QGs action as global internal symmetries. For example, even  though the spin chain models in which the QGs were originally discovered are closely related to the model we study below, in the original works it was used specifically to elucidate the integrable structure of the model, while, as global symmetry of a QFT, a QG is not directly related to integrability -- it is just a finite number of conservation laws in an infinite-dimensional Hilbert space. For some of the other recent developements, see also \cite{Dupuis:2020ndx,Mertens:2020hbs,Collier:2023fwi,Chen:2024unp}.

Let us now summarize here several important papers that have the most similarities with our approach. Conceptually, our thinking is close to that of \cite{Bernard:1990ys}, which studied in details QG symmetry of the sine-Gordon model. However, in that paper, QG is a space-time symmetry, related tightly to the infinite-dimensional symmetry responsible for integrability of the model. Appearance of QG in 2D CFTs was extensively discussed in \cite{Gomez:1990er,Gomez:1996az},
where it was discovered that screening charges appearing in the Coulomb gas formalism of \cite{Dotsenko:1984conformal,Dotsenko:1984four} are related to the QG generators. We develop the Coulomb gas approach to QG in a companion paper \cite{paper2}, and while our methods are largely based on these techniques, there are also several important differences that we emphasize there. As far as the construction of explicit theories with QG as a symmetry is concerned, previous studies were restricted to chiral CFTs, notable examples include \cite{Gaberdiel:1994vv} for WZW models, and \cite{Mack:1991by,MACK1992185} for chiral minimal models. In both cases the emphasis was made on $q$ being root of unity, which also allows for a truncation of a QG to a smaller structure, which was indeed studied in the latter paper. Finally, a version of XXZ spin chain with open boundary conditions that preserves QG symmetry and flows to a boundary CFT in the IR was studied in \cite{Pasquier:1989kd}. This theory is closely related to the main example we study below. 

Why are QGs not ubiquitous, and in most QFTs we know so far we land on usual Lie group continuous symmetries? For weakly coupled theories the answer is based on a Lagrangian intuition: if we consider a Lagrangian constructed of local fields, we expect its symmetry transformations to act on these fields classically. Composition of such actions is also a symmetry, and every infinitesimal transformation can be inverted, thus giving us the defining properties of a group. While most theories we understand well have a Lagrangian description, we also know that there are many theories that are strongly coupled are inherently non-Lagrangian. It is thus natural to look for theories with QG symmetry among such examples. We will consider a theory that is the continuum limit of a certain spin-chain model. Spin chain Hamiltonians are well-defined without a weakly coupled parameter and the fact that they can have a QG as a symmetry is well-established. An important difference between QG generators and the usual Lie algebra elements is how their action is extended from smaller subsystems to a larger system. If the Hilbert space of a system is a tensor product of Hilbert spaces of subsystems, usual symmetry generators will also be realized as a tensor product of the form $X\otimes I+I\otimes X$. The rule for QG is more complicated, and to define the action of a generator it is important to order the subsystems. Ordering of subsystems is natural on a spin chain, but there is no obvious ordering for subsystem in relativistic field theories, at least for those that are space-like separated. As we will see below, natural QFT objects on which the QG can act are not local operators, but rather defect ending operators, that is local operators to which some topological line is attached. These lines allow to define operator ordering even in the Euclidean domain. This does not mean that a QG can never act on a local operator in an interesting way, but we expect that, generically, a transformation would turn it into a defect-ending operator. We will come back to this idea in the conclusion, and hopefully this discussion will appear more concrete to the reader after going through our paper.

Let us finish the introduction by briefly summarizing the content of our work. 
In several parts of the paper we review some previously known techniques, both to make the paper readable for non-experts, and also because we introduce important modifications to some of them. 
In particular, in Section \ref{sec:quantum_group} we introduce the quantum groups $U_q(sl_2)$ and set up our conventions and notations. In Section \ref{sec:results} we explain the properties of general 2D QFTs with $U_q(sl_2)$ global symmetries. In Section \ref{section:example} we introduce our example QFT: a continuous limit of a certain non-local version of the closed XXZ spin chain first studied in \cite{grosse1994quantum} and proceed to solve it using analytic bootstrap methods. The solution is consistent with the properties explained in the previous section. In Section \ref{sec:ising} we consider a limit of our theory where the critical Ising model appears as a subsector. We focus on this limit because it shows that local operators, in this case Ising fermions, can feel the action of QG transformations. Technical details are relegated to several appendices.

In a companion paper \cite{paper2}, we solve the same theory using a different approach: we develop a modified version of the Coulomb gas formalism of \cite{Gomez:1996az}. The Coulomb gas construction makes the structure of defect ending operators and QG generators much more manifest and provides a consistency check of our solution. We also explain the concept of ``hidden QG symmetry'' that appears in minimal models and that is different from a global symmetry.

\section{A Review of Quantum Groups} \label{sec:quantum_group}

In this section we review the definition and the main properties of the quantum group $U_q(sl_2)$, show the difference with the usual groups, and fix all the notations related to the quantum group that are used throughout the text. This section is aimed for people who are not familiar with the notion of quantum groups; experts can skip directly to section \ref{sec:results} where QFTs with quantum group symmetries are discussed. We will discuss only the properties of quantum groups that will play a role in our field theory setup, and we will try to focus on physical intuition rather than mathematical rigor. To the reader who would like to know more about quantum groups, we recommend for example \cite{klimyk2012quantum}. 

\subsection{Definition of \texorpdfstring{$U_q(sl_2)$}{}}
\label{sec:def_Uqsl2}
Before defining the quantum group $U_q(sl_2)$ let us state the idea which stands behind the definition -- why a quantum group can be treated as a deformation of a usual group. Finite dimensional Lie algebras are strictly restricted by a classification in terms of Dynkin diagrams (Cartan matrices) and one cannot simply deform a Lie algebra and preserve all its properties. The trick to overcome this is, for a given Lie algebra $\mathfrak{g}$, to consider a universal enveloping algebra $U(\mathfrak{g})$, which consists of all polynomials made out of generators of $\mathfrak{g}$ factored by the commutation relations. $U(\mathfrak{g})$ is an infinite dimensional object which has a structure of a Hopf algebra, and one is allowed to deform the universal enveloping algebra in such a way that it preserves the Hopf algebra structure. This deformed (or ``quantized'') object is called a quantum group for historical reasons, but is in fact an algebra.

Throughout the text we work only with the $sl_2$ algebra and its deformation, so we provide all the definitions for this particular example, but in principle one can construct a quantum group for any semisimple Lie algebra. The generators of $sl_2(\mathbb{C})$ are usually denoted as $E,F,H$, which satisfy the following commutation relations:
\begin{equation}\label{comm_sl2}
    [H, E] = 2E, \qquad [H, F] = -2F, \qquad [E, F] = H.
\end{equation}
All polynomials made out of $E,F,H$, factored by the commutation relations \eqref{comm_sl2}, form the universal enveloping algebra $U(sl_2)$. For a given finite-dimensional representation, one can easily construct the generators explicitly. 

We will be interested in acting with our generators not only on a single representation but also on the tensor product of two representations. In order to do this, one has to introduce a coproduct $\Delta: U(sl_2) \to U(sl_2) \otimes U(sl_2)$. In the case of $U(sl_2)$, the coproduct takes the familiar form
\begin{equation}
    \Delta(E) = E \otimes 1 + 1 \otimes E, \qquad \Delta(F) = F \otimes 1 + 1 \otimes F, \qquad \Delta(H) = H \otimes 1 + 1 \otimes H.
\end{equation}
The coproduct satisfies a few nice properties: it needs to be compatible with the commutation relations
\begin{equation}\label{compatible}
    \Delta([x,y]) = [\Delta(x),\Delta(y)], \qquad \forall x,y \in U(sl_2).
\end{equation} 
and it needs to be co-associative, so that its action on three and more representations is well-defined \begin{equation}\label{coprod:coasso}
    (\Delta\otimes\mathrm{id})\circ\Delta=(\mathrm{id}\otimes\Delta)\circ\Delta
\end{equation}
There is also an additional map and $S:U(sl_2) \to U(sl_2)$ which is called an antipode (its explicit form is not important for us now). Together with the last map, $U(sl_2)$ forms a Hopf algebra structure.

It appears that one can deform the universal enveloping algebra but preserve the Hopf algebra structure. The deformed object is called a quantum group of $sl_2$, which we denote as $U_q(sl_2)$, and it depends on a deformation parameter $q\in\mathbb{C}$. Its generators satisfy the deformed commutation relations
\begin{equation}\label{comm_rel_Uqsl2}
    [E,F] = \frac{q^{H} - q^{-H}}{q - q^{-1}}, \qquad 
    q^{H}Eq^{-H} = q^{2} E, \qquad
    q^{H}Fq^{-H} = q^{-2}F.
\end{equation}
One can check that in the $q\to1$ limit they become exactly \eqref{comm_sl2}. Generically, matrix elements of $E,F,q^H$ are now some functions of $q$. With the modified commutation relations one is forced to modify the coproduct, such that \eqref{compatible} is satisfied
\begin{equation}\label{coprod}
    \Delta(E) = E \otimes 1 + q^{-H} \otimes E\, , \qquad
    \Delta(F) = F \otimes q^{H} + 1 \otimes F\, , \qquad
    \Delta\left(q^{H}\right) = q^{H} \otimes q^{H}\ .
\end{equation}
This is not the unique way to define a deformed coproduct, however, all the choices lead to equivalent algebras. Our choice appears to be natural in the Coulomb gas formalism \cite{paper2}. Note also that, instead of $H$, the more natural object appearing in the formulas is $q^H$, which itself behaves like a group-like element. One can check that with these definitions the Hopf algebra structure is preserved. 

In the text, we also use the following notations of $q$-deformed numbers and $q$-deformed factorials:
\begin{equation}\label{q_numbers}
 [ n]_q = \frac{q^n-q^{-n}}{q-q^{-1}}
 \end{equation}
 \begin{equation}
 [ n]! =  
 \begin{cases} 
      [n]_q [n-1]_q \ldots [1]_q & n\geq 1 \\
      1 & n = 0
\end{cases}
\end{equation}

\subsection{Finite-dimensional representations for \texorpdfstring{$q$}{q} not a root of unity}
\label{sec:irreps_generic}

In most of this work we will consider $q$ being an arbitrary complex number, but not a root of unity.\footnote{Representation theory for $q$ root of unity is more complicated, due to the fact that the quantum group is no longer semisimple. We will briefly discuss what happens for the case $q=e^{i \pi 3/4}$ in Section \ref{sec:irreps_unity}, which will allow us to make contact with the Ising model CFT. For more details about the representation theory of $U_q(sl_2)$ with $q$ a root of unity, we refer the reader to \cite{klimyk2012quantum}.} In this case, finite-dimensional irreducible representations of $U_q(sl_2)$ are very similar to those of $sl_2$. They are labeled by a half-integer spin $\ell$, and are $(2\ell+1)$ dimensional. We denote these representations as $\mathbb{V}_\ell$. All the states in the representation $\mathbb{V}_\ell$ can be labeled by $\ket{\ell,m}$, where $m \in \{-\ell,\dots,\ell\}$. A highest weight state in our notations is $\ket{\ell,\ell}$, and is annihilated by the raising generator E. All other states can be obtained by acting with the lowering generator F on the highest weight state. Then the action of generators can be written as
\begin{equation}\label{efh_action_canonical}
    \begin{split}
    q^{H} \cdot \ket{\ell,m} &= q^{2m} \ket{\ell,m}\\
    F \cdot \ket{\ell,m} &= \begin{cases}
        f^\ell_m \ket{\ell,m-1}, & \text{if} \;\; m \ne -\ell \\
        0, & \text{if} \;\; m = -\ell
    \end{cases} \\
    E \cdot \ket{\ell,m} &= \begin{cases}
        e^\ell_m \ket{\ell,m+1}, & \text{if} \;\; m \ne \ell \\
        0, & \text{if} \;\; m = \ell 
    \end{cases}
\end{split}
\end{equation}
The matrix elements $e^\ell_m$ and $f^\ell_m$ with the choice of coproduct \eqref{coprod} are given by\footnote{In principle, we still have the freedom of multiplying $e^\ell_m$ by a $\ell$ dependent constant and $f^\ell_m$ by its inverse. At the same time, half-integer powers of $q$ might arise in these and following equations. Throughout this work, we will be mostly interested in $q$ being a complex number, which requires a choice of branch cuts. Different choices of branch cuts amount to redefining $f_{\ell,m}$ and $e_{\ell, m}$ by this freedom.}
\begin{equation}\label{correct_fe}
    \begin{gathered}
        f^{\ell}_{m} = q^{m-1} \sqrt{[\ell+m]_q [\ell-m+1]_q} \\
        e^{\ell}_{m} = q^{-m} \sqrt{[\ell-m]_q [\ell+m+1]_q}
    \end{gathered}
\end{equation}

\begin{example}
To illustrate the formulas let us provide the explicit expression for $E,F,q^H$ in the spin-1/2 representation
\begin{equation}\label{spin1/2repr}
    E = \begin{pmatrix}
        0 & q^{1/2} \\
        0 & 0 
    \end{pmatrix}, \qquad F = \begin{pmatrix}
        0 & 0 \\
        q^{-1/2} & 0
    \end{pmatrix}, \qquad q^H = \begin{pmatrix}
        q & 0 \\
        0 & q^{-1}
    \end{pmatrix},
\end{equation}
and in the spin-1 representation
\begin{equation}
    E = \begin{pmatrix}
        0 & \sqrt{[2]_q} & 0 \\
        0 & 0 & q\sqrt{[2]_q} \\
        0 & 0 & 0
    \end{pmatrix}, \qquad F = \begin{pmatrix}
        0 & 0 & 0 \\
        \sqrt{[2]_q} & 0 & 0 \\
        0 & \frac{1}{q}\sqrt{[2]_q} & 0
    \end{pmatrix}, \qquad q^H = \begin{pmatrix}
        q^2 & 0 & 0 \\
        0 & 1 & 0 \\
        0 & 0 & q^{-2}
    \end{pmatrix}.
\end{equation}
\end{example} 

\subsubsection{Hermitian conjugation and the inner product} \label{sec:conjugation}
Now that we have introduced the ket states $\ket{\ell, m}$, we would like to define the bra states $\bra{\ell, m}$ as well. This, as usual, is done by hermitian conjugating the ket states. However, given that now our expressions will depend on $q$, and $q$ might be complex, the naive complex conjugation fails to give us a sufficiently nice inner product, where e.g. states belonging do different representations of $U_q(sl_2)$ are orthogonal.

Therefore, a different version of Hermitian conjugation is used in this theory \cite{Pasquier:1989kd}. Let us introduce $*$, which transposes and conjugates ket-vectors, with the exception that $q$ is treated as a formal parameter, that is it is not complex conjugated, even when $q$ is not real. $*$ allows us to define the  inner product, which is the usual  sesquilinear form for real $q$. Once the product is computed, we analytically continue in $q$ the final answer. The same operation of complex conjugation acts on the generators of $U_q(sl_2)$, and it has to be compatible with the coproduct
\begin{equation}
    \Delta(x^*) = \Delta(x)^*, \qquad \forall \; x \in U_q(sl_2).
\end{equation}
This is satisfied if we choose\footnote{
The most general choice would be $F^*=\gamma E q^H, E^*=\gamma^{-1} q^{-H} F$ with some non-zero constant $\gamma$. This choice would change the matrix elements \eqref{efh_action_canonical} by some constant.}
\begin{equation}\label{involution}
    F^*= E q^H, \qquad E^* = q^{-H}F, \qquad (q^H)^* = q^H.
\end{equation}
We can check indeed that
\begin{equation}
\begin{aligned}
    \Delta(F^*) &= \Delta(E) \Delta(q^H) = ( E q^H) \otimes q^H + 1 \otimes(Eq^H ) = \Delta(F)^*, \\
    \Delta(E^*) &= \Delta(q^{-H})\Delta(F) = (q^{-H}F) \otimes 1 + q^{-H} \otimes (q^{-H}F) = \Delta(E)^*.
\end{aligned}
\end{equation}
Note that, with the choice of matrix elements \eqref{correct_fe}, Hermitian conjugation $*$ is just a transposition: $F^* = F^T$, $E^* = E^T$.\footnote{We will work with $q$ on the unit circle, but this hermitian conjugation plays an important role for $q \in \mathbb{R}$, where the map $*:U_q(sl_2) \to U_q(sl_2)$ is an involution which makes $U_q(sl_2)$ a Hopf $*$-algebra. Such a map defines a real form of the quantum group. There are two inequivalent choices: $\gamma = 1$ leads to the compact real form $U_q(su_2)$, while $\gamma=-1$ leads to the real form $U_q(su_{1,1})$. }

\subsubsection{Tensor product of representations and quantum Clebsch-Gordan coefficients}
When $q$ is not a root of unity, any finite dimensional representation of $U_q(sl_2)$ is completely reducible, that is, it can be decomposed into a direct sum of irreducible representations. The tensor product of finite dimensional irreducible representations is decomposed into a direct sum over another irreducible representations with the same rule as for $sl_2$
\begin{equation}\label{tensor_prod_generic}
    \mathbb{V}_{\ell_1} \otimes \mathbb{V}_{\ell_2} = \bigoplus_{\ell = |\ell_1-\ell_2|}^{\ell_1+\ell_2} \mathbb{V}_{\ell}\,,
\end{equation}
yet the action of quantum group generators on the tensor product is given by coproduct \eqref{coprod}. The inner product, when we treat $q$ as a formal parameter, allows to define an orthonormal basis for irreducible representations in the tensor product. Let $\ket{\ell,m} \in \mathbb{V}_{\ell}$ denote orthonormal basis states, then we can write
\begin{equation}\label{def:QCG1}
    \ket{\ell,m} = \sum_{m_1,m_2} \QCG{\ell_1,m_1,\ell_2,m_2,\ell,m,q} \ket{\ell_1,m_1} \otimes \ket{\ell_2,m_2}
\end{equation}
which is the definition of the quantum Clebsch-Gordan coefficients. Canonically defined, Clebsch-Gordan coefficients are the same for any choice of coproduct once the inner product and matrix elements of $E$ and $F$ are compatible with it. The matrix of Clebsch-Gordan coefficients is orthogonal, thus the inverse formula has the same coefficients:
\begin{equation}\label{def:QCG}
    \ket{\ell_1,m_1} \otimes \ket{\ell_2,m_2} = \sum_{\ell=|\ell_1-\ell_2|}^{\ell_1+\ell_2} \sum_{m=-\ell}^{\ell} \QCG{\ell_1,m_1,\ell_2,m_2,\ell,m,q} \ket{\ell,m}
\end{equation}
This decomposition will play a role in Section \ref{sec:results}, when we discuss the OPE in the CFT context.

\begin{example}
Let us consider the simplest example, $\ell_1=\ell_2=\frac{1}{2}$, and explicitly compute all the Clebsch-Gordan coefficients. We know that
\begin{equation}
    \mathbb{V}_{\frac{1}{2}} \otimes \mathbb{V}_{\frac{1}{2}} = \mathbb{V}_{1} \oplus \mathbb{V}_{0}
\end{equation}
Let us denote the states in spin-1/2 representation as $\ket{+} = \ket{\frac{1}{2},\frac{1}{2}}$ and $\ket{-} = \ket{\frac{1}{2},-\frac{1}{2}}$. Using the coproduct \eqref{coprod} together with \eqref{correct_fe} we get the following decomposition into irreducible representations \begin{equation} \label{eq:1/2x1/2}
    \begin{split}
        &\ket{1,1} = \ket{+}\otimes\ket{+}, \\
&\ket{1,0} = \frac{1}{\sqrt{1+q^2}} \left( \ket{+}\otimes\ket{-} + q \ket{-}\otimes\ket{+}\right) \qquad \ket{0,0} = \frac{1}{\sqrt{1+q^2}} \left(q \ket{+}\otimes\ket{-} -\ket{-}\otimes\ket{+} \right)\\
        &\ket{1,-1} = \ket{-}\otimes\ket{-}
    \end{split}
\end{equation}
From this example, it can be checked why we defined Hermitian conjugation the way we did in section \ref{sec:conjugation}. Conjugation does not affect $q$, meaning that we have, for example,
\begin{equation}
    \bra{1,0} = \frac{1}{\sqrt{1+q^2}} \bra{+}\otimes\bra{-} + \frac{q}{\sqrt{1+q^2}} \bra{-}\otimes\bra{+}
\end{equation}
and it follows that these states are orthonormal.
Using the definition \eqref{def:QCG1}, we can read the relevant quantum Clebsch-Gordan coefficients from equations \eqref{eq:1/2x1/2}.
\end{example}

Luckily, a general formula for the quantum Clebsch-Gordan coefficients in the case of $U_{q}(sl_2)$ is known.
Remembering the definitions \eqref{q_numbers}, the Clebsch-Gordan coefficients are \cite{QCG}\footnote{In principle, the representation \eqref{efh_action_canonical} and the sesquilinear form defined in section \ref{sec:conjugation} allow an $\ell$-dependent ambiguity $\ket{\ell,m}\rightarrow e^{i\varphi_\ell}\ket{\ell,m}$, which would appear as a $q$-independent phase in the quantum Clebsch-Gordan coefficients. We fix this ambiguity by requiring the quantum Clebsch-Gordan coefficients to be real at real values of $q$, and therefore to be invariant under the conjugation defined in section \ref{sec:conjugation}.}
\begin{equation}\label{qcg_formula}
\begin{aligned}
\QCG{\ell_1,m_1,\ell_2,m_2,\ell_3,m_3,q}
=& \; \delta_{m_1+m_2,m_3} \Delta(\ell_1,\ell_2,\ell_3) q^{\frac 12 (\ell_1+\ell_2-\ell_3)(\ell_1+\ell_2+\ell_3+1)+(\ell_1 m_2 - \ell_2 m_1)}\\
& \left([\ell_1-m_1]![\ell_1+m_1] ![\ell_2-m_2]![\ell_2+m_2]! [\ell_3-m_3]![\ell_3+m_3]! [2\ell_3+1]_q\right)^{1/2}\\
&\sum_{r \in \mathcal{W}} \bigg[ \frac{(-1)^r q^{-r(\ell_1+\ell_2+\ell_3+1)}}{[r]! [\ell_1+\ell_2-\ell_3-r]![\ell_1-m_1-r]! [\ell_2+m_2-r]!}  \\
&\qquad \qquad \qquad \qquad \qquad  \frac{1}{ [\ell_3-\ell_2+m_1+r]! [\ell_3-\ell_1-m_2+r]!}\bigg]
\end{aligned}
\end{equation}
with $\mathcal{W}$ defined so that all factorials have non-negative integer arguments, $\mathcal{W} = \{x | \max (0,-(\ell_3-\ell_2+m_1),-(\ell_3-\ell_1-m_2)) \le x \le \min (\ell_1+\ell_2-\ell_3,\ell_1-m_1,\ell_2+m_2), x \in \mathbb{Z} \}$, and
\begin{equation}
    \Delta(a,b,c) = \left( \frac{[a+b-c]![a-b+c]![-a+b+c]!}{[a+b+c+1]!}\right)^{1/2 }
\end{equation}
In particular, the Clebsch-Gordan coefficients take a very simple form for $\ell_3=m_3=0$. In this case, Clebsch-Gordan coefficients are non-zero only for $\ell_1=\ell_2=\ell$ and $m_1=-m_2=m$\begin{equation}
    \QCG{\ell,m,\ell,-m,0,0,q} =\frac{(-1)^{m-\ell} q^{m}}{\sqrt{[2\ell+1]_q}}\,. \label{eq:QCG_2pf}
\end{equation}

The Clebsch-Gordan coefficients satisfy two orthonormality relations, with the first one following from the orthonormality of \eqref{def:QCG1}
\begin{equation}\label{eq:QCG_orth_1}
    \braket{\ell',m'|\ell,m} = \sum_{m_1,m_2} \QCG{\ell_1,m_1,\ell_2,m_2,\ell,m,q} \QCG{\ell_1,m_1,\ell_2,m_2,\ell',m',q} = \delta_{\ell,\ell'} \delta_{m,m'}\,,
\end{equation}
and the second one following from the orthonormality of \eqref{def:QCG}
\begin{equation}\label{eq:QCG_orth_2}
\begin{aligned}
    \big(\bra{\ell_1,m'_1}\otimes\bra{\ell_2,m'_2}\big) \big(\ket{\ell_1,m_1}\otimes\ket{\ell_2,m_2}\big) &= \sum_{\ell=|\ell_1-\ell_2|}^{\ell_1+\ell_2} \sum_{m=-\ell}^{\ell} \QCG{\ell_1,m'_1,\ell_2,m'_2,\ell,m,q} \QCG{\ell_1,m_1,\ell_2,m_2,\ell,m,q} \\&= \delta_{m_1,m'_1} \delta_{m_2,m'_2}
\end{aligned}
\end{equation}

\subsection{\texorpdfstring{$R$}{R}-matrix}
\label{sec:Rmat}

When we will move on to QFTs with quantum group symmetry, we will have operators transforming under $U_q(sl_2)$. We will want to be able to permute these operators, and in order to do this we need to introduce the universal $R$-matrix.

The coproduct we introduced in \eqref{coprod} acts on two representations of $U_q(sl_2)$, and it acts differently on these two. We could have also defined the coproduct where we swap the two representations, e.g.
\begin{equation}
    (\tau \circ \Delta)(E) = E \otimes q^{-H} + 1 \otimes E
\end{equation}
where $\tau(x \otimes y) = y \otimes x$; one can check that $ (\tau \circ \Delta)$ is a good coproduct, in the sense that it's compatible with the commutation relations and is co-associative.\footnote{If $\Delta(E)$ acts on the tensor product representation $\mathbb{V}_i\otimes\mathbb{V}_j$, then also $(\tau\circ\Delta)(E)$ acts on $\mathbb{V}_i\otimes\mathbb{V}_j$. To be more explicit, if $\Delta(E) = E_i\otimes \mathbb{1}_j +q^{-H_i} \otimes E_j$, then $(\tau \circ \Delta)(E)=E_i\otimes q^{-H_j}+\mathbb{1}_i \otimes E_j$, where the subscript indicates the representation of the generator.} The object that takes us from one coproduct to the other is the universal $R$-matrix, an invertible element in $U_q(sl_2) \otimes U_q(sl_2)$, which is defined by the following properties
\begin{equation}\label{def:Rmatrix}
    \begin{aligned}
        \mathcal{R} \; \Delta(x) \; \mathcal{R}^{-1} &= (\tau \circ \Delta)(x) \\
        (\Delta \otimes 1) (\mathcal{R}) &= \mathcal{R}_{13} \mathcal{R}_{23} \\
        (1 \otimes \Delta) (\mathcal{R}) &= \mathcal{R}_{13} \mathcal{R}_{12}
    \end{aligned}
\end{equation}
$\mathcal{R}_{ij}$ is the $R$-matrix split between $i$-th and $j$-th copy of $U_q(sl_2)$, e.g. if $R$-matrix contains a term $x \otimes y$ then this notation means $(x \otimes y)_{13}\equiv x\otimes1\otimes y$. Similarly, the LHS of the second line is defined via $(\Delta \otimes 1)(x \otimes y) = \Delta(x) \otimes y$ (and similarly for the third line). In QFTs with quantum group symmetry which we discuss in section \ref{sec:results}, swapping coproducts means that we are permuting operators, and $\calR_{ij}$ permutes operators $i$ and $j$. The axioms above imply that $R$-matrix satisfies the quantum Yang-Baxter equation \begin{equation}
    \mathcal{R}_{12} \mathcal{R}_{13} \mathcal{R}_{23} = \mathcal{R}_{23} \mathcal{R}_{13} \mathcal{R}_{12}
\end{equation}
Given the defining properties \eqref{def:Rmatrix}, there are two solutions for $\mathcal{R}$ in $U_q(sl_2)$. The first one is given by\footnote{Equivalently, the formula for the universal $R$-matrix can be written as
\begin{equation}
    \mathcal{R} = q^{\frac{H\otimes H}{2}}\sum_{n=0}^{\infty}\frac{(q-q^{-1})^n\,q^{\frac{1}{2}n(n-1)}}{[n]_q!}(Eq^H)^n\otimes(q^{-H}F)^n = q^{\frac{H\otimes H}{2}}\sum_{n=0}^{\infty}\frac{(q-q^{-1})^n\,q^{\frac{1}{2}n(n-1)}}{[n]_q!}(q^{nH}E^n)\otimes(F^n q^{-nH})
\end{equation}
}
\begin{equation}\label{Rmatrix:option1}
    \mathcal{R}=\left( \sum_{n=0}^{\infty} \frac{(q-q^{-1})^n\,q^{\frac{1}{2}n(n-1)}}{[n]_q!} E^n\otimes F^n \right) q^{\frac{H\otimes H}{2}}\,.
\end{equation}
It can be checked that if $\mathcal{R}$ satisfies the properties \eqref{def:Rmatrix}, then also $\widetilde{\mathcal{R}}=\tau ( \mathcal{R})^{-1}$ does, which then gives a second possible $R$-matrix \begin{equation}\label{Rmatrix:option2}
    \widetilde{\mathcal{R}}=q^{-\frac{H\otimes H}{2}}\,\left( \sum_{n=0}^{\infty} \frac{(q^{-1}-q)^n\,q^{-\frac{1}{2}n(n-1)}}{[n]_q!} F^n\otimes E^n \right).
\end{equation}
We will focus on $\mathcal{R}$, but the following arguments also apply to $\widetilde{\mathcal{R}}$ in a similar way. 

Since the $R$-matrix is an element of $U_q(sl_2)\otimes U_q(sl_2)$, instead of acting on a single representation, it acts on the tensor product of two representations. We denote with $\mathcal{R}$ the universal $R$-matrix as an element of $U_q(sl_2) \otimes U_q(sl_2)$, and with $R_{\ell_1,\ell_2}$ the $R$-matrix evaluated on the tensor product of representations $\mathbb{V}_{\ell_1} \otimes \mathbb{V}_{\ell_2}$. The matrix elements of $R_{\ell_1,\ell_2}$ are defined as
\begin{equation}
    [R_{\ell_1,\ell_2}]_{m_1,m_2}^{m'_1,m'_2} \equiv \bra{\ell_1,m'_1} \otimes \bra{\ell_2,m'_2} R_{\ell_1,\ell_2} \ket{\ell_1,m_1} \otimes \ket{\ell_2,m_2}
\end{equation}

Finally, let us mention some special values of $q$: when $q \to 1$, $\mathcal{R}$ reduces to the identity operator. When $q \to -1$, then $R_{\ell_1,\ell_2}$ reduces to the identity operator when both $\ell_1,\ell_2$ are integers, this is the case for our operators in section \ref{section:example}.

\begin{example}
Let us provide an expression for the $R$-matrix evaluated on the tensor product $\mathbb{V}_{\frac{1}{2}} \otimes \mathbb{V}_{\frac{1}{2}}$. To conveniently represent the $R$-matrix, we treat a pair of indices $(m_1,m_2)$ as a single index $i$ and another pair $(m'_1,m'_2)$ as a single index $j$. So that $R$-matrix can be represented as:
\begin{equation}
    R_{\frac{1}{2}, \frac{1}{2}} = \left(
\begin{array}{cccc}
 \sqrt{q} & 0 & 0 & 0 \\
 0 & \frac{1}{\sqrt{q}} & \frac{q^2-1}{q^{3/2}} & 0 \\
 0 & 0 & \frac{1}{\sqrt{q}} & 0 \\
 0 & 0 & 0 & \sqrt{q} \\
\end{array}
\right)
\end{equation}
Similarly, an expression for the $R$-matrix evaluated on the tensor product $\mathbb{V}_1 \otimes \mathbb{V}_1$:
\begin{equation}
    R_{1,1} = \left(
\begin{array}{ccccccccc}
 q^2 & 0 & 0 & 0 & 0 & 0 & 0 & 0 & 0 \\
 0 & 1 & 0 & \frac{q^4-1}{q^2} & 0 & 0 & 0 & 0 & 0 \\
 0 & 0 & \frac{1}{q^2} & 0 & q-\frac{1}{q^3} & 0 &
   q^2-\frac{1}{q^2}+\frac{1}{q^4}-1 & 0 & 0 \\
 0 & 0 & 0 & 1 & 0 & 0 & 0 & 0 & 0 \\
 0 & 0 & 0 & 0 & 1 & 0 & q-\frac{1}{q^3} & 0 & 0 \\
 0 & 0 & 0 & 0 & 0 & 1 & 0 & \frac{q^4-1}{q^2} & 0 \\
 0 & 0 & 0 & 0 & 0 & 0 & \frac{1}{q^2} & 0 & 0 \\
 0 & 0 & 0 & 0 & 0 & 0 & 0 & 1 & 0 \\
 0 & 0 & 0 & 0 & 0 & 0 & 0 & 0 & q^2 \\
\end{array}
\right)
\end{equation}
\end{example} 
For what concerns the other choice of $R$-matrix, it can be checked that $\widetilde{R}_{\ell_1,\ell_2} = R_{\ell_1,\ell_2}^T(q^{-1})$.

\subsection{Quantum \texorpdfstring{$6j$}{6j}-symbols}
\label{sec:6j}
Last but not least, we review $6j$-symbols; these will play a role in crossing symmetry of the four point function, and will relate e.g. the Virasoro blocks in the s-channel to those in the t-channel.
Let us consider a tensor product of three representations $\mathbb{V}_{\ell_1} \otimes \mathbb{V}_{\ell_2} \otimes \mathbb{V}_{\ell_3}$. First, recall that
\begin{equation}
    \begin{aligned}
        \mathbb{V}_{\ell_1} \otimes \mathbb{V}_{\ell_2} = \bigoplus_{\ell_{12}} d_{\ell_{12}}^{\ell_1,\ell_2} \; \mathbb{V}_{\ell_{12}} = \bigoplus_{\ell_{12}} \mathbb{W}_{\ell_{12}}^{\ell_1,\ell_2} \otimes \mathbb{V}_{\ell_{12}}\\
        \mathbb{V}_{\ell_2} \otimes \mathbb{V}_{\ell_3} = \bigoplus_{\ell_{23}} d_{\ell_{23}}^{\ell_2,\ell_3} \; \mathbb{V}_{\ell_{23}} = \bigoplus_{\ell_{23}} \mathbb{W}_{\ell_{23}}^{\ell_2,\ell_3} \otimes \mathbb{V}_{\ell_{23}}
    \end{aligned}
\end{equation}
where $d_{\ell}^{\ell_1,\ell_2}$ is the number of the irreducible representations $\mathbb{V}_{\ell}$ appearing in the tensor product. In the case of $U_q(sl_2)$, for $q$ not root of unity, the formula above is nothing but \eqref{tensor_prod_generic} where all the degeneracies are equal to 1. However, we want to keep the definition of $6j$-symbols generic. $d_{\ell}^{\ell_1,\ell_2}$ copies of the vector space $\mathbb{V}_{\ell}$ can be rewritten as a tensor product $\mathbb{W}_{\ell}^{\ell_1,\ell_2} \otimes \mathbb{V}_{\ell}$. Returning to the tensor product of three representations, there are two different ways of computing it:
\begin{equation}
    \begin{aligned}
        (\mathbb{V}_{\ell_1} \otimes \mathbb{V}_{\ell_2}) \otimes \mathbb{V}_{\ell_3} = \bigoplus_{\ell_{12},\ell} \mathbb{W}_{\ell_{12}}^{\ell_1,\ell_2} \otimes \mathbb{W}_{\ell}^{\ell_{12},\ell_3} \otimes \mathbb{V}_{\ell} \\
        \mathbb{V}_{\ell_1} \otimes (\mathbb{V}_{\ell_2} \otimes \mathbb{V}_{\ell_3}) = \bigoplus_{\ell_{23},\ell} \mathbb{W}_{\ell}^{\ell_1,\ell_{23}} \otimes \mathbb{W}_{\ell_{23}}^{\ell_2,\ell_3} \otimes \mathbb{V}_{\ell}
    \end{aligned}
\end{equation}
Since the tensor product is associative, there exists an isomorphism between two ways of fusion:
\begin{equation}
    U \begin{bmatrix}
        \ell_1 & \ell_2 \\
        \ell_3 & \ell
    \end{bmatrix}: \bigoplus_{\ell_{12}} \mathbb{W}_{\ell_{12}}^{\ell_1,\ell_2} \otimes \mathbb{W}_{\ell}^{\ell_{12},\ell_3} \to \bigoplus_{\ell_{23}} \mathbb{W}_{\ell}^{\ell_1,\ell_{23}} \otimes \mathbb{W}_{\ell_{23}}^{\ell_2,\ell_3}
\end{equation}
The matrix $U$ is called the Racah matrix, and its matrix elements are called $6j$-symbols\footnote{
In some literature, the matrix elements of Racah matrix are called Racah coefficients, while $6j$-symbols are slightly different because of a different normalization. We follow the notation of \cite{QCG}, and call these objects $6j$-symbols. }
\begin{equation}\label{def:6j}
    \sixj{\ell_1,\ell_2,\ell_{12},\ell_3,\ell,\ell_{23}} =  U_{\ell_{23},\ell_{12}} \begin{bmatrix}
        \ell_1 & \ell_2 \\
        \ell_3 & \ell
    \end{bmatrix}
\end{equation}

\begin{example}
    Let us now compute $6j$-symbols directly from the definition in the simplest case when $\ell_1=\ell_2=\ell_3=\frac{1}{2}$. The associativity of tensor multiplication implies that
    \begin{equation}
        (\mathbb{V}_{\frac{1}{2}} \otimes \mathbb{V}_{\frac{1}{2}}) \otimes \mathbb{V}_{\frac{1}{2}} = \mathbb{V}_{\frac{1}{2}} \otimes (\mathbb{V}_{\frac{1}{2}} \otimes \mathbb{V}_{\frac{1}{2}}) = \mathbb{V}_{\frac{3}{2}} \otimes 2 \; \mathbb{V}_{\frac{1}{2}}
    \end{equation}
    
    Let us consider the first way of tensor multiplication. In this case we have $(\mathbb{V}_{\frac{1}{2}} \otimes \mathbb{V}_{\frac{1}{2}}) \otimes \mathbb{V}_{\frac{1}{2}} = (\mathbb{V}_{1} \otimes \mathbb{V}_{0}) \otimes \mathbb{V}_{\frac{1}{2}}$. The double degeneracy of spin-1/2 representation in the final result means that there are exactly 2 highest weight vectors of spin-1/2 representation. The first one is coming from multiplication of $\mathbb{V}_{0} \otimes \mathbb{V}_{\frac{1}{2}}$:
    \begin{equation}
        v_1 = \ket{0,0} \otimes \ket{+} = \QCG{\frac{1}{2},\frac{1}{2},\frac{1}{2},-\frac{1}{2},0,0,q} \ket{+} \otimes \ket{-} \otimes \ket{+} + \QCG{\frac{1}{2},-\frac{1}{2},\frac{1}{2},\frac{1}{2},0,0,q} \ket{-} \otimes \ket{+} \otimes \ket{+}
    \end{equation}
    where we use the same notations as in previous examples for vectors in spin-1/2 representations: $\ket{+} = \ket{\frac{1}{2},\frac{1}{2}}$ and $\ket{-} = \ket{\frac{1}{2},-\frac{1}{2}}$. The second one is coming from the multiplication of $\mathbb{V}_{1} \otimes \mathbb{V}_{\frac{1}{2}}$:
    \begin{equation}
    \begin{aligned}
        v_2 &= \QCG{1,1,\frac{1}{2},-\frac{1}{2},\frac{1}{2},\frac{1}{2},q} \ket{1,1} \otimes \ket{-} + \QCG{1,0,\frac{1}{2},\frac{1}{2},\frac{1}{2},\frac{1}{2},q} \ket{1,0} \otimes \ket{+} \\
        &= \QCG{1,1,\frac{1}{2},-\frac{1}{2},\frac{1}{2},\frac{1}{2},q} \ket{+} \otimes \ket{+} \otimes \ket{-} + \QCG{1,0,\frac{1}{2},\frac{1}{2},\frac{1}{2},\frac{1}{2},q} \QCG{\frac{1}{2},\frac{1}{2},\frac{1}{2},-\frac{1}{2},1,0,q} \ket{+} \otimes \ket{-} \otimes \ket{+} + \\
        &+\QCG{1,0,\frac{1}{2},\frac{1}{2},\frac{1}{2},\frac{1}{2},q} \QCG{\frac{1}{2},-\frac{1}{2},\frac{1}{2},\frac{1}{2},1,0,q} \ket{-} \otimes \ket{+} \otimes \ket{+}
    \end{aligned}
    \end{equation}
    Both $v_1,v_2$ are annihilated by $E$, have a total $U(1)$ charge equal 1/2, orthogonal to each other and have a norm equal 1. What we have just computed are the vectors
    \begin{equation}
        v_1 \in \mathbb{W}_{0}^{\frac{1}{2},\frac{1}{2}} \otimes \mathbb{W}_{\frac{1}{2}}^{0,\frac{1}{2}}, \qquad v_2 \in \mathbb{W}_{1}^{\frac{1}{2},\frac{1}{2}} \otimes \mathbb{W}_{\frac{1}{2}}^{1,\frac{1}{2}}
    \end{equation}

    Equivalently, we deal with the second way of tensor multiplication: $\mathbb{V}_{\frac{1}{2}} \otimes (\mathbb{V}_{\frac{1}{2}} \otimes \mathbb{V}_{\frac{1}{2}}) = \mathbb{V}_{\frac{1}{2}} \otimes (\mathbb{V}_{1} \otimes \mathbb{V}_{0})$. As before, we compute the two highest weight vectors of spin-1/2 representations appearing in the final result. The first one is coming from multiplication of $\mathbb{V}_{\frac{1}{2}} \otimes \mathbb{V}_{0}$:
    \begin{equation}
        w_1 = \ket{+} \otimes \ket{0,0} = \QCG{\frac{1}{2},\frac{1}{2},\frac{1}{2},-\frac{1}{2},0,0,q} \ket{+} \otimes \ket{+} \otimes \ket{-} + \QCG{\frac{1}{2},-\frac{1}{2},\frac{1}{2},\frac{1}{2},0,0,q} \ket{+} \otimes \ket{-} \otimes \ket{+}
    \end{equation}
    The second one is coming from multiplication of $\mathbb{V}_{\frac{1}{2}} \otimes \mathbb{V}_{1}$:
    \begin{equation}
    \begin{aligned}
        w_2 =& \QCG{\frac{1}{2},-\frac{1}{2},1,1,\frac{1}{2},\frac{1}{2},q} \ket{-} \otimes \ket{1,1} + \QCG{\frac{1}{2},\frac{1}{2},1,0,\frac{1}{2},\frac{1}{2},q} \ket{+} \otimes \ket{1,0} \\
        =& \QCG{\frac{1}{2},-\frac{1}{2},1,1,\frac{1}{2},\frac{1}{2},q} \ket{-} \otimes \ket{+} \otimes \ket{+} + \QCG{\frac{1}{2},\frac{1}{2},1,0,\frac{1}{2},\frac{1}{2},q} \QCG{\frac{1}{2},\frac{1}{2},\frac{1}{2},-\frac{1}{2},1,0,q} \ket{+} \otimes \ket{+} \otimes \ket{-}  \\
        &+\QCG{\frac{1}{2},\frac{1}{2},1,0,\frac{1}{2},\frac{1}{2},q} \QCG{\frac{1}{2},-\frac{1}{2},\frac{1}{2},\frac{1}{2},1,0,q} \ket{+} \otimes \ket{-} \otimes \ket{+}
    \end{aligned}
    \end{equation}
    Again, both $w_1,w_2$ are annihilated by $E$, have a total $U(1)$ charge equal 1/2, orthogonal to each other and have a norm equal 1. What we have just computed are the vectors 
    \begin{equation}
        w_1 \in \mathbb{W}_{\frac{1}{2}}^{\frac{1}{2},0} \otimes \mathbb{W}_{0}^{\frac{1}{2},\frac{1}{2}}, \qquad w_2 \in \mathbb{W}_{\frac{1}{2}}^{\frac{1}{2},1} \otimes \mathbb{W}_{1}^{\frac{1}{2},\frac{1}{2}}
    \end{equation}

    From the equivalence of the vector spaces there must be a transformation $U$ between these two ways of obtaining highest weight vectors:
    \begin{equation}
        \begin{pmatrix}
            w_1 \\
            w_2
        \end{pmatrix} = U \begin{pmatrix}
            v_1 \\
            v_2
        \end{pmatrix}
    \end{equation}
    Indeed, the transformation is given by
    \begin{equation}
            w_1 = \frac{-q}{1+q^2} \; v_1 + \frac{\sqrt{1+q^2+q^4}}{1+q^2} \; v_2, \qquad w_2 = \frac{\sqrt{1+q^2+q^4}}{1+q^2} \; v_1 + \frac{q}{1+q^2} \; v_1,
    \end{equation}
    from which we conclude that
    \begin{equation}
        \begin{aligned}
            \sixj{\frac{1}{2},\frac{1}{2},0,\frac{1}{2},\frac{1}{2},0} = \frac{-q}{1+q^2}, \qquad \qquad  \sixj{\frac{1}{2},\frac{1}{2},1,\frac{1}{2},\frac{1}{2},1} = \frac{q}{1+q^2}\,,\\
            \sixj{\frac{1}{2},\frac{1}{2},1,\frac{1}{2},\frac{1}{2},0} = \sixj{\frac{1}{2},\frac{1}{2},0,\frac{1}{2},\frac{1}{2},1} = \frac{\sqrt{1+q^2+q^4}}{1+q^2}.
        \end{aligned}
    \end{equation}

\end{example}

In the case of $U_q(sl_2)$ there exists a general formula to compute quantum $6j$-symbols \cite{Kirillov:1991ec} \begin{equation}
\begin{aligned}
    \sixj{\ell_1,\ell_2,\ell_{12},\ell_3,\ell,\ell_{23}} =\,& (-1)^{\ell_1+\ell_2+\ell_3+\ell}\sqrt{[2\ell_{12}+1]_q [2\ell_{23}+1]_q}\,\\ & \Delta(\ell_1,\ell_2,\ell_{12})\Delta(\ell_3,\ell,\ell_{12})\Delta({\ell_1,\ell,\ell_{23}})\Delta(\ell_2,\ell_3,\ell_{23})  \\ 
   &  \sum_{z \in \mathcal{W}'}  \bigg[ \frac{(-1)^z [z+1]!}{[z-\ell_1-\ell_2-\ell_{12}]![z-\ell_3-\ell-\ell_{12}]![z-\ell_1-\ell-\ell_{23}]![z-\ell_2-\ell_3-\ell_{23}]!} \\
   & \qquad \quad\frac{1}{[\ell_1+\ell_2+\ell_3+\ell-z]![\ell_1+\ell_3+\ell_{12}+\ell_{23}-z]![\ell_2+\ell+\ell_{12}+\ell_{23}-z]!} \bigg]
\end{aligned}
\end{equation}
where $\mathcal{W}'$ is such that we only have factorials of non-negative integers in the sum, i.e. $\mathcal{W}' = \{x | \max (\ell_1+\ell_2+\ell_{12}, \ell_3+\ell+\ell_{12}, \ell_1+\ell+\ell_{23}, \ell_2+\ell_3+\ell_{23}) \le x \le \min (\ell_1+\ell_2+\ell_3+\ell, \ell_1+\ell_3+\ell_{12}+\ell_{23}, \ell_2+\ell+\ell_{12}+\ell_{23}), x \in \mathbb{Z} \}$.

The relation between $6j$-symbols and Clebsch Gordan coefficients is \cite{Ardonne:2010zu}
\begin{equation}\label{relation:6j3j}
\sixj{\ell_1,\ell_2,\ell_{12},\ell_3,\ell,\ell_{23}} = \sum_{\substack{m_1,m_2,m_3\\m_{12},m_{23}}} \QCG{\ell_1,m_1,\ell_2,m_2,\ell_{12},m_{12},q} \QCG{\ell_{12},m_{12},\ell_3,m_3,\ell,m,q}\QCG{\ell_2,m_2,\ell_3,m_3,\ell_{23},m_{23},q}\QCG{\ell_1,m_1,\ell_{23},m_{23},\ell,m,q}
\end{equation}
There is no sum on $m$, and the whole expression is independent on its value.

Similarly to Clebsch-Gordan coefficients, $6j$-symbols also satisfy orthogonality relations \cite{Kirillov:1991ec}
\begin{equation} \label{eq:6j_orthogonality}
\begin{gathered}
  \sum_i  \sixj{\ell_1,\ell_2,i,\ell_3,\ell_4,\ell} \sixj{\ell_1,\ell_2,i,\ell_3,\ell_4,\ell'} = \delta_{\ell,\ell'} \{\ell_1 \, \ell_4\,  \ell\}\{\ell_2\, \ell_3 \, \ell\} \\
    \sum_\ell  \sixj{\ell_1,\ell_2,i,\ell_3,\ell_4,\ell} \sixj{\ell_1,\ell_2,i',\ell_3,\ell_4,\ell} = \delta_{i,i'} \{\ell_1 \, \ell_2\,  i\}\{\ell_3\, \ell_4 \, i'\} 
  \end{gathered}
\end{equation}
where $\{i \, j\,  k\}$ is the triangular delta
\begin{equation}
    \{i \, j\,  k\} = \begin{cases} 1 \qquad \text{ if } |i-j|\le k \le i+j\\
0 \qquad \text{ otherwise}
    \end{cases}
\end{equation}

\section{QFT with Quantum Group Global Symmetry}\label{sec:results}
In this section we discuss the general properties of QFTs with a quantum group internal global symmetry. We work in two-dimensional Euclidean space and only consider the case of the quantum group \(U_q(sl_2)\).\footnote{In principle the generalization to any Drinfeld-Jimbo type of quantum groups is straightforward.} The statement that the theory has $U_q(sl_2)$ internal global symmetry means
\begin{enumerate}
    \item The $U_q(sl_2)$ generators commute with the space-time symmetry generators;\footnote{For massive QFT, the Coleman-Mandula theorem \cite{Coleman:1967ad} does not apply to theories with quantum group symmetry, and the quantum group can be part of the extension of the Poincaré group \cite{Bernard:1990ys}. For two-dimensional CFT, the Virasoro algebra also has various possible nontrivial extensions. Therefore, it can happen that the quantum group does not commute with the space-time symmetry group. Here we rule out such cases. } 
    \item All the QFT operators \(\mathcal{O}_i(x)\) in the correlation function transform linearly under the quantum group action: $\mathcal{O}_i(x) \longrightarrow X \cdot \mathcal{O}_i(x), \; X \in U_q(sl_2)$. The correlation functions of QFT operators satisfy the Ward identities of the quantum group.
\end{enumerate}
The above constraints provide a quantum group analogue of a well known statement about QFTs with usual symmetry group: correlation functions must be proportional to an invariant tensor under the symmetry group. Now we will explain how to derive $U_q(sl_2)$ Ward identities and provide a general expression for them. 

The action of a generator $X$ on products of operators is obtained by using the coproduct \eqref{coprod}. In the case of two operators \(\mathcal{O}_i(x)\) and \(\mathcal{O}_j(y)\), for example, the action of $X$ gives
\begin{equation}
        \mathcal{O}_i(x) \mathcal{O}_j(y) \xlongrightarrow{X} X\cdot \left( \mathcal{O}_i(x) \mathcal{O}_j(y)  \right) \equiv \Delta(X) \cdot \left( \mathcal{O}_i(x) \mathcal{O}_j(y) \right)\,.
\end{equation}
For multiple operators, the quantum group action is defined by iterating the coproduct:
\begin{equation}
    \mathcal{O}_{i_1}(x_1) \ldots \mathcal{O}_{i_n}(x_n) \xlongrightarrow{X}  X \cdot \left( \mathcal{O}_{i_1}(x_1) \ldots \mathcal{O}_{i_n}(x_n) \right) \equiv \Delta^{n-1}(X) \cdot \left( \mathcal{O}_{i_1}(x_1) \ldots \mathcal{O}_{i_n}(x_n) \right)\,.
\end{equation}
Since the coproduct $\Delta$ satisfies the coassociativity condition \eqref{coprod:coasso}, there is no ambiguity in the definition of \(\Delta^n(X)\). For convenience, we use the notation $X$ for the action of $\Delta^n(X)$. We also assume that the vacuum is in the trivial representation of the quantum group. With the given notations, the Ward identities are formulated as follows:
\begin{equation}\label{QFT:wardid}
    \big\langle X \cdot \left( \mathcal{O}_{i_1}(x_1) \mathcal{O}_{i_2}(x_2) \ldots \mathcal{O}_{i_n}(x_n) \right) \big\rangle = 0, \qquad \text{where} \; X = E,F,H.
\end{equation}
We will also assume that operators are in finite-dimensional representations (see Section \ref{sec:irreps_generic}), and therefore we label every operators by its quantum group quantum numbers $\mathcal{O}_{i,\ell_i,m_i}$. Each single operator transforms under the quantum group action according to the rules in \eqref{efh_action_canonical}, i.e.
\begin{equation}
    H\cdot\mathcal{O}_{i,\ell_i,m_i}=2m_i\,\mathcal{O}_{i,\ell_i,m_i},\quad E\cdot \mathcal{O}_{i,\ell_i,m_i}=e^{\ell_i}_{m_i}\mathcal{O}_{i,\ell_i,m_i+1},\quad F\cdot\mathcal{O}_{i,\ell_i,m_i}=f^{\ell_i}_{m_i}\mathcal{O}_{i,\ell_i,m_i-1}.
\end{equation}
For simplicity, we sometimes use the shorthand notation \(\mathcal{O}_i \equiv \mathcal{O}_{i,\ell_i,m_i}\) when the internal quantum numbers are unimportant.\footnote{We keep all three indices $(i,\ell_i,m_i)$ because generically there might be different operators which have the same quantum numbers $(\ell,m)$.} 

The most immediate Ward identity comes from considering the generator $H$
\begin{equation}\label{u1_charge_cons}
    \big\langle H \cdot \left( \mathcal{O}_{1,\ell_1,m_1}(x_1) \mathcal{O}_{2,\ell_2,m_2}(x_2) \ldots \mathcal{O}_{n,\ell_n,m_n}(x_n) \right) \big\rangle = 0\,.
\end{equation}
Remembering that $H \cdot \calO_{i,\ell_i,m_i} = 2m_i\,\calO_{i,\ell_i,m_i}$, it can be checked that this equation implies the usual $U(1)$ charge conservation: the correlation function is non-zero only if $m_1 + \ldots + m_n = 0$.

The remaining Ward identities can be derived by acting with $F$ on the operators with total charge equals to 1:
\begin{equation}\label{QFT:wardidSimple}
    \big\langle F \cdot \big( \mathcal{O}_{1,\ell_1,m_1}(x_1) \mathcal{O}_{2,\ell_2,m_2}(x_2) \ldots \mathcal{O}_{n,\ell_n,m_n}(x_n) \big) \big\rangle = 0, \qquad \text{for} \; m_1 + \ldots + m_n = 1\,.
\end{equation}
An equivalent set of constraints are obtained by acting with $E$ on the set of operators with the condition $m_1 + \ldots + m_n = -1$. Using that the matrix elements of $F$ are given by \eqref{efh_action_canonical}, the Ward identity \eqref{QFT:wardidSimple} can be written in a more explicit form as
\begin{equation}\label{wardid:Fexplicit}
    \sum_{i=1}^{n}q^{2(m_{i+1}+m_{i+2}+\ldots+m_n)}f_{m_i}^{\ell_i}\big\langle\calO_{1,\ell_1,m_1}(x_1) \ldots \calO_{i,\ell_i, m_i-1}(x_i)\ldots\calO_{n_,\ell_n,m_n}(x_n)\big\rangle=0.
\end{equation}
We can think of the Ward identity \eqref{QFT:wardidSimple} in the following way: for the correlation function to be non-zero, the combination of operators has to overlap with a trivial representation of the quantum group. When $m_1 + \ldots + m_n = 1$, then this combination is definitely not in a trivial representation and the action of $F$ cannot change the irreducible component, which results into zero overlap. 
\begin{example}
    The simplest example of a non-trivial Ward identity is the relation between two-point functions of operators in the spin-\(\frac{1}{2}\) representation of \(U_q(sl_2)\). Let us denote the operators in this representation as $\mathcal{O}_{\pm}$. Recalling our convention \eqref{spin1/2repr}, we know the action of quantum group generators on single operators:
    \begin{equation}
        \begin{split}
        q^H \cdot \mathcal{O}_{\pm} &= q^{\pm 1} \mathcal{O}_\pm, \\
        E \cdot \mathcal{O}_- &= q^{\frac{1}{2}} \mathcal{O}_+, \\
        F \cdot \mathcal{O}_+ &= q^{-\frac{1}{2}} \mathcal{O}_-, \\
        E \cdot \mathcal{O}_+ &= F \cdot \mathcal{O}_- = 0.
        \end{split}
    \end{equation}
    Then the Ward identity \eqref{QFT:wardidSimple} reads as
    \begin{equation}\label{wardid:spin1/2}
        \braket{F \cdot (\mathcal{O}_+(x) \mathcal{O}_+(y))} = q^{-\frac{1}{2}} \braket{\mathcal{O}_+(x) \mathcal{O}_-(y)} + q^{\frac{1}{2}} \braket{\mathcal{O}_-(x) \mathcal{O}_+(y)} = 0,
    \end{equation}
    which tells us that only one out of the two correlation functions is independent.
\end{example}

\subsection{What is different?}
\label{sec:difference}

Now we will explore the consequence of imposing a $U_q(sl_2)$ global symmetry in a QFT and discuss its peculiarities compared to a QFT with ordinary symmetries.
Let us remind the reader that Euclidean unitary QFTs and correlation functions of their local operators should satisfy the Euclidean QFT axioms of \cite{Osterwalder:1973dx}. The one that we are interested in is \begin{itemize}
\item[] \textbf{Locality} (also known as `permutation symmetry' or `local commutativity'): the correlation functions of mutually local operators are invariant under permutations of operators.
\end{itemize}
That is to say that mutually local operators always commute in Euclidean space.
However, the modern approach to QFT considers not just local operators but also non-local operators such as line and surface operators. These operators play a role in our QFTs: we will now see that the upshot of imposing $U_q(sl_2)$ symmetry in a QFT is that the operators $\calO_i(x)$ on which it acts non-trivially necessarily include non-local operators, because they cannot satisfy permutation symmetry.

The violation of the above locality condition can be quickly checked using the Ward identities on the commutator of two operators. Consider the operators transforming under $U_q(sl_2)$ and suppose, for contradiction, that the above locality condition holds. Acting on the commutator of \(\mathcal{O}_i(x)\) and \(\mathcal{O}_j(y)\) with the raising generator $E$ gives:
\begin{equation}\label{eq:nonlocal}
    \begin{split}
        0 = E \cdot ([\mathcal{O}_i(x), \mathcal{O}_j(y)]) =& \left(E \cdot \mathcal{O}_i(x)\right) \mathcal{O}_j(y) + \left(q^{-H} \cdot \mathcal{O}_i(x)\right) \left(E \cdot \mathcal{O}_j(y)\right) \\
        &- \left(E \cdot \mathcal{O}_j(y)\right) \mathcal{O}_i(x) - \left(q^{-H} \cdot \mathcal{O}_j(y)\right) \left(E \cdot \mathcal{O}_i(x)\right) \\
        =& \left(E \cdot \mathcal{O}_i(x)\right) \left(1 - q^{-H}\right) \mathcal{O}_j(y) \\
        &+ \left(q^{-H} - 1\right) \mathcal{O}_i(x) \left(E \cdot \mathcal{O}_j(y)\right).
    \end{split}
\end{equation}
In the first equality, we use the locality assumption \([\mathcal{O}_i(x), \mathcal{O}_j(y)] = 0\) and the linearity of the quantum group action. In the second equality, we expand the commutator and use the definition of the coproduct \eqref{coprod}. In the last equality, we use the locality assumption again.
For the Lie algebra \(sl_2\), where \(q = 1\), the right-hand side of \eqref{eq:nonlocal} vanishes. However, for the quantum group with generic \(q\), it generically does not vanish. This demonstrates the inconsistency between locality and quantum group symmetry. Therefore, in a $U_q(sl_2)$ symmetric theory the operators transforming under the quantum group cannot be mutually local in general.

\begin{example}
    The simplest example where we can see the violation of locality is the two-point function of two operators $\calO_{\pm}$ in the same spin-$\frac{1}{2}$ multiplet of $U_q(sl_2)$, which we considered in \eqref{wardid:spin1/2}. Let $s$ be the space-time spin of $\calO_{\pm}$ (they have the same spin because the quantum group commutes with the rotation group by the assumption of internal symmetry). Then \eqref{wardid:spin1/2} implies that
    \begin{equation}
    \braket{\mathcal{O}_+(x) \mathcal{O}_-(y)} = -q \braket{\mathcal{O}_-(x) \mathcal{O}_+(y)} = (-1)^{2s+1} q \braket{\mathcal{O}_-(y) \mathcal{O}_+(x)}.
\end{equation}
Here the first equality is the rewriting of \eqref{wardid:spin1/2}, and the second equality is a consequence of Ward identity of the rotation group.
If $s$ is not an integer nor half-integer, then locality is violated immediately.\footnote{When the locality condition holds, taking $y-x\rightarrow(y-x)e^{2\pi i}$ does not change the value of the two-point function $\braket{\calO_1(x)\calO_2(y)}$. This implies that $e^{4\pi is}=1$, i.e., $s$ must be integer or half-integer.} If $s$ is an integer or a half-integer, the above equation shows that \([ \mathcal{O}_+(x), \mathcal{O}_-(y) ] \neq 0\) for generic values of \(q\), which also violates locality. Therefore it is a consequence of the Ward identities that the operators in a $U_q(sl_2)$ symmetric theory generically will not be mutually local.
Later, in Section \ref{section:QFT2pt}, we will also see that the spin $s$ is generically non-integer and satisfies some $q$-dependent constraints.
\end{example}

The mildest way in which operators $\calO_i$ can be non-local and still transform non-trivially under the quantum group is if they are defect-ending operators; in other words, they are operators that are attached to topological lines. These lines are topological in the sense that they can be moved around freely, as long as they do not cross other operators or lines, without changing the value of correlation functions. Topological lines are important because they implement symmetry transformations, and they can end on defect-ending operators, which instead are not topological and live at some point $x$ \cite{Chang:2018iay}. An example of this operator is the disorder operator in the two-dimensional Ising model, which comes attached to a topological twist line \cite{Kadanoff:1970kz} and is not mutually local with the spin operator.\footnote{Often topological lines have stronger properties relating to cutting and joining the lines, arising from the fusion categories that the lines form. Such lines are called topological defect lines \cite{Chang:2018iay}. We do not study whether the lines introduced in this work satisfy these properties, and we leave this as a question for the future.} 

It is evident that defect-ending operators attached to topological lines do not generally satisfy permutation symmetry and, therefore, are not mutually local operators. This is because, when computing correlation functions, it is necessary to specify not only the positions of the operators but also the ordering of the lines. Consequently, an ordering convention must be adopted. We fix our convention for the $n$-point function as follows.

\begin{figure}
    \centering
    {\includegraphics[width=1.0\textwidth]{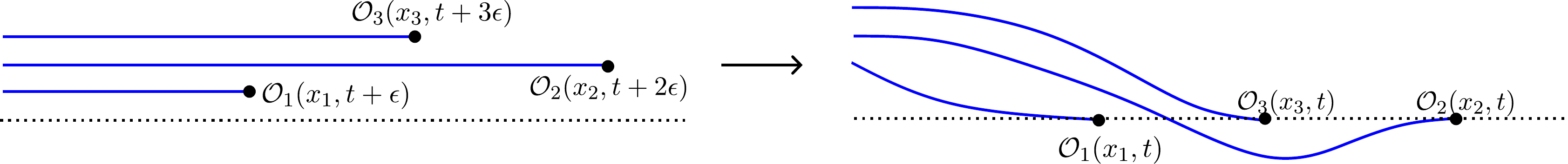} }
    \caption{$\epsilon$-prescription for the configuration $\calO_1(x_1,t) \calO_2(x_2,t)\calO_3(x_3,t)$ in the principal branch. }
    \label{fig:epsilonprescrption}
\end{figure}

Let us consider the $n$-point (Euclidean) equal-time configuration:
$$
(x_1,t),\ (x_2,t),\ \ldots\ ,\ (x_n,t).
$$
Here, the ordering of the spatial coordinates does not matter. The only requirement is that $x_i \neq x_j$ for all $i \neq j$. The $n$-point function in the \textbf{principal branch} is defined by
\begin{equation}
    \braket{\calO_1(x_1,t)\ldots \calO_n(x_n,t)} \equiv \lim_{\epsilon \to 0^+} \braket{\calO_1(x_1,t+\epsilon)\calO_2(x_2,t+2\epsilon) \ldots \calO_n(x_n,t+n \epsilon)}
\end{equation}
with the lines running horizontally to $x = -\infty$ (see figure \ref{fig:epsilonprescrption}). Note that the lines also approach both $\infty$ and the operators horizontally, given that the angle of approach in principle matters and should always be the same; however, in some of the following figures we will not keep track of it in order to make them look simpler.
For fixed $(x_1,x_2,\ldots,x_n)$, the non-equal-time $n$-point functions in the principal branch are obtained by continuously deforming the temporal variables.
For configurations where some spatial variables are equal, i.e., $x_i = x_j$, our principal-branch configuration is defined by approaching this configuration from $x_i < x_j$ with $i < j$. This convention establishes the entire principal branch.\footnote{A formal way to describe our principal branch is that $-\frac{\pi}{2}\leqslant\mathrm{Arg}[x_j-x_i+i(t_j-t_i)]<\frac{3\pi}{2}$ for all $i<j$.}

Other configurations are obtained by continuously deforming the configuration from the principal branch, with the lines remaining attached to the operators. During the deformation, the rule is that the lines must never cross each other. This procedure yields all possible $n$-point configurations $(z_1,z_2,\ldots,z_n)$ with $z_i \neq z_j$ for all $i \neq j$. Here we use the complex coordinates $z_i = x_i + i t_i$.

\begin{figure}
    \centering
    {\includegraphics[width=1.0\textwidth]{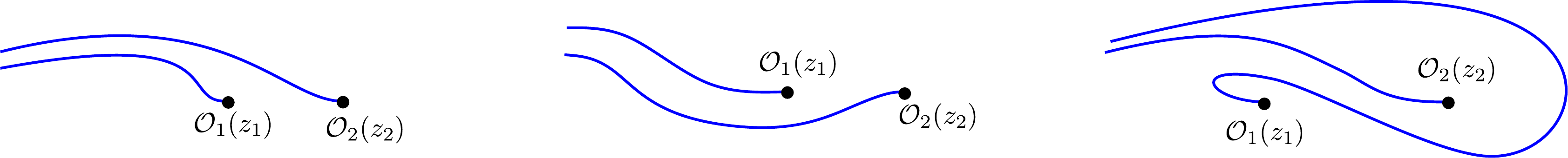} }
    \caption{Three different configurations of topological lines: $\braket{\calO_1(z_1) \calO_2(z_2) \ldots}$ (left), $\braket{\calO_2(z_2) \calO_1(z_1) \ldots}$ (middle) and $\braket{\calO_2(z_2)\calO_1(z_1)  \ldots}_{z_{21}\to z_{21}e^{-2\pi i}}$ (right). {All the lines are attached to the operators horizontally.} The topological lines terminating on other operators in the correlation function are not depicted.}
    \label{fig:ordering}
\end{figure}

To characterize the configurations, we introduce the notation $z_{ij} \to z_{ij}e^{\pm2\pi i}$, which indicates that $z_i$ circles $z_j$ clockwise ($e^{-2\pi i}$) or counterclockwise ($e^{2\pi i}$) but does not circle other points. A general $n$-point function is then denoted using subscripts, e.g.,
$$
\braket{\calO_1(z_1)\ldots\calO_n(z_n)}_{z_{21} \to z_{21}e^{2\pi i},\, z_{53}\to z_{53}e^{-4\pi i},\,\ldots },
$$
where the operations are performed from left to right. The correlation function without a subscript refers to the one in the principal branch. See figure \ref{fig:ordering} for examples.

Figure \ref{fig:ordering} illustrates that the operators do not commute, as the lines cannot generally be crossed without paying some cost. In Section \ref{section:locality}, we will postulate that line crossing can be accomplished using the $R$-matrix. In \cite{paper2}, we will provide a concrete example of constructing the topological lines attached to the operators $\calO_i$.

So far, we've been agnostic about the end point of the topological lines, and just drew them going to infinity horizontally. This will be enough for the study of correlation functions that we carry out in this work. However, this is not a completely satisfactory statement, because for example in a CFT the point at infinity is not a special point. In \cite{paper2} we show an explicit construction of correlation functions where the topological lines end in a point $w$, and show that the value of the correlation function is independent of $w$, and therefore it can be sent to infinity.\footnote{The appearence of point $w$, not sitting at infinity, might suggest that we can move between different orderings of correlation functions simply by moving the operators around. This is not the case: we remind the reader that when working with the topological lines, it's important to remember that the angles of approach are important, both for the points $z_i$ where operators sit, and for the point $w$. A consequence of this is, for example, that we cannot go from the two point function $\braket{\calO_1(z_1) \calO_2(z_2)}$ to $\braket{\calO_2(z_2) \calO_1(z_1)}$ by just rotating $z_2$ counterclockwise around point $z_1$ and $w$ (see the picture below).

    \begin{equation*}
    \includegraphics[width=0.8\linewidth]{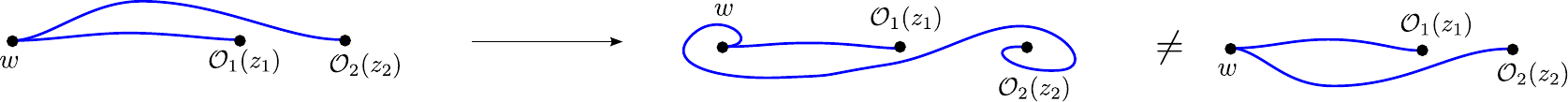}
    \end{equation*}
}

Another option, which is natural from the point of view of spin chains, is that the lines end on another defect line arising from some twisted boundary conditions of the spin chain.  In cases when that other defect line and the junctions where lines join are topological, the two descriptions are equivalent. The reason is that in this case we can freely move all of the junctions to a single point. Upon conformally mapping the cylinder to a plane, this point becomes exactly the special point mentioned in the first option.

\subsection{On the existence of currents }

\begin{figure}[h]
    \centering
    \includegraphics[width = .4 \textwidth]{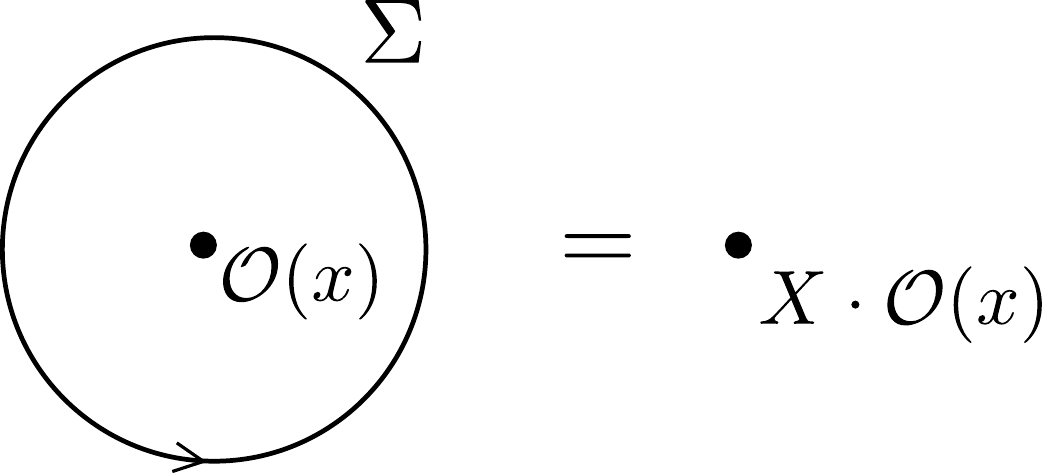}
    \caption{The action of an ordinary global symmetry on local operators.}
    \label{fig:symmetry_op}
\end{figure}

Now let us return to the question of how ordinary symmetries and quantum group symmetries act on operators. The modern approach to symmetries involves implementing a symmetry transformation by considering topological operators that act on operators \cite{Gaiotto:2014kfa}.
In the case of a theory with an ordinary continuous global symmetry,\footnote{In this paper, we consider only the 0-form symmetry as defined in \cite{Gaiotto:2014kfa}.} the action of the symmetry algebra on a local operator is often given by the integral of the local conserved currents: \begin{equation}\label{integral:symmetryop}
    (X\cdot\calO)(x) = \oint_{\Sigma}dy\,j(y)\,\calO(x),
\end{equation}
where the line \(\Sigma\) encircles the point \(x\). Due to the conservation of the current, \(\Sigma\) is topological. Specifically, we can shrink \(\Sigma\) to the point \(x\), as shown in Figure \ref{fig:symmetry_op}. This demonstrates that if $\calO$ is local, then $X \cdot \calO$ is also local. 

\begin{figure}[ht]
    \centering
    \includegraphics[width = .85\textwidth]{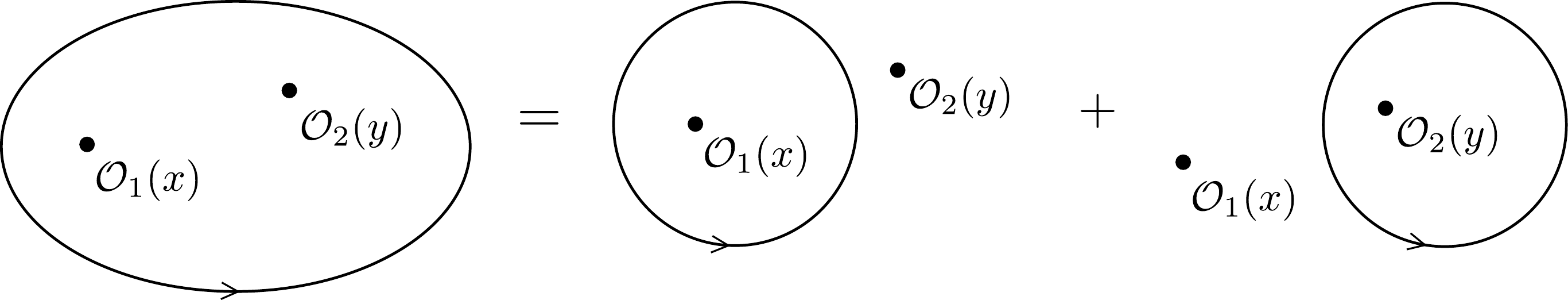}
    \caption{The coproduct $\Delta(X)=X\otimes1+1\otimes X$ for ordinary symmetries acting on local operators.}
    \label{fig:j_coproduct}
\end{figure}

It is then straightforward to see how $X$ acts on two operators: by utilizing the topological nature of $\Sigma$, we can split the integral into two parts, each acting on a single operator, as shown in Figure \ref{fig:j_coproduct}. This implies that a charge, given by the integral of a local conserved current over a codimension-1 surface \eqref{integral:symmetryop}, results in the coproduct {$\Delta(X)=X\otimes1+1\otimes X$}.

\begin{figure}[h]
    \centering
    \includegraphics[width = .9\textwidth]{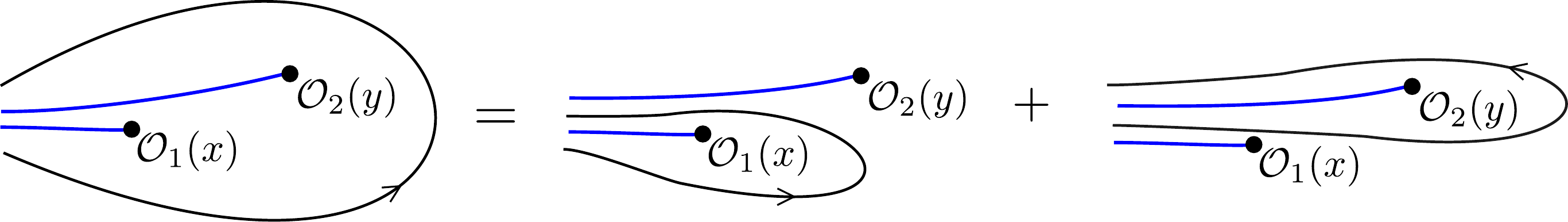}
    \caption{The action of ordinary global symmetry on defect-ending operators.}
    \label{fig:Q_nonlocalop}
\end{figure}

If, instead, we act on defect-ending operators, the situation becomes slightly more complicated: the topological surface cannot close on the operator because the operator is attached to a line. However, we can still split the integral of the current into two parts, so that in both cases, it acts only on one operator and leaves the other unchanged, as shown in Figure \ref{fig:Q_nonlocalop}.

In this paper, we consider a theory with quantum group global symmetry. The action of symmetry generators must be governed by the deformed coproduct (specifically, \eqref{coprod} in the case of \(U_q(sl_2)\)). Based on the previous reasoning, we cannot realize all quantum group generators by integrating local conserved currents over a codimension-1 surface, as this can only produce a coproduct of the form \(\Delta(X)=X\otimes1+1\otimes X\).

For \(U_q(sl_2)\), the coproduct formula \eqref{coprod} indicates that, while it is possible to have a local current that produces the \(H\) generator through the integral \eqref{integral:symmetryop}, it is impossible to have local currents that produce the \(E\) and \(F\) generators.
Still, we would like to conceptualize the action of \(U_q(sl_2)\) as being implemented by some topological line. Interestingly, there are cases where these topological lines can be constructed explicitly by integrating non-local conserved currents \cite{Bernard:1990ys}. Then, the action of the quantum group generators is realized in a manner similar to \eqref{integral:symmetryop}, but with the important distinction that the currents themselves are attached to additional topological lines. 

\begin{figure}
    \centering
    \includegraphics[width =1.0\textwidth]{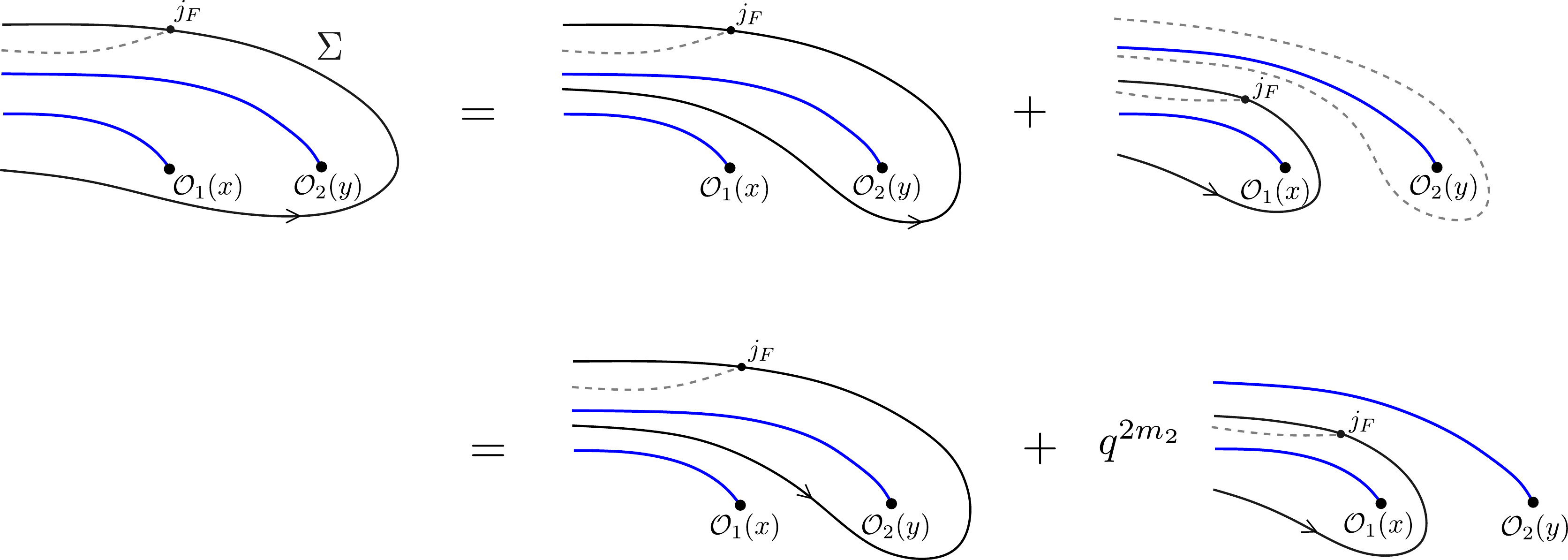}
    \caption{The action of quantum group generator \(F\) on two defect-ending operators, given by the $U_q(sl_2)$ coproduct, $\Delta(F) = 1\otimes F + F \otimes q^H$. The black contour represents the line \(\Sigma\) along which the current operator \(j_F\) is integrated. The dashed line denotes the additional $q^H$-line attached to \(j_F\), while the blue lines are the defect lines attached to \(\calO_1\) and \(\calO_2\).
    }
    \label{fig:coprod_schematic}
\end{figure}

Consider a pair of defect-ending operators \(\calO_1\) and \(\calO_2\). The action of the \(U_q(sl_2)\) generator \(F\) can be schematically represented as shown in Figure \ref{fig:coprod_schematic}. The non-local current $j_F(x)$ is attached to a topological line corresponding to the $U(1)$ element $q^H$. As \(j_F(x)\) moves along the contour, the endpoint of the $q^H$-line moves together with \(j_F(x)\). We observe that the integral can be divided into two parts: one that encircles \(\calO_1\) and another that encircles \(\calO_2\). The contour integral around \(\calO_2\) produces \(\calO_1(F\cdot\calO_2)\). The contour integral around \(\calO_1\) is more intricate because it requires tracking the $q^H$-line to which $j_F(x)$ is attached, which has already encircled \(\calO_2\). Since the $q^H$-line is topological, it can be deformed into the shape shown in Figure \ref{fig:coprod_schematic}, and it schematically yields \((F\cdot\calO_1)(q^{H}\cdot\calO_2)\). This illustrates how a current attached to a topological line realizes the coproduct \(\Delta(F)=F\otimes q^{H}+1\otimes F\). It is crucial that \(\calO_1\) and \(\calO_2\) are defect-ending operators, as this ensures there is no ambiguity in how the integration contour of the current is deformed.\footnote{Conversely, for \(\calO_1\) and \(\calO_2\) being mutually local, a necessary condition is that $H(\calO_1)=H(\calO_2)=0$.}

\begin{figure}
    \centering
    \includegraphics[width =1.0\textwidth]{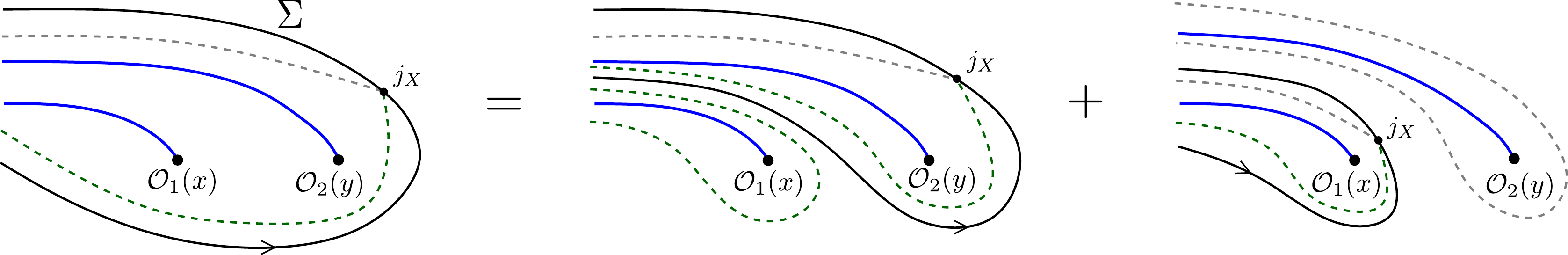}
    \caption{The action of a general quantum group generator \(X\) on two defect-ending operators. The current $j_X$ is integrated along the black contour $\Sigma$. The dashed lines are the topological lines to which $j_X$ is attached.}
    \label{fig:coprod_general}
\end{figure}

We will not use the \(E\) generator in what follows, however, we would like to comment that, in order to realize the coproduct of the \(E\) generator in \eqref{coprod}, the topological line to which the current is attached in Figure \ref{fig:coprod_schematic} must be modified to follow a different path around \(\calO_1\) and \(\calO_2\). The action of a general quantum group generator is depicted in Figure \ref{fig:coprod_general}.

As it is for the usual global symmetries, a requirement that a non-local current whose integral produces a QG symmetry generator exists may be too strong. More generally, all that is required is a topological line \cite{Gaiotto:2014kfa}.  Nevertheless, we believe that it is not enough to have a regular topological line to reproduce the QG structure because it is also necessary to have the $q^H$ line which endpoint is integrated over the topological line. In the rest of this paper we will not rely on the existence of non-local currents; the reader may assume that the generator lines, represented with arrows in all figures, are abstract topological lines. In the companion paper \cite{paper2} we will discuss an example where the lines can be constructed using the Coulomb gas formalism, following the ideas of \cite{Gomez:1990er}. This scenario is somewhere in between: while the action of the lines on operators follows from their free-boson representation, there is no operator in the spectrum of the theory that can be identified as a current. At least, it does not belong to the set of defect-ending operators that naturally appear in the partition function of the theory and that are closed under OPE.

\subsection{Crossing lines: the \texorpdfstring{$R$}{}-matrix}
\label{section:locality}
\subsubsection{Braid locality condition}
As we have seen in Section \ref{sec:difference}, when a system has a quantum group global symmetry, the locality condition $[\calO_i(x),\calO_j(y)]=0$ does not generally hold, resulting in non-local operators. We assume that the operators in the theory are defect-ending operators and that they are almost local in the sense that the defect lines they are attached to are all topological. Under this assumption, it is natural to expect that the violation of the standard locality condition is also topological. To be precise, suppose \(\{\mathcal{O}_i\}\) is a complete operator basis of the theory, where each \(\calO_i\) is attached to a topological line, then a pair of permuted operators \(\calO_j(y)\calO_i(x)\) is expected to be a linear combination of operator pairs \(\calO_k(x)\calO_l(y)\). In other words, we \textbf{assume} a braid locality condition, expressed as follows:
\begin{equation}\label{localityM:general}
    \sum_{k,l}M_{j,i}^{l,k}\,\calO_{k}(x)\calO_{l}(y)=\calO_{j}(y)\calO_{i}(x),
\end{equation}
where \(M_{j,i}^{l,k}\) are complex numbers, and the sum is over all operators in the operator basis \(\{\mathcal{O}_i\}\). \eqref{localityM:general} is called \emph{reordering condition} in \cite{Fröhlich1988}. For ordinary group or Lie algebra symmetries, the braiding matrix \(M\) simplifies to \(M_{j,i}^{l,k}=\pm\delta_{j}^{l}\,\delta_{i}^{k}\), which recovers the Boson/Fermion statistics.

The braid locality condition described above must be consistent with the global symmetry of the theory, meaning that acting with the symmetry generators on both sides should yield the same result. When the theory has quantum group global symmetry, the main subtlety in \eqref{localityM:general} is that the order of the quantum group representations is swapped. Therefore, the braid matrix \(M\) should be designed to effectively restore the original order. As discussed in Section \ref{sec:Rmat}, quantum groups have a natural object related to the swapping of two representations: the \(R\)-matrix. It is no surprise that the modified locality condition involves the \(R\)-matrix. A consistent proposal for \(M\) was presented in \cite{Schomerus:1994ir}, which is the quantum group version of the original proposal by Fröhlich \cite{Fröhlich1988}\footnote{Here, we organize the operators into multiplets of the quantum group, as described after eq.\,\eqref{QFT:wardid}.}
\begin{equation}\label{QFT:localityM}
    \sum_{m_i',m_j'}[M_{j,i}]_{m_jm_i}^{m_j'm_i'}\,\calO_{i,\ell_i,m_i'}(x)\calO_{j,\ell_j,m_j'}(y)=\calO_{j,\ell_j,m_j}(y)\calO_{i,\ell_i,m_i}(x),
\end{equation}
where the matrix \(M_{j,i}\) takes the following form:
\begin{equation}\label{M:general}
    [M_{j,i}]_{m_jm_i}^{m_j'm_i'}=\omega_{j,i}\,[R_{\ell_j,\ell_i}]_{m_jm_i}^{m_j'm_i'}.
\end{equation}
Here, the phase factor \(\omega_{j,i}\) is a complex number that will be further discussed below, and \([R_{\ell_j,\ell_i}]_{m_jm_i}^{m_j'm_i'}\) is the entry of the \(R\)-matrix (introduced in Section \ref{sec:Rmat}), defined as 
\begin{equation}
    \mathcal{R}_{ji}\ket{\ell_j,m_j}\otimes\ket{\ell_i,m_i}\equiv\sum_{m_i',m_j'}[R_{\ell_j,\ell_i}]_{m_jm_i}^{m_j'm_i'}\ket{\ell_j,m_j'}\otimes\ket{\ell_i,m_i'}.
\end{equation}
We postpone the explanation of the consistency between \eqref{QFT:localityM} and the quantum group symmetry to Section \ref{section:localityconsistency}. The underlying assumption of \eqref{QFT:localityM} is that operators in other quantum group multiplets do not appear. This is a reasonable assumption when the quantum group is the maximal non-abelian global symmetry of the theory. Otherwise, the phase factor \(\omega_{j,i}\) should be replaced by a matrix, as the multiplet of the larger non-abelian global symmetry may contain several equivalent quantum group representations.

Now let us expand on the discussion of the phase factor \(\omega_{j,i}\). This phase factor is analogous to the one encountered in the statistics of abelian anyons in two-dimensional systems. Its physical interpretation is that the system might have an additional abelian global symmetry (e.g., \(U(1)\), \(\mathbb{Z}_2\), etc.). One can either absorb \(\omega_{j,i}\) into a redefinition of the \(R\)-matrix by incorporating the extra abelian group into the symmetry algebra, or simply retain it as is.\footnote{The simplest example is when there are fermions in the theory, one can redefine the \(R\)-matrix by absorbing a factor of \((-1)^{F \otimes F}\), i.e., \(R \rightarrow R(-1)^{F \otimes F}\). Here, \(F\) is the fermion number operator, not the \(U_q(sl_2)\) generator. For example, in a free fermion theory without a quantum group, the system has a \(\mathbb{Z}_2\) global symmetry generated by \((-1)^F\). One can verify that \(R = (-1)^{F \otimes F}\) and \(\Delta((-1)^F) = (-1)^F \otimes (-1)^F\) satisfy \eqref{def:Rmatrix}. If we do not absorb \((-1)^{F \otimes F}\) into the \(R\)-matrix, then the effect of this factor will appear in \(\omega_{j,i}\) in \eqref{M:general}. More generally, operators in the theory may be attached to more generic topological lines in addition to quantum group defects. In that case, either the \(R\)-matrix or the phase factor \(\omega_{j,i}\) will be further generalized.} The operators in the system are attached to extra topological lines corresponding to these additional symmetries, producing the phase \(\omega_{j,i}\) when we permute operators \(\calO_i\) and \(\calO_j\). If the system lacks additional abelian symmetry, meaning there are no extra topological lines corresponding to abelian group elements outside the quantum group, then \(\omega_{j,i} \equiv 1\) for all \((i,j)\) pairs.

If we further assume that the QFT has an asymptotic operator product expansion (OPE)
\begin{equation}\label{OPEasymp}
    \calO_i(x)\calO_j(y)\stackrel{\rm asymp}{=\joinrel=}\sum_{k}C_{ijk}(x-y)\calO_k(x),
\end{equation}
then, for the braid locality condition \eqref{QFT:localityM} to be consistent with the OPE, we must have
\begin{equation}\label{omega:relaton}
    \begin{split}
        \omega_{i,j}\,\omega_{i,k}=\omega_{i,l},\quad\omega_{j,i}\,\omega_{k,i}=\omega_{l,i}\quad(l\in j\times k),
    \end{split}
\end{equation}
where by ``$l\in j\times k$" we mean that \(\calO_l\) appears in the OPE of \(\calO_j\calO_k\).\footnote{The same consistency condition appears in the algebraic QFT framework, where the OPE is replaced by the fusion of operators (see eq.\,(4.20) in \cite{Schomerus:1994ir}).} The argument for \eqref{omega:relaton} is postponed to Section \ref{section:localityconsistency}.

\begin{figure}
    \centering
    \includegraphics[width = .8\textwidth]{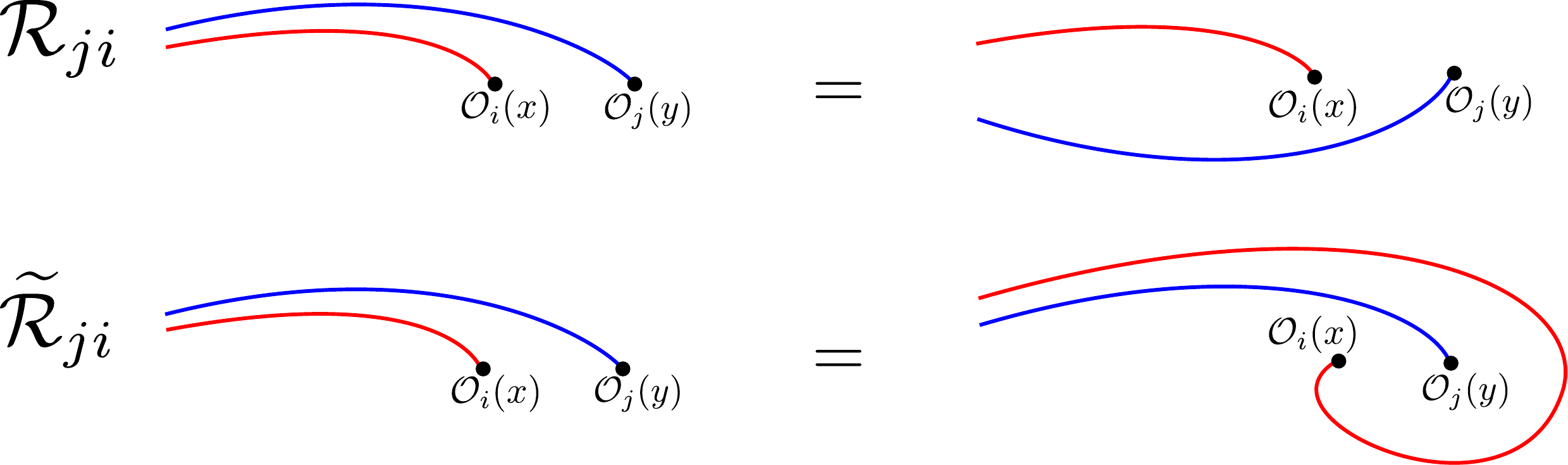}
    \caption{The configurations on both sides of \eqref{QFT:locality}. On the l.h.s., the operators are on the same time slice and ordered in the spatial direction. Other operators are not shown.}
    \label{fig:Rmat1}
\end{figure}

In this paper, we will only consider the case where \(\omega_{j,i}\equiv1\). In other words, we will \textbf{assume} that the braid locality condition takes the following form:
\begin{equation}\label{QFT:locality}
    \sum_{m_i',m_j'}[R_{\ell_j,\ell_i}]_{m_jm_i}^{m_j'm_i'}\,\calO_{i,\ell_i,m_i'}(x)\calO_{j,\ell_j,m_j'}(y)=\calO_{j,\ell_j,m_j}(y)\calO_{i,\ell_i,m_i}(x),
\end{equation}
We emphasize that \(\calO_i\) and \(\calO_j\) in \eqref{QFT:locality} must be a pair of neighboring operators inside the correlation function. Given that the operators are attached to topological lines, it is necessary to specify the configuration of lines on the left-hand side of \eqref{QFT:locality} and how it relates to the configuration on the right-hand side. We assert that there are only two possible ways to relate the line configurations on both sides of \eqref{QFT:locality} that are consistent with the properties of the \(R\)-matrix. The rules for one of these configurations are as follows:
\begin{enumerate}
    \item If, on the l.h.s. of \eqref{QFT:locality}, all the operators are on the same Euclidean time slice and ordered in the spatial direction, then the configuration of \eqref{QFT:locality} is given by Figure \ref{fig:Rmat1}.\footnote{The two plots in figure \ref{fig:Rmat1} are equivalent to each other. One can verify \(\tilde{\mathcal{R}} = \tau(\mathcal{R})^{-1}\) by acting with \(R_{ji}^{-1}\) on both sides of the first plot, swapping the indices \(i \leftrightarrow j\), and then continuously deforming the configuration to the one presented in the second plot.}
    \item For other configurations, the relation between the configurations on both sides of \eqref{QFT:locality} is obtained by continuously deforming both sides (in the same way) from the previous case. It is important to ensure that the lines do not cross during this deformation.
\end{enumerate}

The second possible configuration is given by swapping $\calR_{ji}$ and $\widetilde{\calR}_{ji}$ in Figure \ref{fig:Rmat1}. Recall that the expression for the \(R\)-matrix \(\mathcal{R}\) is given in \eqref{Rmatrix:option1}, and the matrix $\widetilde{\calR} \equiv \tau(\calR)^{-1}$ by \eqref{Rmatrix:option2}. 
Whether we choose the first or the second configuration for \eqref{QFT:locality} depends on the theory, and each theory will consistently select one of them for all the operators. 

We will see in Sections \ref{section:QFTcorrelator} and \ref{section:CFTcorrelator} that the braid locality condition \eqref{QFT:locality} constrains the space-time spins and the OPE coefficients of the theory. For simplicity, we will present the formulas below for the choice of Figure \ref{fig:Rmat1}, but the same arguments also apply for the other possible choice of braid locality condition, leading to very similar conclusions.\footnote{If a theory selects $R$-matrix \eqref{Rmatrix:option2} for the braid locality condition \eqref{QFT:locality}, then all the \(R\)-matrix-related arguments that work for the case of Figure \ref{fig:Rmat1} apply here with the replacement \(\mathcal{R} \to \tilde{\mathcal{R}}\).} The concrete example we will discuss in Section \ref{section:example} corresponds to Figure \ref{fig:Rmat1}.

\subsubsection{Consistency between the locality condition and the quantum group symmetry}\label{section:localityconsistency}
In this subsection, we would like to explain why the locality condition \eqref{QFT:localityM} is consistent with the quantum group symmetry.

The basic requirement for \eqref{QFT:localityM} is that it should be compatible with the coproduct, meaning that for any quantum group generator \(X\), acting with \(\Delta(X)\) on both sides should yield the same result.

It is straightforward to check that this requirement implies the first defining property of the \(R\)-matrix in \eqref{def:Rmatrix}. We begin by acting with \(\Delta(X)\) on both sides of \eqref{QFT:localityM}. The action on the left-hand side is
\begin{equation}\label{locality:consitency1:lhs}
    \begin{split}
        \omega_{j,i}\,[R_{\ell_j,\ell_i}]_{m_j,m_i}^{m_j'm_i'}\,X\left(\calO_{i,m_i'}\calO_{j,m_j'}\right) &= \omega_{j,i}\,[R_{\ell_j,\ell_i}]_{m_j,m_i}^{m_j'm_i'}\,[\Delta(X)_{\ell_i,\ell_j}]_{m_i'm_j'}^{m_i''m_j''}\calO_{i,m_i''}\calO_{j,m_j''} \\
        &\equiv \omega_{j,i}\,\Delta(X)_{ij} \mathcal{R}_{ji}\left(\calO_{i,m_i}\calO_{j,m_j}\right).
    \end{split}
\end{equation}
The action on the right-hand side is
\begin{equation}\label{locality:consitency1:rhs}
    \begin{split}
        X\left(\calO_{j,m_j}\calO_{i,m_i}\right) &= [\Delta(X)_{\ell_j,\ell_i}]_{m_jm_i}^{m_j'm_i'}\calO_{j,m_j'}\calO_{i,m_i'} \\
        &= \omega_{j,i}\,[\Delta(X)_{\ell_j,\ell_i}]_{m_jm_i}^{m_j'm_i'} [R_{\ell_j,\ell_i}]_{m_j'm_i'}^{m_j''m_i''}\,\calO_{i,m_i''}\calO_{j,m_j''} \\
        &\equiv \omega_{j,i}\,\mathcal{R}_{ji}\Delta(X)_{ji}\left(\calO_{i,m_i}\calO_{j,m_j}\right).
    \end{split}
\end{equation}
Here, by \(A_{ij}\), we mean that the left factor of \(A\) is in the representation of \(\calO_i\) and the right factor of \(A\) is in the representation of \(\calO_j\). For example, if \(A = E \otimes F\), then \(A_{ij} = E_iF_j\), and \(\tau(A)_{ij} = A_{ji} = E_jF_i\). The two sides should match for any pair of \(\calO_i\) and \(\calO_j\), which requires that
\begin{equation}\label{Rij:eq1}
    \omega_{j,i}\,\Delta(X)_{ij} \mathcal{R}_{ji}=\omega_{j,i}\,\mathcal{R}_{ji}\Delta(X)_{ji} ,\quad\forall i,j,X.
\end{equation}
We see that the \(\omega_{j,i}\) factors cancel out. Moving \(\mathcal{R}_{ji}\) to the right-hand side of \eqref{Rij:eq1}, we obtain \(\mathcal{R}_{ji}\Delta(X)_{ji}\mathcal{R}_{ji}^{-1}=\Delta(X)_{ij}\). The index-free version of this is precisely the first equation of \eqref{def:Rmatrix}. Therefore, the braid locality condition \eqref{QFT:localityM} is consistent with the quantum group symmetry.

When the theory allows for operator product expansion, there are additional constraints on the braiding matrix \(M_{ji}\), arising from the consistency condition of permuting one operator with two others:
$$\calO_i\calO_j\calO_k\rightarrow\calO_j\calO_k\calO_i\quad \text{or}\quad \calO_i\calO_j\calO_k\rightarrow\calO_k\calO_i\calO_j.$$
Let us focus on the first case, \(\calO_i\calO_j\calO_k\rightarrow\calO_j\calO_k\calO_i\). There are two ways to perform this permutation. The first way is to apply \eqref{QFT:localityM} twice to \(\calO_j\calO_k\calO_i\), which gives
\begin{equation}\label{Rij:eq2lhs}
    \begin{split}
        \calO_{j,\ell_j,m_j}\calO_{k,\ell_k,m_k}\calO_{i,\ell_i,m_i}&=[M_{ki}]_{m_k,m_i}^{m_k',m_i'}\calO_{j,\ell_j,m_j}\calO_{i,\ell_i,m_i'}\calO_{k,\ell_k,m_k'} \\
        &=[M_{ki}]_{m_k,m_i}^{m_k',m_i'}[M_{ji}]_{m_j,m_i'}^{m_j',m_i''}\calO_{i,\ell_i,m_i''}\calO_{j,\ell_j,m_j'}\calO_{k,\ell_k,m_k'} \\
        &\equiv \omega_{ki}\omega_{ji}\mathcal{R}_{ji}\mathcal{R}_{ki}\left(\calO_{i,\ell_i,m_i}\calO_{j,\ell_j,m_j}\calO_{k,\ell_k,m_k}\right).
    \end{split}
\end{equation}
The second way is to use the OPE \eqref{OPEasymp} for \(\calO_j\calO_k\) and then apply \eqref{QFT:localityM}. This gives an asymptotic expansion:
\begin{equation}\label{Rij:eq2rhs}
    \calO_j\calO_k\calO_i\stackrel{\rm asymp}{=\joinrel=}\sum_{l}C_{jkl}\calO_l\calO_i=\sum_{l}C_{jkl}\,\omega_{li}\,\mathcal{R}_{li}\left(\calO_i\calO_l\right)=\mathcal{R}_{(jk),i}\left(\sum_{l}\omega_{li}C_{jkl}\calO_i\calO_l\right).
\end{equation}
Here, we use the fact that \(C_{jkl}\calO_l\) transforms under the tensor product representation \(\mathbb{V}_{\ell_j}\otimes\mathbb{V}_{\ell_k}\) of the quantum group, which implies \(C_{jkl}\mathcal{R}_{li}=\mathcal{R}_{(jk),i}C_{jkl}\) (both \(C\) and \(\mathcal{R}\) have many tensor indices). \(\mathcal{R}_{(jk),i}\) means that the first factor of \(\mathcal{R}\) acts on \(\calO_j\calO_k\) via the tensor product representation, and the second factor of \(\mathcal{R}\) acts on \(\calO_i\).

Now, we want \eqref{Rij:eq2lhs} and \eqref{Rij:eq2rhs} to be consistent at the level of asymptotic expansions, which requires that
\begin{equation}\label{Rij:eq2}
    \omega_{ji}\,\omega_{ki}\,\mathcal{R}_{ji}\mathcal{R}_{ki}=\omega_{li}\,\mathcal{R}_{(jk),i},\quad\forall i,j,k\ \&\ l\in j\times k.
\end{equation}
Without the \(\omega\) factors, \eqref{Rij:eq2} is exactly the second equation of \eqref{def:Rmatrix} satisfied by the \(R\)-matrix. With the \(\omega\) factors, \eqref{Rij:eq2} requires that \(\omega_{ji}\omega_{ki}=\omega_{li}\), which is one of the constraints on the \(\omega\) factors in \eqref{omega:relaton}. A similar argument applies to the other case of permutation, \(\calO_i\calO_j\calO_k\rightarrow\calO_k\calO_i\calO_j\). Therefore, the braid locality condition \eqref{QFT:localityM}, together with the constraint \eqref{omega:relaton}, is consistent with the OPE \eqref{OPEasymp}.

\begin{figure}[ht]
    \centering
    \includegraphics[width = .8\textwidth]{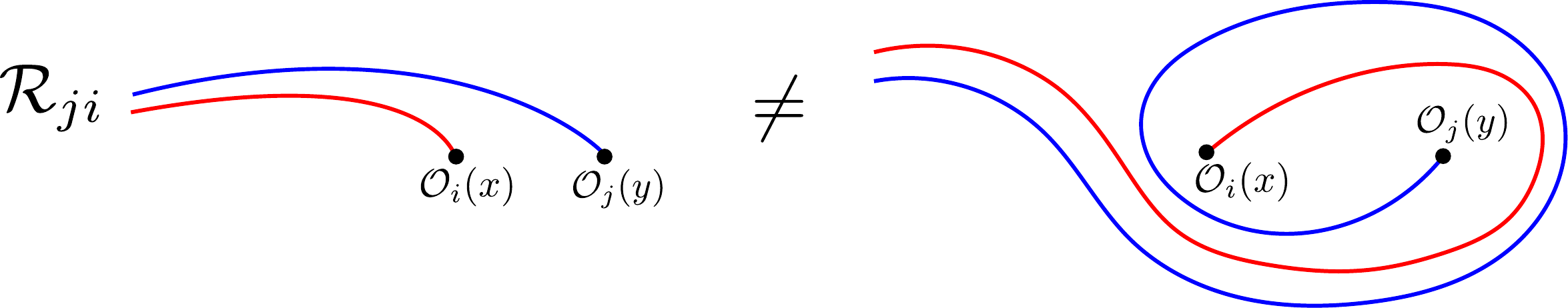}
    \caption{The ``wrong'' permutation. Here, we set \(\omega \equiv 1\) for simplicity.}
    \label{fig:Rmat_wrong}
\end{figure}

\begin{remark}
    In the above argument, it is essential that the permutation ``minimally" changes the field configuration (see Rules 1 and 2 below eq.\,\eqref{QFT:locality}), corresponding to Figure \ref{fig:Rmat1}. It follows from Rule 1 and Rule 2 that the permutation of three operators can be done in two different ways, as shown in Figure \ref{fig:Rmat3O}, which explains why \eqref{Rij:eq2lhs} and \eqref{Rij:eq2rhs} can match (at least as asymptotic expansions).
    \begin{figure}[ht]
    \centering
    \includegraphics[width = 0.9 \textwidth]{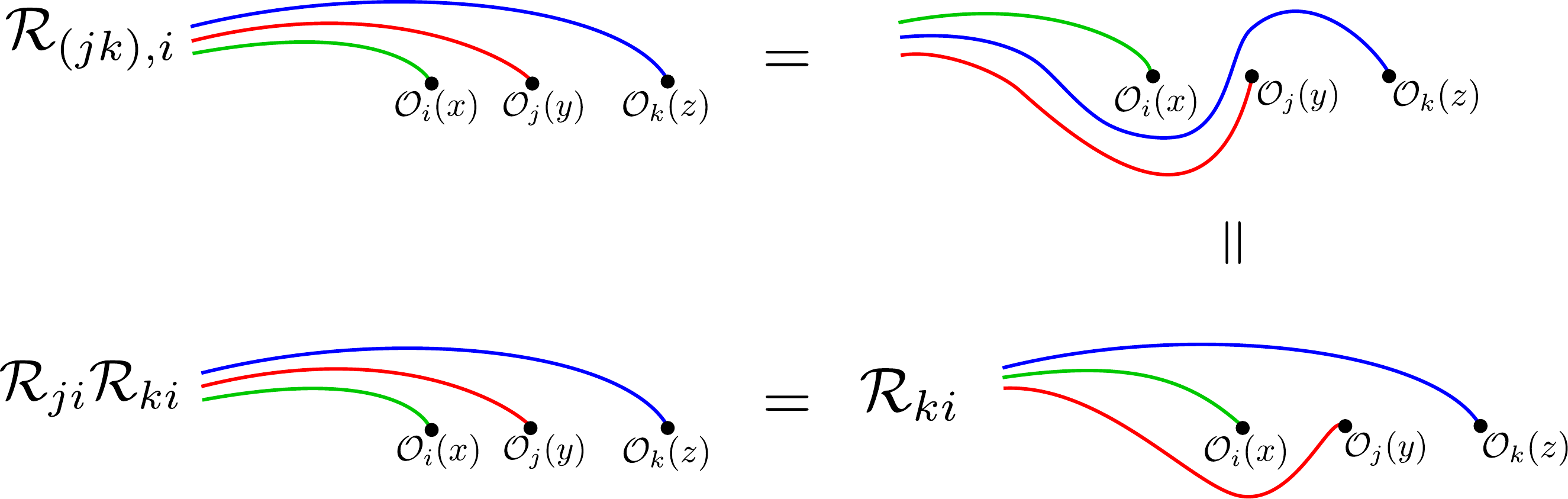}
    \caption{The permutation between \(\calO_i\) and the \((\calO_j\calO_k)\) pair. Here, we set \(\omega \equiv 1\) for simplicity.}
    \label{fig:Rmat3O}
\end{figure}
    
    Suppose we use other configurations for the permutation, for example, the one shown in Figure \ref{fig:Rmat_wrong}. In that case, \eqref{QFT:localityM} is still consistent with the quantum group symmetry by the first defining property of the \(R\)-matrix. However, the second and third defining properties of the \(R\)-matrix no longer imply the consistency between \eqref{QFT:localityM} and the OPE \eqref{OPEasymp}. The reason is that for the rules using the incorrect configuration, the two-step permutation \(\calO_i\calO_j\calO_k \rightarrow \calO_j\calO_i\calO_k\rightarrow \calO_j\calO_k\calO_i\) is not the same as the permutation \(\calO_i(\calO_j\calO_k) \rightarrow (\calO_j\calO_k)\calO_i\), where the \((\calO_j\calO_k)\) pair is regarded as a composite operator. This discrepancy arises because the fundamental group of a sphere with two punctures is not abelian. Consequently, \eqref{Rij:eq2lhs} does not equal \eqref{Rij:eq2rhs}, even as asymptotic expansions.
\end{remark}

\subsection{Structure of QFT correlation functions}\label{section:QFTcorrelator}
In our work we consider operators transforming in finite-dimensional representations of the quantum group. We choose an operator basis according to the spacetime symmetry group $\mathbb{R}^2 \rtimes SO(2)$ and the \(U_q(sl_2)\) global symmetry. Each operator has as quantum numbers the space-time spin \(s\) and \((\ell, m)\), which corresponds to the vector $\ket{\ell,m}$ in the spin-\(\ell\) representation of \(U_q(sl_2)\).

A general correlation function in the above basis takes the form
\begin{equation}\label{QFTcorr:generalform}
    \braket{\mathcal{O}_{1,\ell_1,m_1}(z_1) \ldots \mathcal{O}_{n,\ell_n,m_n}(z_n)} = z_{21}^{-\sum_{k=1}^{n} s_k} \sum_{r=1}^{N} F_r\left(\left|z_{21}\right|, \frac{\left|z_{21}\right| z_{31}}{z_{21}}, \ldots, \frac{\left|z_{21}\right| z_{n1}}{z_{21}}\right) (T_r)_{m_1 \ldots m_n}.
\end{equation}
Here \(z_{ij} \equiv z_i - z_j\); the sum is taken over trivial representations appearing in the tensor product \(\mathbb{V}_{\ell_1} \otimes \ldots \otimes \mathbb{V}_{\ell_n}\). Each \(T_r\) is a \(U_q(sl_2)\)-invariant tensor corresponding to $r$-th trivial representation, and \(N\) is the total number of trivial representations. The form factors \(F_r\) are manifestly translation and rotation invariant, but they can be multi-valued functions of the last \(n-2\) variables.

\subsubsection{Two-point functions, constraint on the space-time spin}\label{section:QFT2pt}
For two-point functions, the invariant tensor \(T_r\) in \eqref{QFTcorr:generalform} corresponds to the trivial representation in \(\mathbb{V}_{\ell_1} \otimes \mathbb{V}_{\ell_2}\). This occurs only when \(\ell_1 = \ell_2 = \ell\) and there is only one trivial representation, which gives the unique tensor:
\begin{equation}
\label{eq:canonical2pt}
    T_{m_1 m_2}(\ell) = \QCG{\ell, m_1, \ell, m_2, 0, 0, q} = \frac{(-1)^{m_1 - \ell} q^{m_1}}{\sqrt{[2\ell + 1]_q}} \delta_{m_1 + m_2, 0}.
\end{equation}
Here the square bracket \([\ldots]_q\) denotes the quantum Clebsch-Gordan coefficients given in \eqref{qcg_formula}. Practically, in this case, one can also derive the tensor structure \eqref{eq:canonical2pt} using Ward identities \eqref{u1_charge_cons} and \eqref{wardid:Fexplicit}. Moreover, the function \(F_r\) in \eqref{QFTcorr:generalform} reduces to a single-valued function. Thus, the two-point function is given by:
\begin{equation}\label{QFT:2ptform}
    \braket{\mathcal{O}_{1,\ell_1,m_1}(z_1) \mathcal{O}_{2,\ell_2,m_2}(z_2)} = (-1)^{m_1 - \ell_1} q^{m_1} z_{21}^{-s_1 - s_2} F(\left| z_{21} \right|) \delta_{\ell_1, \ell_2} \delta_{m_1 + m_2, 0}.
\end{equation}
Here we have ignored the unimportant overall factors by absorbing them in the definition of the operators. The form factor $F(\abs{z_{21}})$ is model dependent.

Now we would like to argue that given any operator $\mathcal{O}$, its space-time spin $s$ is fixed by its quantum group spin $\ell$, up to an integer. Let us consider the two-point function $\braket{\mathcal{O}_{\ell,m}\mathcal{O}_{\ell,-m}}$. Using the braid locality condition \eqref{QFT:locality}, we have
\begin{equation}\label{QFT:2ptlocality0}
    \sum_{m'}[R_{\ell,\ell}]_{-m,m}^{-m',m'}\braket{\mathcal{O}_{\ell,m'}(z_1)\mathcal{O}_{\ell,-m'}(z_2)}=\braket{\mathcal{O}_{\ell,-m}(z_2)\mathcal{O}_{\ell,m}(z_1)}_{}.
\end{equation}
Recall that in principle there are two possible options on how the operators are permuted: figure \ref{fig:Rmat1} and the same with $\calR \leftrightarrow \widetilde \calR$ swapped. The choice of how to permute operators is theory dependent, and different choices will lead to different consequences. Here let us focus on the case of figure \ref{fig:Rmat1} since it corresponds to the theory we will study below. Then, by picking $m=-\ell$, the only non-vanishing component of the $R$-matrix is given by $[R_{\ell,\ell}]_{\ell,-\ell}^{\ell,-\ell}=q^{-2\ell^2}$ (see eq.\,\eqref{Rmatrix:option1}). So we get
\begin{equation}\label{QFT:2ptlocality1}
    \begin{split}
        q^{-2\ell^2}\braket{\mathcal{O}_{\ell,-\ell}(z_1)\mathcal{O}_{\ell,\ell}(z_2)}=\braket{\mathcal{O}_{\ell,\ell}(z_2)\mathcal{O}_{\ell,-\ell}(z_1)}. \\
    \end{split}
\end{equation}
Plugging \eqref{QFT:2ptform} into \eqref{QFT:2ptlocality1} we get\footnote{This expression has an additional sign ambiguity when $\ell$ is half integer. In fact, this can be traced back to a similar ambiguity in the definition of the $\mathcal{R}$-matrix (\ref{Rmatrix:option1} - \ref{Rmatrix:option2}) when the eigenvalue of both operators under $H$ is odd. From the algebraic perspective, either choice of branch for $q^{-2\ell^2}$ is allowed. That means that in a theory with quantum group symmetry, any operator of half-integer (quantum group) spin will have to be consistent with one of the two choices.}
\begin{equation} 
    q^{-2\ell^2}(-1)^{2\ell}q^{-\ell}z_{21}^{-2s}=q^{\ell}\left(z_{21}e^{i\pi}\right)^{-2s},
\end{equation}
which implies the relation\footnote{Similarly, if we were in a theory where the braid locality condition \eqref{QFT:locality} was the other possible one, given by swapping $\calR$ and $\widetilde \calR$, then by repeating the above analysis, we get
\begin{equation}\label{spin:option2}
    s=-\frac{\ell(\ell+1)}{\pi i}\log(q)+\ell+\mathbb{Z}.
\end{equation}
Comparing \eqref{spin:option1} and \eqref{spin:option2}, we see that the $q$-dependent part differ by a minus sign. Therefore, the spectrum of a theory will tell us which braid locality condition we should take. }
\begin{equation}\label{spin:option1}
    s=\frac{\ell(\ell+1)}{\pi i}\log(q)-\ell+\mathbb{Z}.
\end{equation}
We see that the space-time spin is fixed up to an integer ambiguity. The integer ambiguity is not surprising in the case of CFTs: a primary operator and its Virasoro descendants transform under the same $U_q(sl_2)$ representation, but their space-time spins differ by integers.

\subsubsection{Three-point functions, quantum Clebsch-Gordan coefficients}\label{section:QFT3pt}
For the tensor product of three $U_q(sl_2)$ representations, $\mathbb{V}_{\ell_1}\otimes\mathbb{V}_{\ell_2}\otimes\mathbb{V}_{\ell_3}$, a single trivial representation appears in the decomposition if and only if the following two conditions are satisfied: (a) $\ell_1+\ell_2+\ell_3\in\mathbb{Z}$, and (b) $\ell_1$, $\ell_2$ and $\ell_3$ satisfy the triangle inequalities. The corresponding $U_q(sl_2)$-invariant tensor is
 \begin{equation}
    T_{m_1m_2m_3}=\QCG{\ell_1,m_1,\ell_2,m_2,\ell_3,m_1+m_2,q}\QCG{\ell_3,m_1+m_2,\ell_3,m_3,0,0,q},
\end{equation}
where $[\ldots]_q$ is the quantum Clebsch-Gordan coefficient. One can also derive this tensor structure using Ward identities \eqref{u1_charge_cons} and \eqref{wardid:Fexplicit}. Then by \eqref{QFTcorr:generalform}, a general three-point function is given by
\begin{equation}\label{QFT:3ptform}
    \braket{\mathcal{O}_{1,\ell_1,m_1}(z_1) \mathcal{O}_{2,\ell_2,m_2}(z_2)\mathcal{O}_{3,\ell_3,m_3}(z_3)}=z_{21}^{-s_1-s_2-s_3} F\left(\left|z_{21}\right|, \frac{\left|z_{21}\right| z_{31}}{z_{21}}\right)T_{m_1m_2m_3}.
\end{equation}
Here the form factor $F\left(\left|z_{21}\right|, \frac{\left|z_{21}\right| z_{31}}{z_{21}}\right)$ is a multi-valued function of the second variable, and it is independent of $m_i$'s.

\subsection{Structure of CFT correlation functions}\label{section:CFTcorrelator}
In this subsection, we discuss the general properties of CFT correlation functions with $U_q(sl_2)$ global symmetry. Since CFT is a special case of QFT, the QFT properties discussed in the previous subsections still hold. We focus on the correlation functions of primary operators, since they encode all the CFT data.\footnote{The situation becomes more complicated for $q$ root of unity, but we postpone the relevant discussion to Section \ref{sec:ising}.} 

By $U_q(sl_2)$ global symmetry, the formula \eqref{def:QCG} implies that the OPE between two operators $\calO_i$ and $\calO_j$ takes the following form
\begin{equation}\label{CFT:OPE}
    \calO_{i,\ell_i,m_i}(z_1) \calO_{j,\ell_j,m_j}(z_2)= \sum_{k} \frac{C_{ijk}}{z_{21}^{h_{ijk}}\bz_{21}^{\hb_{ijk}}} \QCG{\ell_i,m_i,\ell_j,m_j,\ell_k,m_k,q} \left(\calO_{k,\ell_k,m_k}(z_1)+\ldots \right).
\end{equation}
$C_{ijk}$ is the OPE coefficient, which encodes the dynamical information and does not depend on $m_i$, $m_j$ and $m_k$, and $h_{ijk}\equiv h_i+h_j-h_k$. Here ``$+\ldots$" denotes the Virasoro descendant terms, which are kinematically fixed by conformal symmetry. 

Any $n$-point function can be computed by iterating the OPE \eqref{CFT:OPE}, and we will now use this to work out how correlation functions of the CFT with quantum group global symmetry look like. Because of the $U(1)$ subalgebra of $U_q(sl_2)$, correlation functions are non-zero only if the total $U(1)$ charge adds up to zero, $\sum_i m_i = 0$.

\subsubsection{Two-point functions}\label{section:CFT2pt}
As described in Section \ref{section:QFT2pt}, the two-point functions will be non-zero only for operators with the same \(U_q(sl_2)\) spin. Now, by conformal symmetry, the operators are also required to have the same scaling dimension. We normalize operators so that their two-point function is
\begin{equation} \label{eq:2pf}
    \braket{\mathcal{O}_{i,\ell_i,m_i}(0) \mathcal{O}_{j,\ell_j,m_j}(z,\bar{z})} = \delta_{ij} \QCG{\ell_i, m_i, \ell_j, m_j, 0, 0, q} \frac{1}{z^{2h_i} \bar{z}^{2\bar{h}_i}}.
\end{equation}
Since for generic $h_i$ and $\bar{h}_i$ the function $z^{2h_i} \bar{z}^{2\bar{h}_i}$ is a multi-valued, we need to fix the convention. Here we choose the standard principal branch for the CFT two-point function: $z^{2h_i}$ and $\bar{z}^{2\bar{h}_i}$ are positive when $z,\bar{z}>0$ in the principal branch.
Comparing this to the QFT two-point function \eqref{QFT:2ptform}, we see that the form factor is fixed by conformal symmetry.

The space-time spin of $\mathcal{O}_i$ is given by $s_i=h_i-\bar{h}_i$. The argument in section \ref{section:QFT2pt} shows that
\begin{equation}\label{CFT:spinrelation}
    h_i-\bar{h}_i=\left[\frac{\ell_i(\ell_i+1)}{\pi i}\log(q)-\ell_i+n_i\right],
\end{equation}
where $n_i\in\mathbb{Z}$ is a theory-dependent integer.\footnote{If the theory chooses the braid locality condition \eqref{QFT:locality} given by figure \ref{fig:Rmat1} with $\calR \leftrightarrow \widetilde\calR$, then it would give an overall minus sign in \eqref{CFT:spinrelation}.}

\subsubsection{Three-point functions, constraints on the OPE coefficients}\label{section:CFT3pt}
When considering three-point functions, we can first perform the OPE between two operators, and then we are left with a two-point function. What we obtain is
\begin{equation}\label{CFT:3pt1}
\begin{split}
    \braket{\mathcal{O}_i(z_1, \bar{z}_1) \mathcal{O}_j(z_2, \bar{z}_2) \mathcal{O}_k(z_3, \bar{z}_3)} &= C_{ijk} \QCG{\ell_i, m_i, \ell_j, m_j, \ell_k, -m_k, q} \QCG{\ell_k, -m_k, \ell_k, m_k, 0, 0, q} \\
    &\quad \times \frac{1}{z_{21}^{h_{ijk}} z_{32}^{h_{jki}} z_{31}^{h_{ikj}}} \frac{1}{\bar{z}_{21}^{\bar{h}_{ijk}} \bar{z}_{32}^{\bar{h}_{jki}} \bar{z}_{31}^{\bar{h}_{ikj}}}.
\end{split}
\end{equation}
In this case, we have first performed the OPE between \(\mathcal{O}_i\) and \(\mathcal{O}_j\). Comparing this to the QFT three-point function \eqref{QFT:3ptform}, we see that the form factor is fixed by conformal symmetry.

There are multiple equivalent ways to compute the three-point function. For example, we could also first perform the OPE between \(\mathcal{O}_j\) and \(\mathcal{O}_k\), then we get
\begin{equation}\label{CFT:3pt2}
    \begin{split}
        \braket{\mathcal{O}_i(z_1, \bar{z}_1) \mathcal{O}_j(z_2, \bar{z}_2) \mathcal{O}_k(z_3, \bar{z}_3)} &= C_{jki} \QCG{\ell_j, m_j, \ell_k, m_k, \ell_i, -m_i, q} \QCG{\ell_i, m_i, \ell_i, -m_i, 0, 0, q} \\
        &\quad \times \frac{1}{z_{21}^{h_{ijk}} z_{32}^{h_{jki}} z_{31}^{h_{ikj}}} \frac{1}{\bar{z}_{21}^{\bar{h}_{ijk}} \bar{z}_{32}^{\bar{h}_{jki}} \bar{z}_{31}^{\bar{h}_{ikj}}}.
    \end{split}
\end{equation}
Eqs.\,\eqref{CFT:3pt1} and \eqref{CFT:3pt2} imply a relation between the OPE coefficients \(C_{ijk}\) and \(C_{jki}\). This can be derived using the properties of quantum Clebsch-Gordan coefficients:
\begin{equation}
\begin{aligned}
    \QCG{\ell_2, m_2, \ell_3, m_3, \ell_1, -m_1, q} \QCG{\ell_1, m_1, \ell_1, -m_1, 0, 0, q} 
    &= (-1)^{\ell_1 - \ell_3 + m_2} q^{m_2} \left(\frac{[2\ell_1 + 1]_q}{[2\ell_3 + 1]_q}\right)^{1/2} \QCG{\ell_1, m_1, \ell_2, m_2, \ell_3, -m_3, q} \cdot \\& \qquad \qquad \qquad \qquad \qquad \qquad \qquad \qquad \qquad \cdot \QCG{\ell_1, m_1, \ell_1, -m_1, 0, 0, q} \\ 
    &= (-1)^{-\ell_3 + m_1 + m_2} \frac{q^{m_1 + m_2}}{\sqrt{[2\ell_3 + 1]_q}} \QCG{\ell_1, m_1, \ell_2, m_2, \ell_3, -m_3, q} \\ 
    &= \QCG{\ell_1, m_1, \ell_2, m_2, \ell_3, -m_3, q} \QCG{\ell_3, -m_3, \ell_3, m_3, 0, 0, q},
\end{aligned}
\end{equation}
where we have used the symmetry properties of the QCG coefficients \eqref{eq:QCG_symmetries}, some of their explicit values \eqref{eq:QCG_2pf}, and the neutrality condition \(m_1 + m_2 + m_3 = 0\). Thus, we obtain
\begin{equation} \label{eq:OPE_cyclic}
    C_{ijk} = C_{jki} (= C_{kij}).
\end{equation}
We know that there are \(3! = 6\) possible orderings of \(ijk\). The above relations state that even permutations of the indices do not change the OPE coefficient. For odd permutations, it suffices to study the relation between \(C_{ijk}\) and \(C_{jik}\). This can be done by using the braid locality condition \eqref{QFT:locality}. We act with $\mathcal{R}_{ji}$ on \(\mathcal{O}_i \mathcal{O}_j\) in \eqref{CFT:3pt1}:
\begin{equation}
    \braket{\mathcal{R}_{ji} \cdot \left( \mathcal{O}_i(z_1, \bar{z}_1) \mathcal{O}_j(z_2, \bar{z}_2) \right) \mathcal{O}_k(z_3, \bar{z}_3)} = \braket{\mathcal{O}_j(z_2, \bar{z}_2) \mathcal{O}_i(z_1, \bar{z}_1) \mathcal{O}_k(z_3, \bar{z}_3)}.
\end{equation}
This gives the relation\footnote{Here in the l.h.s. we put the operators such that they are spatially ordered on the same time slice: $z_1<z_2<z_3$, and we choose the principal branch for the configuration of topological lines. Then we can use figure \ref{fig:Rmat1}.}
\begin{equation}
    C_{ijk} \sum_{m_i', m_j'} [R_{\ell_j, \ell_i}]_{m_j, m_i}^{m_j', m_i'} \QCG{\ell_i, m_i', \ell_j, m_j', \ell_k, m_i + m_j, q} = C_{jik} \QCG{\ell_j, m_j, \ell_i, m_i, \ell_k, m_i + m_j, q} e^{-i\pi(h_{ijk} - \bar{h}_{ijk})}.
\end{equation}
Consequently, by \eqref{CFT:spinrelation} and \eqref{eq:Rmat_QCG}, the above relation implies
\begin{equation}
    C_{jik} = (-1)^{n_i + n_j + n_k} C_{ijk},
\end{equation}
where \(n_i\) is the integer in \eqref{CFT:spinrelation}.\footnote{Using \eqref{eq:Rmattilde_QCG}, one can show that the conclusion remains the same if the theory chooses the other possible braid locality condition, given by swapping $\calR \leftrightarrow \widetilde \calR$ in Figure \ref{fig:Rmat1}.}

In summary, \(C_{ijk}\) is invariant under even permutations of \(ijk\), and is multiplied by a factor of \((-1)^{n_i + n_j + n_k}\) under odd permutations of \(ijk\):
\begin{equation}\label{CFT:OPErelation}
    C_{ijk} = (-1)^{P(\sigma)(n_i + n_j + n_k)} C_{\sigma(i) \sigma(j) \sigma(k)}, \quad \sigma \in S_3, \quad P(\sigma) := \begin{cases}
        0, & \sigma \ \mathrm{even}, \\
        1, & \sigma \ \mathrm{odd}. \\
    \end{cases}
\end{equation}
\begin{remark}
    The relations \eqref{CFT:OPErelation} lead to a selection rule for the OPE between two identical operators:
    \[
    C_{iik} = (-1)^{n_k} C_{iik},
    \]
    which implies that only operators with even \(n_k\) appear in the OPE. 
    
    Consider the special case where \(q = 1\). The above selection rule and \eqref{CFT:spinrelation} indicate that only operators with spin \(s_k = \ell_k + n_k\) (where \(n_k\) is even) appear in the OPE. For instance, when \(\calO_i\) is a scalar operator (\(s_i = 0\)) with \(\ell_i = 1\), the operators that can appear in the OPE of \(\calO_i \calO_i\) will have $s_k$ even if $\ell_k =0,2$ and $s_k$ odd if $\ell_k =1$.
This selection rule is consistent with the usual conformal bootstrap framework \cite{Rattazzi:2010yc,Kos:2013tga}.

\end{remark}

\subsubsection{Four point functions, quantum \texorpdfstring{$6j$}{6j}-symbols and crossing symmetry}
For the four-point functions, there are generally many invariant tensors corresponding to the trivial representations appearing in the tensor product $\left(\mathbb{V}_{\ell_1} \otimes \mathbb{V}_{\ell_2} \otimes \mathbb{V}_{\ell_3} \otimes \mathbb{V}_{\ell_4}\right)$. There are two natural ways to count independent \(U_q(sl_2)\)-invariant tensors. One way is to decompose \(\mathbb{V}_{\ell_1} \otimes \mathbb{V}_{\ell_2}\) into irreducible representations \(\mathbb{V}_\ell\) first, then find the trivial representation in \(\mathbb{V}_\ell \otimes \mathbb{V}_{\ell_3} \otimes \mathbb{V}_{\ell_4}\) for each \(\ell\). For this decomposition, the contribution to the form factor in \eqref{QFTcorr:generalform} comes from spin-\(\ell\) operators in the OPE of \(\mathcal{O}_1 \mathcal{O}_2\). This corresponds to the s-channel expansion of the CFT four-point function, where we obtain 
\begin{equation}
\begin{aligned} \label{eq:4pf_s}
    \braket{\mathcal{O}_1(0) \mathcal{O}_2(z, \bar{z}) \mathcal{O}_3(1) \mathcal{O}_4(\infty)} &= \sum_{\ell} F^{(s)}_\ell(z, \bar{z}) T_{\ell}^{(s)}, \\
    F^{(s)}_\ell(z, \bar{z}) &= \sum_{j \in [\ell]} C_{12j} C_{34j} \mathcal{F}^{(s)}_{h_j}(z) \bar{\mathcal{F}}^{(s)}_{\bar{h}_j}(\bar{z}), \\
    T_\ell^{(s)} &= \sum_{m_{12}} \QCG{\ell_1, m_1, \ell_2, m_2, \ell, m_{12}, q} \QCG{\ell, m_{12}, \ell_3, m_3, \ell_4, -m_4, q} \QCG{\ell_4, -m_4, \ell_4, m_4, 0, 0, q},
\end{aligned}
\end{equation}
where \(T_{\ell}^{(s)}\) is the \(U_q(sl_2)\)-invariant tensor that exchanges the spin-\(\ell\) representation in the s-channel, \(F_{\ell}^{(s)}\) is the form factor, \(\sum_{j \in [\ell]}\) means summing over primary operators \(\mathcal{O}_j\) in the spin-\(\ell\) representation of \(U_q(sl_2)\), and \(\mathcal{F}_{h_j}^{(s)}\) (\(\bar{\mathcal{F}}_{\bar{h}_j}^{(s)}\)) is the chiral (anti-chiral) Virasoro block related to the exchange of the Virasoro multiplet of operator \(\mathcal{O}_j\). Here we used eq.\,\eqref{CFT:OPErelation}, which implies $C_{j34}=C_{34j}$.

Another way is to decompose \(\mathbb{V}_{\ell_2} \otimes \mathbb{V}_{\ell_3}\) into irreducible representations \(\mathbb{V}_\ell\) first, then find the trivial representation in \(\mathbb{V}_{\ell_1} \otimes \mathbb{V}_\ell \otimes \mathbb{V}_{\ell_4}\). This decomposition corresponds to the t-channel expansion of the CFT four-point function, and gives:
\begin{equation}
\begin{aligned} \label{eq:4pf_t}
    \braket{\mathcal{O}_1(0) \mathcal{O}_2(z, \bar{z}) \mathcal{O}_3(1) \mathcal{O}_4(\infty)} &= \sum_{\ell} F_{\ell}^{(t)}(z, \bar{z}) T_{\ell}^{(t)}, \\
    F_{\ell}^{(t)}(z, \bar{z}) &= \sum_{k \in [\ell]} C_{23k} C_{41k} \mathcal{F}^{(t)}_{h_k}(z) \bar{\mathcal{F}}^{(t)}_{\bar{h}_k}(\bar{z}), \\
    T_{\ell}^{(t)} &= \sum_{m_{23}} \QCG{\ell_2, m_2, \ell_3, m_3, \ell, m_{23}, q} \QCG{\ell_1, m_1, \ell, m_{23}, \ell_4, -m_4, q} \QCG{\ell_4, -m_4, \ell_4, m_4, 0, 0, q}.
\end{aligned}
\end{equation}
Here we used eq.\,\eqref{CFT:OPErelation} again. The blocks in the $t$-channel are related to the blocks in the $s$-channel by $\calF^{(t)}_h (z)= \calF^{(s)}_h (1-z)|_{h_1 \leftrightarrow h_3}$. 

The CFT four-point function has another expansion, namely the u-channel expansion, which corresponds to taking the OPE between \(\mathcal{O}_1\) and \(\mathcal{O}_3\). In this paper, we will not discuss the u-channel expansion for two reasons. First, since \(\mathcal{O}_1\) and \(\mathcal{O}_3\) are not neighboring operators in the correlation function, taking the \(\mathcal{O}_1 \mathcal{O}_3\) OPE requires permuting \(\mathcal{O}_2\) and \(\mathcal{O}_3\) first. This permutation involves the \(R\)-matrix and makes the computation more complicated than the s- and t-channel expansions (although it can be done in principle). Second, as a consequence of the locality condition \eqref{QFT:locality}, the consistency between the s- and t-channel expansions implies the consistency between the u-channel expansion and the other expansions.\footnote{Suppose we already know that the four-point function satisfies s-t crossing symmetry. Then using the locality condition \eqref{QFT:locality} we have
\begin{equation}
    \begin{split}
        \braket{\mathcal{O}_1(0) \mathcal{R}_{32}\left(\mathcal{O}_2(z, \bar{z}) \mathcal{O}_3(1) \right)\mathcal{O}_4(\infty)} =&\braket{\mathcal{O}_1(0)\mathcal{O}_3(1) \mathcal{O}_2(z, \bar{z})  \mathcal{O}_4(\infty)}. \\
    \end{split}
\end{equation}
The s- and t- channel expansions of the right-hand side are actually the u- and t-channel expansions of the left-hand side. Therefore, the s-t crossing symmetry + locality condition implies all the other crossing symmetries.} Therefore, studying the u-channel expansion will not provide additional insights into the theory.

Now let us discuss the crossing symmetry between the s- and t-channel expansions. By \eqref{eq:4pf_s} and \eqref{eq:4pf_t}, we have the consistency condition:
\begin{equation}\label{crossing:nonexplicit}
    \sum_{\ell} F^{(s)}_\ell(z, \bar{z}) T_{\ell}^{(s)} = \sum_{\ell'} F_{\ell'}^{(t)}(z, \bar{z}) T_{\ell'}^{(t)}.
\end{equation}
This gives \(N\) independent constraints, where \(N\) is the number of independent $U_q(sl_2)$-invariant tensors. To make the constraints explicit, we recall that \(\{T_{\ell}^{(s)}\}\) and \(\{T_{\ell'}^{(t)}\}\) are two bases that expand the same space of \(U_q(sl_2)\)-invariant tensors, as we have seen in section \ref{sec:6j}. These two bases are related by an \(N \times N\) invertible matrix whose entries are just the $6j$-symbols:\footnote{To work out this explicit expression, we multiply both sides of \eqref{TTrelation} by \(\QCG{\ell_1, m_1, \ell_2, m_2, \ell, m_{12}, q} \QCG{\ell, m_{12}, \ell_3, m_3, \ell_4, -m_4, q}\) and then sum over \(m_1\), \(m_2\), \(m_3\), and \(m_{12}\). Then, using the relation \eqref{relation:6j3j} on the left-hand side and the orthogonality properties of quantum Clebsch-Gordon coefficients  \eqref{eq:QCG_orth_1} on the right-hand side, we recover \eqref{TTrelation}.}
\begin{equation}\label{TTrelation}
    T_{\ell'}^{(t)} = \sum_{\ell} \sixj{\ell_1, \ell_2, \ell, \ell_3, \ell_4, \ell'} T_{\ell}^{(s)},
\end{equation}
Using \eqref{crossing:nonexplicit} and \eqref{TTrelation}, we get the explicit crossing symmetry between the s- and t-channel expansions:
\begin{equation}\label{crossing:explicit1}
    F^{(s)}_\ell(z, \bar{z}) = \sum_{\ell'} \sixj{\ell_1, \ell_2, \ell, \ell_3, \ell_4, \ell'} F_{\ell'}^{(t)}(z, \bar{z}), \quad \forall \ell.
\end{equation}
Using the orthogonality relations \eqref{eq:6j_orthogonality} of the \(6j\) symbols, the s-t crossing symmetry can also be expressed as:
\begin{equation}\label{crossing:explicit2}
    F^{(t)}_{\ell'}(z, \bar{z}) = \sum_{\ell} \sixj{\ell_1, \ell_2, \ell, \ell_3, \ell_4, \ell'} F_{\ell}^{(s)}(z, \bar{z}), \quad \forall \ell'.
\end{equation}
This crossing property of the four-point functions is closely related to crossing properties of Virasoro blocks that we will discuss in the next section.

\section{An example: closed XXZ with non-local terms }\label{section:example}

Now that we've described what a $U_q(sl_2)$ symmetric CFT should look like, we will turn our attention to one explicit example which satisfies all the properties listed in Section \ref{sec:results}. We will focus on a $U_q(sl_2)$ symmetric spin chain \cite{grosse1994quantum} which, in its continuum limit, is described by a CFT which inherits the quantum group symmetry \cite{pallua1998closed}. The partition function and therefore the spectrum of the theory are known, so it remains to determine the OPE coefficients, which in turn determine all the correlation functions of the theory. This will also serve as a cross check that quantum group symmetric CFTs, despite their peculiarities, can be well defined; the fact that XXZ$_q$ is crossing symmetric is evidence of this.

In order to study this theory, we would like to present two methods, which lead to the same results.
In this section we achieve this, modulo a sign ambiguity, by using a bootstrap approach and studying the crossing symmetry of four point functions. In \cite{paper2}, we discuss an alternative method to study this CFT, using a Coulomb Gas construction. In principle, for solving the theory it suffices to apply only one of the two methods. However, we would like to discuss both of them because each of them has its own advantage: the bootstrap method is a more straightforward way of computing the OPE coefficients, while the Coulomb gas method gives us a very intuitive physical picture of what a CFT with quantum group symmetry looks like.

\subsection{Spin chain formulation}
\label{sec:spinchain}

The most straightforward way to construct a theory with quantum group symmetry is to consider a spin chain. Once quantum group generators $E, F, q^H$ are given on a single lattice site one can construct the global quantum group generators on the full spin chain explicitly using the coproduct \eqref{coprod}. However, constructing a physically interesting Hamiltonian that would be invariant under quantum group symmetry requires some effort. Also, we would like to work with closed spin chains so that the theory in the continuum limit is defined on a cylinder and we do not need to worry about the boundary. For example, one could try to start with the well known closed XXX spin chain that is invariant under global $SU(2)$ symmetry and try to deform it to the $U_q(sl_2)$ invariant spin chain. A naive deformation to the standard closed XXZ spin chain preserves only a $U(1)$ global symmetry \cite{Franchini_2017}.\footnote{Strictly speaking, the global symmetry of the closed XXZ spin chain is $O(2) = U(1) \rtimes \mathbb{Z}_2$; and the global symmetry of the closed XXX spin chain is $SO(3)$ (if one defines a symmetry as transformations of operators instead of states in the Hilbert space \cite{Cheng:2022sgb}). However, throughout the text we will use the historical notations of $U(1)$ and $SU(2)$ respectively.} The main obstruction to the invariance of the closed spin chain under the quantum group is the interaction term between the last and the first spins which breaks the symmetry. One way to recover the $U_q(sl_2)$ symmetry is to consider an open XXZ spin chain with special boundary, i.e. imaginary surface terms, conditions as it was shown in \cite{Pasquier:1989kd}. In turn, for the closed XXZ one has to replace the coupling of the last and the first spins with a ``non-local'' term \cite{martin1991potts,grosse1994quantum}.

We consider a closed spin chain of XXZ type modified by a non-local term following \cite{grosse1994quantum} and study its continuum limit.\footnote{Let us remind the reader what is the thermodynamic limit of the standard closed XXZ spin chain. The spin chain is given by the Hamiltonian \cite{baxter2016exactly}
\begin{equation*}
    \mathcal{H}^{\text{XXZ}} = -\frac{1}{2} \sum_{i=1}^{L} \bigg( \sigma_{i}^{x}\sigma_{i+1}^{x} + \sigma_{i}^{y}\sigma_{i+1}^{y} + \Delta \sigma_{i}^{z}\sigma_{i+1}^{z} \bigg).
\end{equation*}
For $-1\le\Delta<1$, the continuum limit of the theory is described by a compact boson CFT with $c=1$. The radius of the boson varies with $\Delta$. This result is well studied both analytically \cite{kadanoff1979correlation,lukyanov1998low} and numerically \cite{alcaraz1987conformal,Alcaraz:1987zr}, for a more recent discussion see \cite{Cheng:2022sgb}. The first special value of the coupling is $\Delta=1$ at which the spin chain becomes a ferromagnetic XXX chain (with our choice of sign) with highly degenerate ground state (which is no longer described by a CFT in the continuum limit). The second special value is $\Delta=-1$, the spin chain is equivalent to an antiferromagnetic XXX chain (one can obtain a Hamiltonian with all ``+'' signs after unitary rotation of every other spin by $U=e^{i \pi\sigma^{z}/2}$ which leaves the eigenvalues invariant). The latter flows to a compact boson CFT with $c=1$ and self-dual radius of the boson ($SU(2)_1$ WZW theory) with inherited $SU(2)$ symmetry. For the other values of $\Delta$ the spin chain spectrum is gapped. See for example \cite{Franchini_2017} for the detailed description.} In what follows we refer to this model as XXZ$_q$ spin chain. The spin chain is defined on an even number $L$ of lattice sites with the Hamiltonian given by
\begin{equation}\label{hamilt_XXZ_quantum_closed}
    \mathcal{H} = Lq - \sum_{i=1}^{L-1} R_i - R_0,
\end{equation}
where $R_i$ for $i \in 1, \ldots, L-1$ are defined as
\begin{equation}\label{R_i_explicit}
    R_{i} = \frac{1}{2} \bigg[ \sigma_{i}^{x} \sigma_{i+1}^{x} + \sigma_{i}^{y} \sigma_{i+1}^{y} + \frac{q+q^{-1}}{2} \big( \sigma_{i}^{z}\sigma_{i+1}^{z} + 1 \big) - \frac{q-q^{-1}}{2} \big( \sigma_{i}^{z} - \sigma_{i+1}^{z} - 2 \big) \bigg]
\end{equation}
and represent nearest-neighbour interactions. $\sigma_i^a$ are the standard Pauli matrices at the $i$-th lattice site. For now $q$ can be any complex number; later, when considering the continuum limit, we will specify $q$ to be on the unit circle.
For $i \in 1, \ldots L-1$, the $R_i$ satisfy the Hecke algebra relations \cite{Jones:1987dy}
\begin{equation}\label{Hecke_rels}
    R_i R_{i+1} R_i = R_{i+1} R_i R_{i+1}, \qquad \qquad R_i^2 = (q-q^{-1}) R_i + 1.
\end{equation}
As a consequence, $R_{i}^{-1} = R_i - (q-q^{-1})$.

What is left to do is to define $R_0$. 
Were we to set this term to $0$, we would end up with the open spin chain Hamiltonian of \cite{Pasquier:1989kd}.
$R_0$ must be different from \eqref{R_i_explicit} with the identification $\sigma_{L+i} = \sigma_{i}$, as this would give us the usual XXZ spin chain with periodic boundary conditions (PBC), which we know is not $U_q(sl_2)$ symmetric. Instead, in order to define $R_L$, we define the operator $G=R_1 R_2 \ldots R_{L-1}$, which acts like a one-site lattice translation on the $R_i$, as can be checked by using \eqref{Hecke_rels}
\begin{equation}
    G R_i G^{-1} = R_{i+1}, \qquad i=1,\ldots,L-2 \label{eq:G_translation}
\end{equation}
Then we use this property in order to define the operator $R_0$ as \cite{grosse1994quantum}
\begin{equation}
        R_0 \equiv R_L = G R_{L-1} G^{-1}.
\end{equation}
With this equation the property \eqref{eq:G_translation} becomes true for any $i$ with the identification $R_{L+i} = R_i$,\footnote{The cases of $i=1,\ldots,L-2$ are already mentioned and the case of $i=L-1$ is by definition. Here the nontrivial check is the case of $i=L$.} and $G$ acts like a translation operator on the $R_i$ and commutes with the Hamiltonian \eqref{hamilt_XXZ_quantum_closed}. An explicit formula for $R_0$ in terms of sigma matrices looks nothing like \eqref{R_i_explicit} and is pretty cumbersome. However, the main peculiarity of $R_0$ is that it is highly non-local:
it contains $L$-site interactions between all the sites in the spin chain.

This way, we end up with a $U_q(sl_2)$ invariant Hamiltonian. To see this let us introduce the global generators of $U_q(sl_2)$ on the lattice. We start by defining generators $E, F, q^H$ on a single spin site and then use the coproduct \eqref{coprod} to extend them on the full spin chain: 
\begin{equation}\label{gen_sl2q_lattice}
    \begin{gathered}
        q^{H} = q^{\sigma^{z}} \otimes \dots \otimes q^{\sigma^{z}} \\
        E = \sum_{i=1}^{L} q^{-\sigma^{z}} \otimes \dots \otimes q^{-\sigma^{z}} \otimes \sigma_{i}^{+} q^{-\frac{\sigma^z}{2}} 
        \otimes 1 \otimes \dots \otimes 1 \\
        F = \sum_{i=1}^{L} 1 \otimes \dots \otimes 1 \otimes q^{\frac{\sigma^{z}}{2}} 
        \sigma_{i}^{-} \otimes q^{\sigma^{z}} \otimes \dots \otimes q^{\sigma^{z}}
    \end{gathered}
\end{equation}
Here $\sigma^{\pm}_i = \frac{1}{2}(\sigma_i^x \pm i\sigma_i^y)$. The $U_q(sl_2)$ generators defined in this way commute with the Hamiltonian for every $L$. Schematically, it can be written as 
\begin{equation}
    [\mathcal{H}, U_q(sl_2)] = 0
\end{equation}

There is a particular reason why the Hamiltonian chosen in such way commutes with the quantum group. It follows from a Hilbert space structure of the spin chain and the structure of operators acting on it. Everything that commutes with $U_q(sl_2)$ in the space of operators acting on the Hilbert space is called a centralizer. In our case the centralizer is a Temperley-Lieb (TL) algebra, the elements of which can be expressed via $R_i$'s.\footnote{The Temperley-Lieb algebra has generators $e_i$ which satisfy the following relations \cite{temperley1971relations}
\begin{equation*}
    e_i^2 = (q+q^{-1})e_i, \qquad e_i e_{i\pm1} e_i = e_i\,, \qquad [e_i,e_j]=0 \text{   if   } |i-j|\ge2\,.
\end{equation*}
These generators are obtained from the Hecke algebra elements \eqref{R_i_explicit} by
\begin{equation*}
    e_i = q-R_i\,,
\end{equation*}
which, together with \eqref{Hecke_rels}, imply the TL relations. Note that generically Hecke algebra has different dimensionality from the TL algebra, however, since the Hilbert space is made out of spin-1/2 representations of $U_q(sl_2)$ not all elements of Hecke algebra are linearly independent. Thus, in the case of $U_q(sl_2)$ the proper quotient of the Hecke algebra gives a TL algebra \cite{Jones:1987dy}. 
}$^, \,$\footnote{Sometimes, to distinguish between open and closed boundary conditions it is proposed to consider TL or Jones-TL algebras respectively, which differ by existence of a translation generator. For precise details on these algebras see for example \cite{read2007enlarged}.}
Then, since the Hamiltonian \eqref{hamilt_XXZ_quantum_closed} is expressed in terms of elements of the TL algebra it must commute with the quantum group for any length of the spin chain $L$. In the continuum limit, for $q=e^{i\varphi}$, the TL algebra becomes the Virasoro algebra which commutes with $U_q(sl_2)$ \cite{read2007enlarged, koo1994representations, read2007associative}. The structure of the Hilbert space is also determined by $U_q(sl_2)$ and TL. For generic $q$ the Hilbert space of the spin chain can be decomposed into a following direct sum
\begin{equation}\label{Schur_Weyl_2}
    \underbrace{\mathbb{V}_\frac{1}{2} \otimes \ldots \otimes \mathbb{V}_\frac{1}{2}}_{L \; \text{times}} = \bigoplus_{j} \mathbb{W}_j \otimes \mathbb{V}_j,
\end{equation}
where $\mathbb{W}_j$ denotes irreducible representations of TL, and $\mathbb{V}_j$ stands for those of $U_q(sl_2)$. Here $j$ goes over all representations allowed by the theory. In the continuum limit $\mathbb{V}_j$ survive and $\mathbb{W}_j$ become Virasoro representations. Such a structure of a Hilbert space and operators on it is known as a (quantum generalization of) Schur-Weyl duality \cite{Jimbo:1985vd}, which is a statement about dual structure of $U_q(sl_2)$ and TL. When $q$ is a root of unity the quantum group is no longer semisimple, and the structure of the Hilbert space is different from \eqref{Schur_Weyl_2}. We study it in detail in section \ref{sec:ising}.

For generic values of $q \in \mathbb{C}$, it can easily be checked that the Hamiltonian \eqref{hamilt_XXZ_quantum_closed} is not Hermitian $\mathcal{H}^{\dagger} \ne \mathcal{H}$, so it follows that the theory is not unitary.\footnote{One can notice that the Hamiltonian is also not symmetric $\mathcal{H}^T \ne \mathcal{H}$. This means that the left and right eigenvectors are related in a non-trivial manner, and defining a nice inner product on the lattice requires some work. Given that we are ultimately interested in the continuum theory, we do not pay attention to this obstacle.} The spectrum for generic $q$ is complicated, and the eigenvalues of \eqref{hamilt_XXZ_quantum_closed} are complex. The situation becomes nicer in two particular cases: $q$ real and $q$ on the unit circle; for both these cases the energy levels are either real or come in complex conjugated pairs. When $q$ is real, then all the matrix elements of the Hamiltonian are real, which immediately implies the statement. For the case $q=e^{i\varphi}$ one can introduce the operator \begin{equation}
    \mathcal{X} = \prod_{i=1}^{L} \sigma_i^x
\end{equation}
that acts on the Hamiltonian as
\begin{equation}\label{PT_sym}
    \mathcal{X} \mathcal{H} \mathcal{X}^{-1} = \mathcal{H}^T\left( q^{-1} \right) \equiv \mathcal{H}^\dagger.
\end{equation}
The Hamiltonian is self-adjoint under the $\mathcal{X}$ metric instead of the standard metric. Hamiltonians of such type are known in the literature as $PT$-symmetric (or sometimes called pseudo-Hermitian), see  for example \cite{bender2007making,mannheim2013pt}. The property \eqref{PT_sym} implies the statement about the eigenvalues, given that the similarity transformation does not change the characteristic equation and we have the following sequence of equalities \begin{equation}
    \det(\mathcal{H}-\lambda \mathbb{1}) = \det(\mathcal{H}^\dagger-\lambda \mathbb{1}) = \det((\mathcal{H}^*)^T - \lambda \mathbb{1}) = \det(\mathcal{H}^* - \lambda\mathbb{1}) = 0,
\end{equation}
where the last equality was obtained by the fact that transposition can also be performed by a similarity transformation. This implies that the characteristic equation is real, thus the spectrum is either real or comes in complex conjugated pairs. Furthermore, the explicit solution in \cite{grosse1994quantum, pallua1998closed} indicate that for $q=e^{i\varphi}$ the spectrum is only real. 
As usual, everything becomes more complicated when $q$ is a root of unity: in this case the Hamiltonian cannot be diagonalized, but only brought to a Jordan form. From the continuum theory point of view, this will play a role in Section \ref{sec:ising}.

Let us now comment on certain limits of the Hamiltonian \eqref{hamilt_XXZ_quantum_closed}. In the $q \to 1$ limit the Hamiltonian becomes the ferromagnetic XXX Hamiltonian (because of the minus sign in front in our convention), and the quantum group becomes just $SU(2)$. This theory has highly degenerate ground state with magnon excitations and is not described by a CFT in the continuum limit. 

From the point of view of CFTs, the limit $q \to -1$ is more interesting, and \eqref{hamilt_XXZ_quantum_closed} becomes
\begin{equation}
    \mathcal{H}(q=-1) = \frac{L}{2} + \frac{1}{2} \sum_{i=1}^{L} \bigg( -\sigma_{i}^{x}\sigma_{i+1}^{x} -\sigma_{i}^{y}\sigma_{i+1}^{y} + \sigma_{i}^{z}\sigma_{i+1}^{z} \bigg)
\end{equation}
Note that the sum is now up to $L$ instead of $L-1$ in \eqref{hamilt_XXZ_quantum_closed}. This happens because, for even number of spins $L$, the non-local $R_0$ term becomes the standard local interaction term between the first and the last spin. One can apply the unitary transformation
\begin{equation}\label{U_sign_change}
    U = \exp \bigg( i \frac{\pi}{2} \sum_{j=1}^{L} j\sigma_j^{z} \bigg),
\end{equation}
which rotates every other spin around $z$ axis. This changes the signs in front of $x$ and $y$ interaction terms and we get the standard antiferromagnetic XXX Hamiltonian
\begin{equation}\label{AFM_XXX}
    U\mathcal{H}(q=-1)U^{-1} = \frac{L}{2} + \frac{1}{2} \sum_{i=1}^{L} \bigg( \sigma_{i}^{x}\sigma_{i+1}^{x} + \sigma_{i}^{y}\sigma_{i+1}^{y} + \sigma_{i}^{z}\sigma_{i+1}^{z} \bigg)
\end{equation}
which flows to a compact boson of self-dual radius with $c=1$ CFT ($SU(2)_1$ WZW theory) in the continuum limit. Recalling that we study the spin chain on even number of spin sites, the Hilbert space would only contain representations of integer spin for which $(-1)^H=1$ holds. Thus, the quantum group commutation relations \eqref{comm_rel_Uqsl2} in this limit become
\begin{equation}
    [E,F] = -H, \qquad [H,E] = 2E, \qquad [H,F] = -2F,
\end{equation}
After the redefinition $E \to iE, F \to iF$ one gets the standard $su_2$ commutation relations.

Finally, let us mention that the open XXZ spin chain with special boundary conditions invariant under quantum group symmetry considered in \cite{Pasquier:1989kd} is related to \eqref{hamilt_XXZ_quantum_closed} without the $R_0$ term by applying the transformation \eqref{U_sign_change} together with changing $q \to \frac{1}{q}$.

\subsection{Partition function and spectrum}
\label{sec:partFn}
In this section we review the results of \cite{pallua1998closed} on the derivation of the spectrum of the XXZ$_q$ model \eqref{hamilt_XXZ_quantum_closed} in the continuum limit. Let us take $q$ to be on the unit circle: in this case the spin chain is gapless and flows to a CFT in the continuum limit. The spin chain is integrable by the Bethe ansatz \cite{grosse1994quantum}. Let us also choose a parametrization 
\begin{equation}\label{q_param}
    q=e^{i\varphi}=e^{i\pi \frac{\mu}{\mu+1}}, \qquad \mu>0.
\end{equation}
In this parametrization, the ground state is unique and is a singlet under the quantum group for $\frac{\pi}{2} < \varphi < \pi$, which in terms of $\mu$ reads as $\mu>1$. All our continuum calculations fall into these limits.\footnote{Let us mention that for $0<\varphi<\frac{\pi}{2}$ the quantum group spin of the ground state is non-zero and depends on $\varphi$, which means that the ground state is degenerate.} It is known \cite{Alcaraz:1987zr,alcaraz1987conformal} that, for the closed critical spin chain, the ground state energy scales with the length $L$ as
\begin{equation}
    E_0^\mu(L) = L e_{\infty}^\mu - \frac{\pi \zeta^\mu c}{6L} + O(L^{-2}),
\end{equation}
where $c$ is the central charge of the corresponding conformal field theory; $e_\infty^\mu$ is a ``vacuum energy density", it depends only on $\mu$ and can be computed exactly in the $L\to \infty$ limit with the help of the Bethe ansatz computations \cite{grosse1994quantum}; $\zeta^\mu$ is a ``speed of sound" on the spin chain and can also be computed by Bethe ansatz: $\zeta^\mu = \pi \frac{\sin(\pi - \varphi)}{\pi - \varphi}$

What is important from the CFT point of view is that in the parametrization \eqref{q_param} the central charge for our model is given by \cite{grosse1994quantum}
\begin{equation}\label{central_charge}
    c = 1 - \frac{6}{\mu(\mu+1)}.
\end{equation}
This formula is widely used to parametrize the central charge of the unitary minimal models $\mathcal{M}_{\mu,\mu+1}$ when $\mu$ is an integer \cite{DiFrancesco:1997nk}. However, in our case we do not restrict to an integer $\mu$; moreover, the case of integer $\mu$ means $q$ being the root of unity, which leads to the more complicated quantum group structure, see Section \ref{sec:ising}.

For the excited states, the scaling behavior is also known to be \cite{Alcaraz:1987zr}
\begin{equation}
    E_i^\mu = E_0^\mu(L) + \frac{2\pi\zeta^\mu}{L} (\Delta_i^\mu + \mathbb{N}) + O(L^{-2})
\end{equation}
from which one can find the scaling dimensions $\Delta_i^\mu = h^\mu_i+\bar{h}^\mu_i$ of the primary operators of the theory. We ommit the dependence on $\mu$ from the notation $h^\mu_i$ and $\bar{h}^\mu_i$ in the rest of this note.

\begin{figure}[h] \centering
\includegraphics[height=0.3\textheight]{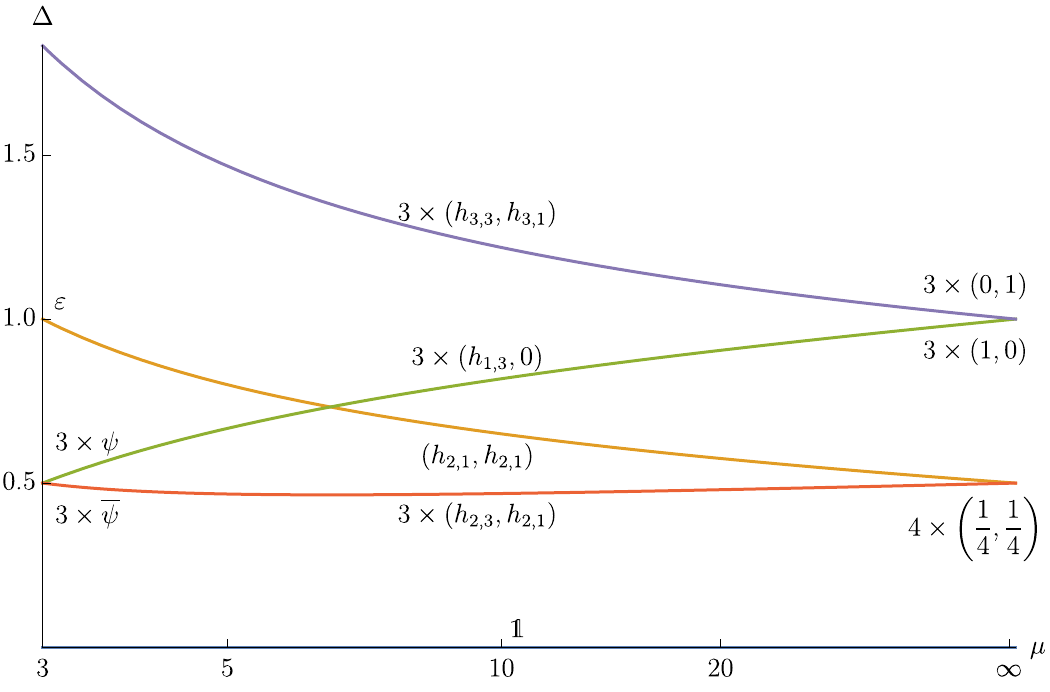}
\caption{Energy plot $\Delta(\mu)$}
\label{plot_delta_of_mu}
\end{figure}

\begin{figure}[h] \centering
\includegraphics[height=0.3\textheight]{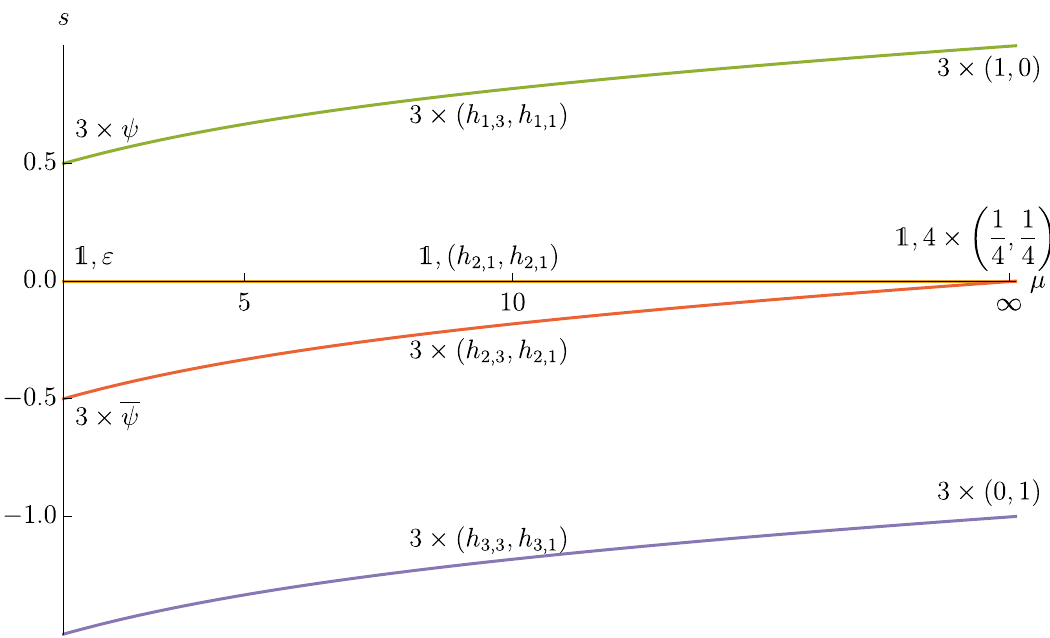}
\caption{Spacetime spin plot $s(\mu)$}
\label{plot_spin_of_mu}
\end{figure}

Let us now proceed to the partition function of the XXZ$_q$ spin chain computed in \cite{pallua1998closed}. The spectrum of the theory consists only of degenerate $c<1$ Virasoro primary fields and their descendants. We use the standard conformal dimensions for these fields from Kac table \cite{DiFrancesco:1997nk}, labeled by two integers $r,s$:
\begin{equation}\label{Kac_dim}
    h_{r,s} = \frac{[r(\mu+1)-s\mu]^2 - 1}{4\mu(\mu+1)}.
\end{equation}
And the characters associated with the degenerate Virasoro representations are given by (using the notation $p = e^{2\pi i \tau}, \bar{p} = e^{2\pi i \bar{\tau}}$)
\begin{equation}\label{degenerate_char}
    \chi_{r,s}(\tau) = \frac{p^{\frac{1}{4\mu(\mu+1)}}}{\eta(\tau)} \left( p^{h_{r,s}} - p^{h_{r,-s}} \right),
\end{equation}
where $\eta(\tau)$ is the Dedekind eta function. Then the torus partition function of the XXZ$_q$ CFT is expressed as \begin{equation}\label{part_fn}
    Z (\tau, \bar{\tau}) = \sum_{\ell=0}^{\infty} (2\ell+1) \sum_{r=1}^{\infty} \chi_{r,2\ell+1}(\tau) \chi_{r,1}(\bar{\tau})
\end{equation}
The $(2\ell+1)$ degeneracies correspond to the quantum group degeneracies of representation of spin $\ell$. Note that there are infinitely many primary operators of a given spin $\ell$. Also note that $\ell$ can only be an integer number, which is a consequence of taking only an even number of spin sites $L$. To develop some intuition about the spectrum we plot the several lowest energy states as a function of $\mu$ in Figure \ref{plot_delta_of_mu} and their Lorentz spin as a function of $\mu$ in Figure \ref{plot_spin_of_mu}. 

Let us mention a few intriguing properties of the partition function. First, the theory is clearly not chiral symmetric, as can be observed already from the first excited states in an expansion of \eqref{part_fn}. As explained in Section \ref{sec:results}, this is to be expected from the properties of quantum group symmetric QFTs: see e.g. \eqref{spin:option1}, which shows that, for generic $q$, an operator with spin $s$ and an operator with spin $-s$ cannot both transform in the same spin $\ell$ representation of $U_q(sl_2)$.
Second, the partition function is not modular invariant, as can be easily checked numerically. The reason for this is that the Hamiltonian \eqref{hamilt_XXZ_quantum_closed} is not translationally invariant.\footnote{We mentioned earlier that the Hamiltonian \eqref{hamilt_XXZ_quantum_closed} commutes with the operator $G=R_1 \ldots R_{L_1}$, which acts as a one-site translation operator on the Hecke algebra elements $R_i$ \eqref{eq:G_translation}. However, $G$ does not act like a one-site translation on generic operators, as can be checked e.g. from the fact that $G^L\neq 1$.}

From the partition function \eqref{part_fn}, we can read off the operator content of the theory. The generic primary operators corresponding to the states of dimension $(h_{r,2\ell+1},h_{r,1})$ are in representation of spin-$\ell$ under the quantum group. We denote them as
\begin{equation}\label{general_ops}
    W_{r,2\ell+1}^m \longleftrightarrow (h_{r,2\ell+1},h_{r,1}),
\end{equation}
where the index $m \in \{-\ell, \ldots, \ell\}$ is the index transformed by $U_q(sl_2)$. It can be checked that the relation between spacetime spin $s=h-\bar{h}$ and $U_q(sl_2)$ spin $\ell$ exactly agrees with \eqref{spin:option1}. Notice that there is also a purely chiral subsector of a theory which consists of operators $W^m_{1,2\ell+1}$ of conformal dimension $(h_{1,2\ell+1},h_{1,1}=0)$. These fields also transform in the spin-$\ell$ representation of the quantum group, however, for these fields there is a single primary field for a given $\ell$ (up to quantum group degeneracies).

We will study OPE coefficients using crossing symmetry in the case of chiral operators $W_{1,2\ell+1}^m$ and general operators \eqref{general_ops} respectively in Section \ref{sec:chiral_OPE_coeffs} and Section \ref{sec:general_OPE_coeffs}. 

\subsubsection*{Compact boson of self-dual radius, \texorpdfstring{$\mu \to \infty$}{}}
Now let us study the limits of the partition function and first focus on $\mu \to \infty$, which corresponds to the $q \to -1$ limit for the lattice Hamiltonian. First of all, notice that in this limit the central charge \eqref{central_charge} becomes $c=1$. Second, from \eqref{Kac_dim} one can see that $h_{r,s} \to \frac{(r-s)^2}{4}$. Then the partition function \eqref{part_fn} in this limit becomes exactly the partition function of $SU(2)_{1}$ WZW theory as predicted from the lattice manipulations \eqref{AFM_XXX}. Note that at this limit the chirality invariance is restored. Let us provide an explicit resummation of the partition function in this limit and prove this claim.

Recall that partition function of $SU(2)_1$ WZW theory can be written as
\begin{equation}\label{su2level1_partFn}
    Z_{\widehat{su(2)}_1}(\tau,\bar{\tau}) = \widehat{\chi}_{0}(\tau) \widehat{\chi}_{0}(\bar{\tau}) + \widehat{\chi}_{\frac{1}{2}}(\tau) \widehat{\chi}_{\frac{1}{2}}(\bar{\tau})
\end{equation}
where $\widehat{\chi}_i(\tau)$ are the characters of $\widehat{su(2)}$ given by:
\begin{equation}
    \begin{aligned}
        \widehat{\chi}_{0}(\tau) &= \frac{1}{\eta(\tau)} \sum_{j=0}^{\infty} (2j+1) \bigg( p^{j^2} - p^{(j+1)^2} \bigg) \\
        \widehat{\chi}_{\frac{1}{2}}(\tau) &= \frac{1}{\eta(\bar{\tau})} \sum_{j=0}^{\infty} (2j+2) \bigg( p^{\left(j+\frac{1}{2}\right)^2} - p^{\left(j+\frac{3}{2}\right)^2} \bigg)
    \end{aligned}
\end{equation}
The partition function \eqref{part_fn} in the $\mu \to \infty$ limit becomes
\begin{equation}
    Z_{\mu \to \infty}(\tau,\bar{\tau}) = \frac{1}{|\eta(\tau)|^2} \sum_{r=1}^{\infty} \sum_{\ell=0}^{\infty} (2\ell+1) \bigg( p^{\frac{(r-2l-1)^2}{4}} - p^{\frac{(r+2l+1)^2}{4}} \bigg) \bigg( \bar{p}^{\frac{(r-1)^2}{4}} - \bar{p}^{\frac{(r+1)^2}{4}} \bigg).
\end{equation}
Let us consider only odd $r = 2k+1$, from the sum, then we get
\begin{equation}
    Z_{\mu \to \infty}^{\text{odd}}(\tau,\bar{\tau}) = \frac{1}{\eta(\bar{\tau})} \sum_{k=0}^{\infty} \bigg( \bar{p}^{k^2} - \bar{p}^{(k+1)^2} \bigg) \underbrace{\frac{1}{\eta(\tau)} \sum_{\ell=0}^{\infty} (2\ell+1) \bigg( p^{(\ell-k)^2} - p^{(\ell+k+1)^2} \bigg)}_{\zeta_k(\tau)}
\end{equation}
Now it is easy to notice that 
\begin{equation}
    \zeta_k(\tau) = (2k+1) \big[ 1+2p+2p^4+2p^9+\ldots \big] = (2k+1)\widehat{\chi}_0(\tau),
\end{equation}
which provides that $Z_{\mu \to \infty}^{\text{odd}}(\tau,\bar{\tau}) = \widehat{\chi}_{0}(\tau) \widehat{\chi}_{0}(\bar{\tau})$. Repeating the same computations for even $r=2k+2$ we get the second part of the partition function: $Z_{\mu \to \infty}^{\text{even}}(\tau,\bar{\tau}) = \widehat{\chi}_{\frac{1}{2}}(\tau) \widehat{\chi}_{\frac{1}{2}}(\bar{\tau})$. Finally, the total partition function of the XXZ$_q$ spin chain in the limit $\mu \to \infty$, which is the sum of even and odd parts, gives exactly the partition function \eqref{su2level1_partFn} of $SU(2)_1$ WZW theory.

Let us now track which states become the first light operators in the $\mu \to \infty$ limit. The four operators of dimension $\left( \frac{1}{4},\frac{1}{4} \right)$ in the WZW theory fall into $1+3$ for finite $\mu$:
\begin{equation}
    1 \times \left( h_{2,1}, h_{2,1} \right) + 3 \times \left( h_{2,3}, h_{2,1} \right) \to 4\times \left( \frac{1}{4}, \frac{1}{4} \right)
\end{equation}
Whereas both chiral and antichiral currents stay in spin-1 representation:
\begin{equation}
    3\times \left( h_{1,3}, h_{1,1} \right) \to 3\times\left( 1, 0 \right), \qquad\qquad 3\times \left( h_{3,3}, h_{3,1} \right) \to 3\times\left( 0, 1 \right)
\end{equation}
But for finite $\mu$, of course, these are spin-1 representations of a quantum group. These results confirm the lattice expectations: the quantum group in $\mu \to \infty$ limit becomes the lattice $SU(2)_{\text{lat}}$, which in its turn is known to become a diagonal embedding into the symmetry group $SU(2)_L \times SU(2)_R$ of WZW theory.

\subsubsection*{Ising limit, \texorpdfstring{$\mu = 3$}{}}
Another important limit which we study is $\mu \to 3$. The central charge \eqref{central_charge} in this limit is $c=\frac{1}{2}$. Let us first look at the first few terms of the partition function in this limit: \begin{equation}\label{ising_partFn}
    Z_{\mu=3}(p,\bar{p}) = p^{\frac{1}{48}} \bigg[ 1 + \bigg( 3p^{\frac{1}{2}} + 3\bar{p}^{\frac{1}{2}} \bigg) +  \bigg( p^{\frac{1}{2}}\bar{p}^{\frac{1}{2}} \bigg) + \dots \bigg]
\end{equation}
Up to this order the partition function looks chiraly symmetric, however, in the higher order terms the chiral invariance is broken. Let us recall that the local Ising operators labeled by their conformal dimensions are
\begin{equation}
    \mathbb{1} = \left(0,0\right), \qquad \varepsilon = \left(\frac{1}{2},\frac{1}{2}\right), \qquad \sigma = \left(\frac{1}{16},\frac{1}{16}\right),
\end{equation}
while the $\mathbb{Z}_2$ twisted Ising subsector contains the fields
\begin{equation}
    \mu = \left(\frac{1}{16},\frac{1}{16}\right), \qquad \psi = \left( \frac{1}{2},0 \right), \qquad \overline{\psi} = \left(0,\frac{1}{2}\right).
\end{equation}
It is natural to conclude that \eqref{ising_partFn} contains fields $\mathbb{1}, \varepsilon, \psi, \overline{\psi}$. However, the fields of dimension $(1/2,0)$ and $(0,1/2)$ are three-fold degenerate, because they belong to spin-1 $U_q(sl_2)$ multiplets. In the section \ref{sec:ising} we comment on the constructing $\psi, \overline{\psi}$ out of these multiplets. For now
let us just keep track of the operators which become the above mentioned fields in the $\mu \to 3$ limit:
\begin{equation}\label{isingOps}
    \begin{split}
        \left( h_{1,1}, h_{1,1} \right) \to& \;\; \mathbb{1} = (0,0) \\
        \left( h_{2,1}, h_{2,1} \right) \to& \;\; \varepsilon = \left(\frac{1}{2},\frac{1}{2}\right) \\
        3\times\left( h_{1,3}, h_{1,1} \right) \to& \;\; 3\times\left(\frac{1}{2},0\right) \\
        3\times\left( h_{2,3}, h_{2,1} \right) \to& \;\; 3\times \left(0,\frac{1}{2}\right)
    \end{split}
\end{equation}
In \cite{pallua1996minimal} the limits of $\mu$ to the other integer numbers were considered, where the fields of other minimal models are recovered. Note that for $\mu=3$, the quantum group parameter $q=e^{i\pi \frac{3}{4}}$ becomes a root of unity. The structure of higher dimensional $U_q(sl_2)$ representations as well as Virasoro representations is much more complicated and will be discussed in section \ref{sec:ising}.

\subsection{OPE coefficients from crossing symmetry} \label{sec:OPE_xing}

From the partition function \eqref{part_fn} we know the scaling dimension of operators in the theory. What we are missing, however, are the OPE coefficients of these operators.
In order to fix this, we can use crossing symmetry of four point functions.
Given that the operators of the theory are degenerate, e.g. their scaling dimensions are given by \eqref{Kac_dim} with integer $r$ and $s$ for any $\mu$, these operators satisfy BPZ differential equations. This makes the job of computing Virasoro blocks much easier. 

One way to see this is that the operartor $W_{r,s}$ has a null descendant at level $r s$ in the holomorphic sector, which means that its correlation functions satisfy a degree $r s$ differential equation \`a la BPZ \cite{Belavin:1984vu}. The operator $W_{r,s}$ satisfies\begin{equation}
    \left(\sum_{N=1}^{rs} \sum_{\substack{p_1 \geqslant \ldots \geqslant p_N \geqslant 1 \\ p_1 + \ldots + p_N = rs}} A_{p_1, \ldots, p_N} \, L_{-p_1} L_{-p_2} \ldots L_{-p_N} \right) W_{r,s} = 0
\end{equation}
where the coefficients \(A_{p_1, \ldots, p_N}\) are fixed (up to an overall constant factor) by the Virasoro algebra.\footnote{The explicit form of the BPZ operator is unimportant in this work. We would like to comment that it has been worked out in the case of \(r=1\) or \(s=1\) by Benoit and Saint-Aubin \cite{Benoit:1988aw}. For general \((r, s)\), to the authors' knowledge, it remains an open problem. However, for practical computations, there exists an algorithm developed by Bauer, Di Francesco, Itzykson and Zuber \cite{Bauer:1991qm,Bauer:1991ai}.}

The consequence is that correlation functions with an insertion of $W_{rs}$ satisfy a $rs$ order differential equation \cite{DiFrancesco:1997nk}
\begin{equation}
   \mathcal{L}_{r,s} \braket{W_{r,s}(z) \calO_1(x_1) \ldots \calO_n(x_n)} = 0
\end{equation}
given by
\begin{equation}\label{BPZ_operator}
    \mathcal{L}_{r,s} = \sum_{N=1}^{rs} \sum_{\substack{p_1 \geqslant \ldots \geqslant p_N \geqslant 1 \\ p_1 + \ldots + p_N = rs}} A_{p_1, \ldots, p_N} \, \calL_{-p_1} \calL_{-p_2} \ldots \calL_{-p_N}
\end{equation}
where 
\begin{equation}\label{def:DiffVir}
\mathcal{L}_{-p} = \sum_{i} \left[ \frac{(p-1)h_{i}}{(x_i - z)^{p}} - \frac{1}{(x_i - z)^{p-1}} \frac{\partial}{\partial x_i} \right],
\end{equation}
Similarly, correlation functions of $W_{r,s}$ also satisfy a differential equation of order $r$ in the antiholomorphic sector, which follows from acting with the differential operator $\overline{\mathcal L}_{r,1}$ on the correlation functions.
This allows us, in principle, to obtain the Virasoro blocks and then fix the OPE coefficients of the theory; in practice, the behaviour of Virasoro blocks under crossing for $c<1$ is known \cite{Dotsenko:1984four}, and this made the explicit computation of the Virasoro blocks unnecessary. Just by using a few known properties of Virasoro blocks under crossing, we will be able to find the OPE coefficients up to a discrete phase: we will show that the square of OPE coefficients of our theory are related to the OPE coefficients of generalized minimal models with a $\pm$ ambiguity.\footnote{The unitary series of minimal models $\mathcal{M}_{\mu,\mu+1}$ is labeled by an integer $\mu$. Throughout this work, by minimal models we generically consider generalized minimal models, i.e. $\mathcal{M}_{\mu,\mu+1}$ with $\mu \in \mathbb{R}$.}

An important consequence of having degenerate operators, which can for example be seen from the solutions to the BPZ equations in an OPE limit, is that the fusion rules are fixed. Looking only at e.g. the holomorphic sector of the theory, we have that the OPE of two operators with holomorphic weight $h_{r_1,s_1}$ and $h_{r_2,s_2}$ can only exchange
\begin{equation}\label{eq:MM_fusion_rule}
    h_{r_1,s_1} \times h_{r_2,s_2} = \sum_{\substack{r_3=\abs{r_1-r_2}+1 \\ r_1+r_2+r_3\ {\rm odd}}}^{r_1+r_2-1}\,\sum_{\substack{s_3=\abs{s_1-s_2}+1 \\ s_1+s_2+s_3\ {\rm odd}}}^{s_1+s_2-1} h_{r_3,s_3}\,.
\end{equation}
Combining the holomorphic and antiholomorphic sectors, we see that our CFT has the fusion rules
\begin{equation} \label{eq:Wfusion}
    W_{r_1,s_1} \times W_{r_2,s_2} = \sum_{\substack{r_3=\abs{r_1-r_2}+1 \\ r_1+r_2+r_3\ {\rm odd}}}^{r_1+r_2-1}\,\sum_{\substack{s_3=\abs{s_1-s_2}+1 \\ s_1+s_2+s_3\ {\rm odd}}}^{s_1+s_2-1} W_{r_3,s_3}\,.
\end{equation}
It's important to remember that operator $W_{r,s}$ transforms in a spin $\ell = \frac{s-1}2$ representation of $U_q(sl_2)$, and these fusion rules agree with the decomposition rules of tensor product of $U_q(sl_2)$ representation.

Finally, we will denote OPE coefficients by the Kac labels of operators, so that e.g. the OPE \eqref{CFT:OPE} in our theory is
\begin{equation}\label{OPE_W}
    W_{r_1,s_1}^{m_1}(z_1) W_{r_2,s_2}^{m_2}(z_2)= \sum_{\substack{\text{allowed } \\r_3,s_3}} \sum_{m_3} \frac{C_{(r_1,s_1),(r_2,s_2),(r_3,s_3)}}{z_{21}^{h_{123}} \bz_{21}^{\hb_{123}}} \QCG{\frac{s_1-1}2,m_1,\frac{s_2-1}2,m_2,\frac{s_3-1}2,m_3,q} \left(W_{r_3,s_3}^{m_3}(z_1)+\ldots \right).
\end{equation}
where again we use the shorthand notation $h_{123} = h_1+h_2-h_3=h_{r_1,s_1}+h_{r_2,s_2}-h_{r_3,s_3}$.

\subsubsection{Relations of OPE coefficients}
Before going through with the bootstrap analysis, let us determine the relation of OPE coefficients under the permutation of operator orderings. This relation is given in \eqref{CFT:OPErelation}; it remains to determine the numbers $n_i$ for the case of the XXZ$_q$ CFT. These are fixed straightforwardly by \eqref{CFT:spinrelation} and \eqref{Kac_dim}. We get that the spacetime spin of operator $W_{r,s}$ is
\begin{equation}\label{XXZ:spin}
    h_{r,s}-h_{r,1}=\frac{\ell(\ell+1)}{\pi i}\log(q)- r\ell\quad(s=2\ell+1).
\end{equation}
where we use that $q=e^{i\pi\frac{\mu}{\mu+1}}$ in the XXZ$_q$ CFT. The spin formula \eqref{XXZ:spin} is consistent with \eqref{CFT:spinrelation} which comes from the assumption that the braiding matrix $M$ in the locality condition \eqref{QFT:localityM} is the same as the $R$-matrix of the quantum group, \eqref{QFT:locality}. Comparing \eqref{XXZ:spin} with \eqref{CFT:spinrelation} we get
\begin{equation}
    n_{r,s}=-(r-1)\ell,\quad(s=2\ell+1).
\end{equation}
Then using \eqref{CFT:OPErelation}, we conclude that the OPE coefficients of the XXZ$_q$ CFT satisfy following relations
\begin{equation}\label{relation:XXZOPE}
    \begin{split}
        C_{(r_1,s_1),(r_2,s_2),(r_3,s_3)}=C_{(r_3,s_3),(r_1,s_1),(r_2,s_2)}=(-1)^{(r_1-1)\ell_1+(r_2-1)\ell_2+(r_3-1)\ell_3}C_{(r_2,s_2),(r_1,s_1),(r_3,s_3)}. 
    \end{split}
\end{equation}
These relations will be important later when we determine the values of the OPE coefficients.

\subsubsection{Chiral operators and crossing kernels}
\label{sec:chiral_OPE_coeffs}

The situation becomes particularly simple when studying the degenerate chiral operators of the theory, $W_{1,s}^m$. These operators have weights $(h_{1,s},0)$ and transform in the the spin $2\ell+1$ representation of $U_q(sl_2)$.
There are two main simplifications: the first is that
the antiholomorphic dependence of the correlation function becomes trivial, $\bar\calF(\bz) = 1$, and equation \eqref{crossing:explicit2} just becomes a statement about the crossing properties of Virasoro blocks. The second one is that the OPE of chiral operators only contains chiral primaries, and given that we have only one chiral operator per spin $\ell \in \mathbb{N}$ representation of the quantum group, we will exchange only one operator per allowed spin in the correlation function.

Indeed, in the chiral case, the fusion rules \eqref{eq:Wfusion} become
\begin{equation}
    W^{m_1}_{1,s_1} \times W^{m_2}_{1,s_2}\sim \sum_{\substack{s=1+|s_1-s_2| \\s+s_1+s_2 = 1 \text{ mod } 2}} ^{s_1+s_2-1}  W^{m_1+m_2}_{1,s}\,.
\end{equation}
Again, the fact that this condition is compatible with the fusion rules of the $U_q(sl_2)$ representations \eqref{tensor_prod_generic} follows from the relation between spacetime spin and $U_q(sl_2)$ spin \eqref{spin:option1}.

Let us now consider the four point function 
$
    \braket{W_{1,s_1}(0)W_{1,s_2}(z) W_{1,s_3}(1) W_{1,s_4}(\infty) }\,.
$
Given that the antiholomorphic blocks are trivial and we exchange one operator per $U_q(sl_2)$ spin, the crossing equations \eqref{crossing:explicit2}reduce to

\begin{equation} \label{eq:crossing_6j}
    \calF^{(t)}_{1,s'}(z) = \sum_s \frac{C_{(1,s_1),(1,s_2),(1,s)}C_{(1,s_3),(1,s_4),(1,s   )}}{C_{(1,s_2),(1,s_3),(1,s')}C_{(1,s_4),(1,s_1),(1,s')}} \sixj{\frac{s_1-1}{2},\frac{s_2-1}{2},\frac{s-1}{2},\frac{s_3-1}{2},\frac{s_4-1}{2},\frac{s'-1}{2}}_q \calF^{(s)}_{1,s}(z)\,.
\end{equation}
For the Virasoro blocks, we avoid writing explicitly dependence on the external dimensions $h_i$, and use $\calF^{(s)}$ for the $s$-channel blocks, and $\calF^{(t)}$ for the $t$-channel ones. They are related by
$\calF^{(t)}_h(z) =  \calF^{(s)}_h(1-z)|_{h_1 \leftrightarrow h_3}$. {Since the exchanged operators are all in the Kac table, we use the notation $\calF^{(s/t)}_{r,s}(z)\equiv\calF^{(s/t)}_{h_{r,s}}(z)$.} Finally, we've used the cyclic property of OPE coefficients \eqref{eq:OPE_cyclic}, $C_{ijk} = C_{jki}$.

Therefore we find that the crossing kernel for Virasoro blocks for degenerate operator is simply a $6j$ symbol times a ratio of OPE coefficients. Notice that the chiral blocks $\calF_j$ are just given by the solution of a differential equation which relies only on the degeneracy of an operator, and therefore are the same that appear in minimal models with the same central charge. Now we will fix the OPE coefficients using \eqref{eq:crossing_6j}.

The most straightforward way to fix the OPE coefficients would be to solve numerically the BPZ equation in the $s$ and $t$-channel, and then impose \eqref{eq:crossing_6j}. However, our life is made easy by the fact that the behavior of $c<1$ blocks under crossing was worked out in \cite{Furlan:1989ra}, see also \cite{esterlis2016closure}, and we can use this together with \eqref{eq:crossing_6j} to fix the OPE coefficients.
This result can be repackaged in the formula \cite{Cremmer:1994kc,Teschner:1995ga}
\begin{equation}\label{fusion:singleQG}
        \calF^{(t)}_{1,s'} (z) = \sum_s  \frac{g_{(1,s_1),(1,s_2)}^{(1,s)} g_{(1,s),(1,s_3)}^{(1,s_4)} }{g_{(1,s_2),(1,s_3)}^{(1,s')} g_{(1,s_1),(1,s')}^{(1,s_4)} }\sixj{\frac{s_1-1}{2},\frac{s_2-1}{2},\frac{s-1}{2},\frac{s_3-1}{2},\frac{s_4-1}{2},\frac{s'-1}{2}}_q  \calF^{(s)}_{1,s}(z)\,.
\end{equation}
where $g$'s are functions that can be deduced from the explicit results in the literature \cite{Dotsenko:1984four,Furlan:1989ra,esterlis2016closure}, or found explicitly in \cite{Cremmer:1994kc}.

Rather than working with the explicit formulas for $g$, we will focus on the simple case of $s_3 = s_1$ and $s_4 = s_2$.
In Appendix \ref{app:fusion_kernel_OPE}, we show that, using crossing of diagonal minimal models, the $g$'s are related to the OPE coefficients of minimal models $C^{\rm MM}$, which were found in \cite{Dotsenko:1984four} (see also \cite{poghossian2014two,esterlis2016closure})\footnote{The fusion rules of generalized minimal models are precisely \eqref{eq:MM_fusion_rule} \cite{Zamolodchikov:2005fy}. When dealing with the explicit expressions of $C^{\rm MM}$ found in the literature, one sometimes has to be careful to consider only $r_i, s_i$ allowed by the fusion rules.} by the formula
\begin{equation}
    \left(C^{\rm{MM}}_{(r_1,s_1),(r_2,s_2),(r,s)}\right)^2 = \beta_{12} \left(g_{(r_1,s_1),(r_2,s_2)}^{(r,s)}g_{(r_1,s_1),(r,s)}^{(r_2,s_2)} \right)^2\, \label{eq:g_to_C}
\end{equation}
with $\beta_{12}$ an unimportant constant that cannot depend on $(r,s)$. Putting together \eqref{eq:crossing_6j} and \eqref{fusion:singleQG} for this choice of operators, we obtain
\begin{equation}
\begin{aligned}\sum_s \left(\frac{C_{(1,s_1),(1,s_2),(1,s)}}{C_{(1,s_2),(1,s_3),(1,s')}} \right)^2 &\sixj{\frac{s_1-1}{2},\frac{s_2-1}{2},\frac{s-1}{2},\frac{s_1-1}{2},\frac{s_2-1}{2},\frac{s'-1}{2}}_q  \calF^{(s)}_{1,s}(z) =\\& = \sum_s  \frac{g_{(1,s_1),(1,s_2)}^{(1,s)} g_{(1,s),(1,s_1)}^{(1,s_2)} }{g_{(1,s_2),(1,s_1)}^{(1,s')} g_{(1,s_1),(1,s')}^{(1,s_2)} }\sixj{\frac{s_1-1}{2},\frac{s_2-1}{2},\frac{s-1}{2},\frac{s_1-1}{2},\frac{s_2-1}{2},\frac{s'-1}{2}}_q   \calF^{(s)}_{(1,s)}(z)\,.
\end{aligned}
\end{equation}

This means that crossing fixes the OPE coefficients of chiral operators in XXZ$_q$ to be proportional to a product of $g's$
\begin{equation}
    \left(C_{(1,s_1),(1,s_2),(1,s_3)}\right)^2 \sim g_{(1,s_1),(1,s_2)}^{(1,s_3)}g_{(1,s_1),(1,s_3)}^{(1,s_2)} 
\end{equation}
To make contact with the minimal model OPE coefficients, it's simpler to work with the fourth power of the OPE coefficients, and we find the relation
\begin{equation}
\left(C_{(1,s_1),(1,s_2),(1,s_3)}\right)^4 = \gamma_{12} \left(C^{\rm MM}_{(1,s_1),(1,s_2),(1,s_3)}\right)^2
\end{equation}
with $\gamma_{12}$ an undetermined constant.
Then it's enough to notice the invariance under permutation of the operators of both $C$ for chiral operators, see \eqref{relation:XXZOPE}, and $C^{\rm MM}$, to determine that $\gamma_{12}$ can only be an overall constant $\gamma_{12} = \gamma$. Then it's enough to require proper normalization of operators $C_{(1,s),(1,s),(1,1)} = 1$ to fix $\gamma=1$.

To summarize, we have shown that for chiral operators in the XXZ$_q$ CFT, their OPE coefficient are related to the usual minimal model ones by a square root
\begin{equation}
    \left(C_{(1,s_1),(1,s_2),(1,s_3)}\right)^2 = \pm C^{\rm MM}_{(1,s_1),(1,s_2),(1,s_3)}\,. \label{eq:C_as_MM_chiral}
\end{equation}
This result, together with the explicit form of the OPE \eqref{CFT:OPE}, is almost enough to allow us to construct all $n$-point functions for the chiral operator of the theory. Our argument does not allow us to determine the sign of the square of the OPE coefficient as should be expected: given any choice of $i,j$, and $k$, we could always redefine operators in minimal models so that $C_{ijk}^{\rm MM} \to -C_{ijk}^{\rm MM} $. However, fixing the physical relative signs, and the overall sign of $\left(C_{(1,s_1),(1,s_2),(1,s_3)}\right)^2$ is a task that we will complete in \cite{paper2}, using different techniques.\footnote{Determining the sign of the square of an OPE coefficient is a trivial task in a unitary theory by reflection positivity, but XXZ$_q$ is a non-unitary theory.}

Let us remark one important fact: while the fusion kernel of Virasoro blocks for $c<1$ \eqref{fusion:singleQG} has long been known, our theory gives another explanation for the appearance of $6j$-symbols in the crossing kernel. 
Indeed, a person that only knows about minimal model might be surprised by this fact: minimal models are not $U_q(sl_2)$ symmetric, so why should an object like $U_q(sl_2)$ $6j$-symbols appear?\footnote{The categorical language of \cite{Moore:1988qv} explains the appearance of 6$j$-symbols as objects solving the pentagon identity. Here we explain it from the group-theoretical point of view. 
In the case of $s_1 = 2$, the fusion kernel is a $2 \times 2$ matrix, and the two independent Virasoro blocks satisfy a second-order hypergeometric differential equation. It can be explicitly verified that in this case, \eqref{fusion:singleQG} holds. For general $s_1$, the validity of \eqref{fusion:singleQG} follows recursively from the pentagon equation, as discussed in \cite[Theorem 8.1]{Teschner:1995ga}.} However, given the same $c$ and $h_i$, both minimal models and the theories we are studying have the same Virasoro blocks. So we can say that $U_q(sl_2)$ $6j$-symbols appear in minimal model crossing kernels for Virasoro blocks because the theory we are studying exists and is crossing symmetric.\footnote{A different way of understanding the appearance of $6j$-symbols in conformal blocks is through the coset construction of minimal models \cite{Pawelkiewicz:2013wga}.}

We also have checked explicitly that with this choices of OPE coefficients correlation functions satisfy crossing for many correlation functions with $W_{1,3}$ and $W_{1,5}$ inserted by computing the blocks explicitly; this allows to uniquely fix the sign ambiguity \eqref{eq:C_as_MM_chiral} for these operators. This further supports our result for the chiral OPE coefficient.\footnote{When doing this sort of computations one encounters several branch points for the $6j$-symbols as $q$ moves along the unit circle. How we choose to go around these branch points then might give us different overall phases for the OPE coefficients. In our case, for the $6j$-symbols we start from $q=1$ and keep $q$ inside the unit circle by giving a small positive imaginary part to $\mu \to \mu + i 0^+$.}

\subsubsection{Non-chiral operators}
\label{sec:general_OPE_coeffs}
The situation for non-chiral operators becomes more complicated: equation \eqref{crossing:explicit1} involves products of holomorphic and antiholomorphic blocks, and more than one operator is exchanged per representation of $U_q(sl_2)$. However, OPE coefficients can be still fixed by using the crossing properties of Virasoro blocks, modulo the usual sign ambiguity.

The crossing equation \eqref{crossing:explicit2} in this case gives
\begin{equation} \label{eq:crossing_Wrs}
    \begin{split}
        &\sum_{r'}C_{(r_2,s_2),(r_3,s_3),(r',s')}C_{(r_4,s_4),(r_1,s_1),(r',s')}\mathcal{F}^{(t)}_{r',s'}(z)\mathcal{F}^{(t)}_{r',1}(\bar{z})\\
        &=\sum_{s}\sixj{\frac{s_1-1}{2},\frac{s_2-1}{2},\frac{s-1}{2},\frac{s_3-1}{2},\frac{s_4-1}{2},\frac{s'-1}{2}}_{q} \,\sum_{r}C_{(r_1,s_1),(r_2,s_2),(r,s)}C_{(r_3,s_3),(r_4,s_4),(r,s)}\mathcal{F}^{(s)}_{r,s}(z)\mathcal{F}^{(s)}_{r,1}(\bar{z}) \,,
    \end{split}
\end{equation}
where the sum over $r$ and $s$ is over values allowed by the fusion rules. Throughout this section we write explicitly the dependence of $6j$-symbols on $q$; the reason why this is necessary will be clear shortly. 
Note that given that our operators are degenerate, all sums are finite sums, so we can regard them as function of independent complex $z$ and $\bar{z}$.

Contrary to the chiral case, crossing equations don't tell us anything directly on the behavior of Virasoro blocks under crossing. However, \eqref{eq:crossing_Wrs} is still solvable using known result of the fusion kernel of the Virasoro blocks for general $(r,s)$, from e.g. \cite{Teschner:1995ga,Cremmer:1994kc}
\begin{equation}
\begin{aligned} \label{eq:crossing_F}
    \calF_{r',s'}^{(t)} (z) =& \sum_{r,s} \frac{g_{(r_1,s_1),(r_2,s_2)}^{(r,s)}g_{(r,s),(r_3,s_3)}^{(r_4,s_4)}}{g_{(r_2,s_2),(r_3,s_3)}^{(r',s')}g_{(r_1,s_1),(r',s')}^{(r_4,s_4)}}(-1)^{f(r_i,s_i,r,s,r',s')} \\ & \qquad \qquad \qquad \qquad \times \sixj{\frac{r_1-1}{2},\frac{r_2-1}{2},\frac{r-1}{2},\frac{r_3-1}{2},\frac{r_4-1}{2},\frac{r'-1}{2}}_{\tilde{q}} \sixj{\frac{s_1-1}{2},\frac{s_2-1}{2},\frac{s-1}{2},\frac{s_3-1}{2},\frac{s_4-1}{2},\frac{s'-1}{2}}_{q} \calF_{r,s}^{(s)} (z)
    \end{aligned}
\end{equation}
with
\begin{equation}
    f(r_i,s_i,r,s,r',s') = \frac 12 (s_2-1)(r+r'-r_2-r_4) + \frac 12 (r_2-1)(s+s'-s_2-s_4)
\end{equation}
where we remind the reader that in our case, $s_i$, $s$ and $s'$ are all odd.
While $q=e^{i \pi \frac{\mu}{\mu+1}}$, $\tilde q$ is defined as $\tilde{q}=e^{i \pi \frac{\mu+1}{\mu}}$. The function $g$'s are related to the minimal model OPE coefficients as in \eqref{eq:g_to_C}.

We will focus again on the case $(r_3,s_3) = (r_1,s_1)$ and $(r_4,s_4) = (r_2,s_2)$. Starting from \eqref{eq:crossing_Wrs}, we first use the crossing properties of the block $\calF^{(s)}_{r,1}$, which can be written by setting $s=s_i=1$ in \eqref{eq:crossing_F} and using the orthogonality relations \eqref{eq:6j_orthogonality}, as well as the symmetry properties of $g$,
\begin{equation}\label{fusion:r1}
    \begin{split}
        \mathcal{F}_{r,1}^{(s)}(\bar{z})=\sum_{r'}
        \frac{g_{(r_1,1),(r_2,1)}^{(r',1)}g_{(r_1,1),(r',1)}^{(r_2,1)}}{g_{(r_1,1),(r_2,2)}^{(r,1)}g_{(r_1,1),(r,1)}^{(r_2,1)}}
        \sixj{\frac{r_1-1}{2},\frac{r_2-1}{2},\frac{r-1}{2},\frac{r_1-1}{2},\frac{r_2-1}{2},\frac{r'-1}{2}}_{\tilde{q}}\mathcal{F}_{r',1}^{(t)},(\bar{z})
    \end{split}
\end{equation}

Now we plug in this equation into \eqref{eq:crossing_Wrs}, which together with the permutation properties of OPE coefficients \eqref{relation:XXZOPE}, gives
\begin{equation}
\begin{aligned}
   \sum_{r'} & \left(C_{(r_1,s_1),(r_2,s_2),(r',s')}\right)^2\mathcal{F}^{(t)}_{r',s'}(z)\mathcal{F}^{(t)}_{r',1}(\bar{z})= \\ &=\sum_{r,r',s} \left(C_{(r_1,s_1),(r_2,s_2),(r,s)}\right)^2 \frac{g_{(r_1,1),(r_2,1)}^{(r',1)}g_{(r_1,1),(r',1)}^{(r_2,1)}}{g_{(r_1,1),(r_2,2)}^{(r,1)}g_{(r_1,1),(r,1)}^{(r_2,1)}}\sixj{\frac{r_1-1}{2},\frac{r_2-1}{2},\frac{r-1}{2},\frac{r_1-1}{2},\frac{r_2-1}{2},\frac{r'-1}{2}}_{\tilde{q}}\\ & \qquad\qquad\qquad\qquad \qquad\qquad\qquad\qquad\qquad \times \sixj{\frac{s_1-1}{2},\frac{s_2-1}{2},\frac{s-1}{2},\frac{s_1-1}{2},\frac{s_2-1}{2},\frac{s'-1}{2}}_q 
         \calF_{r,s}^{(s)}(z) \calF_{r',1}^{(t)}(\bz)
\end{aligned}
\end{equation}
Comparing the coefficients of $\calF_{r',1}^{(t)}(\bz)$, we find the equality
\begin{equation}
\begin{aligned}
    \mathcal{F}^{(t)}_{r',s'}(z) = \sum_{r,s}\left(\frac{C_{(r_1,s_1),(r_2,s_2),(r,s)}}{C_{(r_1,s_1),(r_2,s_2),(r',s')}}\right)^2 &\frac{g_{(r_1,1),(r_2,1)}^{(r',1)}g_{(r_1,1),(r',1)}^{(r_2,1)}}{g_{(r_1,1),(r_2,2)}^{(r,1)}g_{(r_1,1),(r,1)}^{(r_2,1)}} \sixj{\frac{r_1-1}{2},\frac{r_2-1}{2},\frac{r-1}{2},\frac{r_1-1}{2},\frac{r_2-1}{2},\frac{r'-1}{2}}_{\tilde{q}} \\ & \qquad\qquad\qquad \times \sixj{\frac{s_1-1}{2},\frac{s_2-1}{2},\frac{s-1}{2},\frac{s_1-1}{2},\frac{s_2-1}{2},\frac{s'-1}{2}}_q  \calF_{r,s}^{(s)}(z)
    \end{aligned}
\end{equation}
Comparing this to \eqref{eq:crossing_F}, we fix the OPE coefficients in terms of the $g_s$ modulo a constant that depends on the external operators
\begin{equation}
(C_{(r_1,s_1),(r_2,s_2),(r,s)})^2 \sim (-1)^{\frac12 (s_2-1)(r-r_2)+ \frac12 (r_2-1)(s-s_2)}g_{(r_1,s_1),(r_2,s_2)}^{(r,s)} g_{(r_1,1),(r_2,1)}^{(r,1)} g_{(r_1,s_1),(r,s)}^{(r_2,s_2)} g_{(r_1,1),(r,1)}^{(r_2,1)}
\end{equation}
Squaring this relation, remembering that all $s_i$'s are odd XXZ$_q$, and substituting \eqref{eq:g_to_C} we find 
\begin{equation}
    \left(C_{(r_1,s_1),(r_2,s_2),(r_3,s_3)}\right)^4 = \left(C^{\rm MM}_{(r_1,s_1),(r_2,s_2),(r_3,s_3)} C^{\rm MM}_{(r_1,1),(r_2,1),(r_3,1)} \right)^2
\end{equation}
where, as before, the overall constant is excluded by the requirement of permutation symmetry of the squared OPE coefficients of XXZ$_q$ and by the proper normalization of two point functions. Therefore we obtain the generalization of \eqref{eq:C_as_MM_chiral}
\begin{equation}
    \left(C_{(r_1,s_1),(r_2,s_2),(r_3,s_3)}\right)^2 =\pm C^{\rm MM}_{(r_1,s_1),(r_2,s_2),(r_3,s_3)} C^{\rm MM}_{(r_1,1),(r_2,1),(r_3,1)}\,. \label{eq:C_as_MM_general}
\end{equation}
Using the power of crossing symmetry, we have almost completely fixed the OPE coefficients. To confirm our predictions, we also computed several correlation functions of the operator $W_{2,3}$ using the BPZ equations, and find that crossing symmetry fixes the OPE coefficients to be \eqref{eq:C_as_MM_general}. This allows us to fix the sign as well, but is hard to compute for operators with higher $r$ and $s$'s.
 We will fix the sign of all OPE coefficients using different techniques in the companion paper \cite{paper2}.

We stress that the study of quantum group symmetric CFTs gives an explanation of the appearance of $U_q(sl_2)$ $6j$-symbols in the fusion kernel for general operators with weights $h_{r,s}$, even in theories with no quantum group symmetries.
As we already mentioned in the previous section, in the case of chiral operators, crossing symmetry of the XXZ$_q$ CFT is a direct explanation of the appearance of $U_q(sl_2)$ $6j$-symbols in the fusion kernels of Virasoro blocks for degenerate operators with weights $h_{1,s}$.  The Coulomb gas techniques we elaborate on in \cite{paper2} allow us to study a theory different than XXZ$_q$, symmetric under two quantum groups. Crossing symmetry of this theory then becomes the condition \eqref{eq:crossing_F} for general operators with weight $h_{r,s}$, and therefore gives us an explanation for the appearance of $6j$-symbols in the fusion kernel in general degenerate Virasoro blocks.

\section{Ising limit and local unitary subsectors}
\label{sec:ising}

So far the discussion of the XXZ$_q$ CFT has focused on the case of generic $q$, i.e. $q$ not a root of unity. However, the case of $q$ root of unity is interesting because it will allow us to make contact with the unitary series of minimal models, as well as to think about a broader question of whether a QG can act on a local and unitary theory. Indeed, we remind the reader that the minimal model $\mathcal{M}_{\mu,\mu+1}$ has central charge $c= 1-\frac{6}{\mu(\mu+1)}$, where $\mu$ is a positive integer $\mu \ge 3$; this coincides with the central charge of the XXZ$_q$ CFT with $q = e^{i \pi \frac{\mu}{\mu+1}} $. A more concrete reason to expect a connection between these two theories comes from the lattice, where it was shown that the XXZ$_q$ lattice model for $q$ root of unity has a subsector which is directly related to some twisting of unitary minimal models \cite{grosse1994quantum,pallua1996minimal}. Here we will focus on the same question from the point of view of the CFT, and show that in the simplest case $\mu \to 3$, the XXZ$_q$ CFT develops a subsector which reproduces the fermionic formulation of the Ising model. In particular, we show agreement of both the spectrum and correlation functions, up to six points.

Let's start from the lattice model: it is known that \eqref{hamilt_XXZ_quantum_closed} with $\mu = 3$, and therefore $q =  q_I \equiv e^{i \pi 3/4} $,  and $N=2L$ sites has a close connection with the $L$ site Ising model with some special boundary conditions \cite{grosse1994quantum}\footnote{This Hamiltonian in \cite{grosse1994quantum} is defined with an overall minus sign because the choice of $q$ is different than the choice we make here.}
\begin{equation} \label{eq:HIsing}
    H_{\rm Ising} = \frac{1}{\sqrt{2} } \sum_{i=1}^{L-1}  \sigma_i^x \sigma_{i+1}^x +\frac{1}{\sqrt{2}} \sum_{i=1}^{L} \sigma_i^z +\frac{1}{\sqrt{2}}\Gamma  \sigma_L^x \sigma_1^x + \sqrt{2}L
\end{equation}
with
\begin{equation}
    \Gamma = \prod_{i=1}^L \sigma_i^z
\end{equation}
The operator $\Gamma$ measures the $\mathbb{Z}_2$ parity of a state, and therefore the Hamiltonian \eqref{eq:HIsing} is a Ising model with periodic boundary conditions {(PBC)} for $\mathbb{Z}_2$ even states and antiperiodic boundary conditions {(APBC)} for $\mathbb{Z}_2$ odd ones.   

The precise relation between the two lattice models, as explained in \cite{grosse1994quantum}, is that all the eigenvalues of \eqref{eq:HIsing} with $L$ sites are also eigenvalues of \eqref{hamilt_XXZ_quantum_closed} with $\mu = 3$ with $N=2L$ sites. This cannot be a one to one correspondence given that \eqref{hamilt_XXZ_quantum_closed} acts on $2L$ sites, and therefore has a much larger Hilbert space. However, there is a recipe \cite{Alcaraz:2000hv} on how to project the XXZ$_q$ spin chain Hilbert space down to a subsector which has the same eigenvalues of the Ising one.\ This coincidence between energy levels might look like magic at first, but is based on the fact that the Hamiltonians of both lattice models can be expressed as sums of generators of the same Temperley-Lieb algebra \cite{Alcaraz:1987zr,grosse1994quantum}. 

\begin{figure}
    \centering
    \includegraphics[width = .75\textwidth]{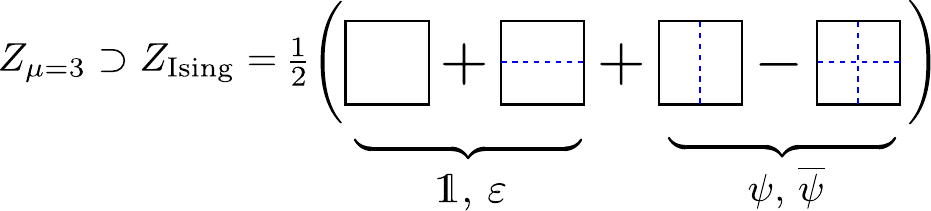}
    \caption{The partition function of the Ising subsector of XXZ$_q$. The blue line represent the topological line $\eta$ implementing the $\mathbb{Z}_2$ twist, and we imagine time flowing vertically. The $\eta$ line inserted vertically means we are considering operators of the $\mathbb{Z}_2$ twisted sector, while the $\eta$ line inserted horizontally gives a minus sign to the $\mathbb{Z}_2$ odd operators.}
    \label{fig:ZIsing}
\end{figure}

Based on these spin chain arguments, we also expect a connection between the quantum group symmetric CFT in the $\mu \to 3$ limit and the Ising CFT with a specific twisting. If we have a critical spin chain on a circle, by the state-operator correspondence, its energy spectrum in the continuum limit gives the operator spectrum of the CFT describing its critical point. In our case, the spin chain has some twisted boundary conditions, so we expect to recover the operator spectrum of the Ising CFT with some special twisting. Given that the energy levels of twisted Ising are a subset of the XXZ$_q$ energy levels, we also expect the operator spectrum of the twisted Ising CFT to form a subsector of the operator spectrum of the XXZ$_q$ CFT, see figure \ref{fig:ZIsing}. In particular, given that we have PBC for $\mathbb{Z}_2$ even states, we expect to recover the $\mathbb{Z}_2$ even operators from the untwisted sector of the CFT, $\mathbb{1}$ and $\vareps$; the APBC for $\mathbb{Z}_2$ odd states will instead give us the $\mathbb{Z}_2$ odd operators from the $\mathbb{Z}_2$ twisted sector of the CFT, the fermion $\psi$ and $\bar \psi$.\footnote{These type of twistings of partition functions have appeared also in the context of the Arf invariant, see e.g. \cite{Kapustin:2017jrc,Lee:2018eqa,Karch:2019lnn}.} All in all, we recover the fermionic formulation of the Ising model.

We will now take the $\mu \to 3$ limit of  our CFT. This is not as straightforward as one might hope:
as often happens in non-unitary CFTs, the limit $\mu$ becoming an integer will give rise to some divergences in correlation functions that can be made finite by constructing appropriate logarithimic multiplets \cite{Cardy:2013rqg}. We will first deal with these divergences and study the logarithmic CFT arising in the $\mu \to 3$ limit, and then identify the Ising subsector of the theory, which is unitary and for which we conjecture that logarithms drop out, a fact that we check in several correlation functions. Besides indecomposable representation of the Virasoro algebra, related to the logarithmic nature of the CFT at $\mu = 3$, we also find indecomposable representations of the quantum group, which can happen when $q$ is a root of unity.

Let us also mention that the situation is similar to that in logarithmic minimal models \cite{Pearce:2006sz}, where a logarithmic CFT admits a closed subsector which corresponds to a (in this case untwisted) minimal model.

\subsection{Logarithmic multiplets} \label{sec:log_mult}
As $\mu \to 3$, correlation functions develop divergences that can be cured by forming logarithmic multiplets. We will now study in detail one of the logarithmic multiplets arising in this limit, the one appearing from mixing of operators in the Verma module of $W_{1,5}$ and $W_{1,3}$. The reason to study this specific logarithmic multiplet is that it will allow us to identify the Ising operator $\psi$ later in Section \ref{sec:psi}.

An important concept here will be the difference between null states and zero-norm operators. We call an operator zero-norm if its two-point function with itself is zero; we call instead an operator null if all of its correlation functions, with all other operators of the theory, vanish. We need to make this distinction because our theory is not unitary: in a unitary theory, operators with zero norm must necessarily be null. In our case, instead, we will see that we have operators which have zero norm but are not null, and will play a role in logarithmic multiplets.

An example of the difference between zero-norm and null operators can be seen in the descendants of $W_{1,3}$. This operator has dimension $(h_{1,3},0)$, meaning that, for any value of $\mu$, it has a null descendant at level three in its holomorphic sector. We can conclude that this is the case by checking that this operator is missing from the partition function for all values of $\mu$. Something more, however, happens at $\mu = 3$. Given that $h_{1,3}|_{\mu=3}=h_{2,1}|_{\mu=3}$, its level two descendant becomes zero-norm (but not null) at $\mu = 3$.

To see this, we consider the level two descendant
\begin{equation} \label{eq:Ydef}
    Y^m = \left(L_{-2}-\frac{3}{2(1+2 h_{1,3})} L_{-1}^2 \right) W^{m}_{1,3}
\end{equation}
where the relative coefficient between the two terms was chosen so that $Y^m$ is a quasiprimary for generic $\mu$, $L_1 Y^m = 0$.

We can see that indeed the two-point function of this operator vanishes in the limit $\mu \to 3$. Identifying $\mu = 3 +\epsilon$,
\begin{equation}
    \braket{Y^m(x) Y^{-m}(0)} = \QCG{1,m,1,-m,0,0,q} \frac{\mathcal{N}_Y}{x^{2(h_{1,3}+2)}} = \frac{35}{96}\frac{1}{x^5} \epsilon  \QCG{1,m,1,-m,0,0,q_I} + O(\epsilon^2)
\end{equation}
with $\mathcal{N}_Y = \frac{(\mu -3) (3 \mu -2) (3 \mu +1)}{2 \mu  (\mu +1) (3 \mu -1)}$ and we remind the reader that $q_I =  e^{i \pi 3/4}$.\footnote{We remind the reader that Clebsch-Gordan coefficients tend to have branch points on the unit circle. We work by starting from $q=1$ and continuing it inside the unit disc until $q_I$.}
Using the results about OPE coefficient for chiral operators \eqref{eq:C_as_MM_chiral}, we can check that three-point function $\braket{YYY}$ actually remains finite in this limit.
\begin{equation}
    \braket{YYY}\sim O(1)
\end{equation}
Indeed, the operator $Y^m$ is zero-norm, given the vanishing of the two-point function, but is not null, because some of its other correlation functions are not zero.
The fact that the operator $Y^m$ is not identically zero means that we cannot mod it out of our theory. We also are prevented from naively normalizing $Y^m$ so that it has a unit two-point function for $\mu \to 3$, because then the three-point function would diverge.

This might look inconvenient, but by itself is not a problem in a non-unitary theory. What actually is a problem is when we go and look at higher point functions, where we actually find divergences. As an example, in the $\mu \to 3$ limit the $W_{1,3}$ four point function diverges because of the contribution of the exchange of the operator $W_{1,5}$. A similar reasoning applies to correlation functions of operator $W_{1,5}$.

The right procedure to deal with these divergences is to form a logarithmic multiplet out of $Y^m$ and $W_{1,5}^m$. We define the following operators,
\begin{equation}
\begin{aligned}
    A^m &= \lim_{\epsilon \to 0}   \frac{(-1)^m\alpha_1}{\sqrt{\epsilon}} W_{1,5}^m+ \frac{\alpha_2}{\epsilon}Y^m\\
    B^m &=\lim_{\epsilon \to 0}   \alpha_3 Y^m \label{eq:ABdef}
\end{aligned}
\end{equation}
where $m=-1,0,1$. The fate of $W_{1,5}^{\pm 2}$ is less interesting and will be discussed later.

We need to choose the coefficients $\alpha_i$ so that we have no divergences in the two-point function of operators $A^m$. Note that the requirement that the $\alpha_i$ coefficients do not depend on $m$ is non-trivial, and the fact that divergences in $\braket{A_m A_{-m}}$ are cancelled for every value of $m$ relies strongly on the fact that some QCG coefficients are proportional to each other at  $q=q_I$ 
\begin{equation}
    \QCG{1,m,1,-m,0,0,q_I} \sim \QCG{2,m,2,-m,0,0,q_I}\,, \qquad m=-1,0,1\,.
\end{equation}

The combinations $A_m$ and $B_m$, for appropriately chosen $\alpha_j$, have finite correlation functions in the $\mu \to 3$ limit. The price to pay is that the dilation operator $L_0$ cannot be diagonalized anymore \cite{Gurarie:1993xq,Cardy:2013rqg}
\begin{equation}
\begin{aligned}
    L_0  A^m &= h_{1,5} \frac{(-1)^m\alpha_1}{\sqrt{\epsilon}} W_{1,5}^m+ (h_{1,3}+2) \frac{\alpha_2}{\epsilon}Y^m = h_{1,5} A^m + \frac{h_{1,3}+2-h_{1,5}}{\eps} \alpha_2 Y^m =\\&= 
    h_{1,5} A^m + \frac{h_{1,3}+2-h_{1,5}}{\eps} \frac{\alpha_2}{\alpha_3} B^m\,.
    \end{aligned}
\end{equation}
where the last expression is finite because the dimensions of operators $W_{1,5}$ and $Y$ match at $\mu = 3$, $h_{1,3}+2-h_{1,5} \sim O(\eps)$.
We make the choice\footnote{The subleading term in $\alpha_1$ is chosen so that the $A_m$ two-point function has no terms without a logarithm. Different choices of this subleading term are allowed and correspond to the usual redefinition $A \to A + c B$, which does not change the structure of the Jordan block.}
\begin{equation}
    \alpha_1 = \sqrt{2} e^{i \pi 3/4}\,, \qquad  \alpha_2 = -8 \sqrt{\frac{3}{35}}-\frac{(193+210 \pi ) \epsilon }{70 \sqrt{105}}\,, \qquad \alpha_3 = 2 \sqrt{\frac{3}{35}}
\end{equation}
so that we recover a nice normalization for the action of the Virasoro group on the logarithmic multiplet\begin{align}
    L_0 \begin{pmatrix}
        A^m \\ B^m 
    \end{pmatrix} = \begin{pmatrix}
        5/2 & 1 \\ 0 &5/2
    \end{pmatrix}\begin{pmatrix}
         A^m \\ B^m
    \end{pmatrix}\ ,
\end{align}
and the following two-point functions
\begin{equation} \label{eq:AB_2pf}
\begin{aligned}
    \braket{A^m (0) A^{-m}(x)} &= \frac{\log x}{x^5} \QCG{1,m,1,-m,0,0,q_I}\ ,\\
    \braket{A^m (0) B^{-m}(x)} &= -\frac{1}{2x^5} \QCG{1,m,1,-m,0,0,q_I}\ ,\\
    \braket{B^m (0) B^{-m}(x)} &= 0\ .
\end{aligned}
\end{equation}

One can now notice that operators $A_m$ are not primaries,\footnote{Sometimes in the literature the concept of indecomposability parameter of staggered modules is discussed \cite{Gurarie:2004ce}. While in some normalization that would be the coefficient of the $\braket{AB}$ or the $\braket{AA}$ \cite{Vasseur:2011fi}, in our choice \eqref{eq:AB_2pf}, the indecomposability parameter can be read from the action of $L_2$ on $A^m$.}
\begin{equation}
    L_{2} A^m =-\frac{1}{4}\sqrt{\frac{35}{3}} W^m_{1,3}\ ,
\end{equation}
but at the same time they also are not descendants. Conversely, $B$ is both a descendant, see \eqref{eq:Ydef}, and a primary
\begin{equation}
    L_{n>0} B^m = 0 \end{equation}
where the $n=2$ case can be checked using the definition \eqref{eq:ABdef}, and other cases follow straightforwardly from construction; $B$ is an example of zero-norm but non null operator. This happens when the logarithmic multiplet is formed by mixing an operator which is a primary and one that is a descendant in \eqref{eq:ABdef}. This logarithmic multiplet is then called a level two staggered logarithmic module \cite{Gaberdiel:1996kx,Rohsiepe:1996qj,Gorbenko:2020xya}. 

\begin{figure}\centering
    \qquad
    \subfloat{{\includegraphics[width=4cm,valign=c]{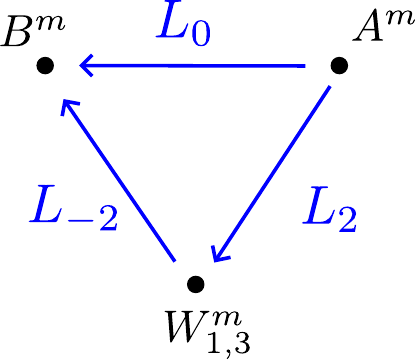} }}\qquad \qquad \qquad\qquad
    \subfloat{{\includegraphics[width=6cm,valign=c]{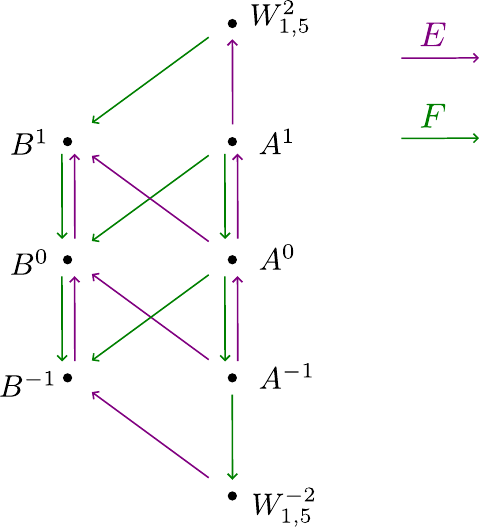} }}\caption{Left: action of the Virasoro algebra on the logarithmic multiplet. Right: action of the Quantum Group on the same operators}\label{fig:indecomposable_each}\end{figure}

The structure of the staggered module for the Virasoro algebra is sketched in figure \ref{fig:indecomposable_each}.
One can verify, via nontrivial computations, that all three-point functions of the operators defined above are finite. 

\subsection{Indecomposability of the quantum group action for \texorpdfstring{$q$}{q} a root of unity}
\label{sec:irreps_unity}

We've seen that two operators with the same scaling dimension can combine to form a logarithmic multiplet. However, for $\mu \neq 3$, these operators transform under different representations of $U_q(sl_2)$. How is this compatible with the quantum group structure of our theory? We will see that, as $q$ becomes a root of unity at $\mu = 3$, we get more complicated representations of the quantum group, which themselves turn out to be indecomposable but reducible, so that the final result is consistent under the action of both quantum group and Virasoro.

The reason why representation theory for $U_q(sl_2)$ is much more complicated when $q$ is a root of unity is that, if $q^n=\pm1$ for some integer $n$, the generators $E,F$ become nilpotent
\begin{equation}
    E^n = F^n = 0.
\end{equation}
It follows that the generic $q$ representations $\mathbb{V}_{\ell}$ remain irreducible at $q$ root of unity only for small enough spin $\ell$; namely, the dimensionality of representation must be $2\ell+1\le n$. If $\ell$ is larger than that, the representation $\mathbb{V}_{\ell}$ will be reducible because acting with $F$ on the highest weight state $n$ times gives zero. Furthermore, the quantum group when $q$ is a root of unity is no longer semisimple. Thus, we are not surprised that the tensor product of representations is not always decomposable into a direct sum of irreducible representations.

In this section, we only focus on what is relevant for the connection of XXZ$_q$ with the Ising model.
More details about representation theory of $U_q(sl_2)$ for $q$ root of unity can be found e.g. in \cite{klimyk2012quantum}.
In our case, we will be interested in what happens to the representation $\mathbb{V}_2$, under which the operator $W_{1,5}$ transforms, when $q \to q_I$. In this limit, this representation combines with the spin $1$ representation $\mathbb{V}_1$ to make a larger representation which we denote as $(\mathbb{V}_2,\mathbb{V}_1)$, with the following properties: (1) it is not a highest weight representation, (2) it is reducible but indecomposable representation, (3) some states in the representation are of zero norm under the inner product defined in section \ref{sec:irreps_generic}. We now will show by an explicit computation that the operators $A^m$, $B^m$ and $W_{1,5}^{\pm 2}$ all transform in this representation $(\mathbb{V}_2,\mathbb{V}_1)$. The structure of this representation is shown in Figure \ref{fig:indecomposable_each}.

In order to study this,
we act with the quantum group generators $E$, $F$ and $H$ on the operators $A^m$ and $B^m$. To start, given that we're summing operators with the same eigenvalue under $H$, everything looks as usual from the point of view of the global $U(1)$, and $A^m$ and $B^m$ are eigenoperators of $H$ with eigenvalue $2m$.

Let's now pay attention to the generators $E$ and $F$. As a reminder, the operator $W_{r,s}$ transforms in a spin $\ell = \frac{s-1}{2}$ of $U_q(sl_2)$, therefore the action of $E$ and $F$ for generic $\mu$ is
\begin{equation}
\begin{aligned}
    E W^m_{r,s}(x) &= e_{\frac{s-1}{2},m} W^{m+1}_{r,s}\\
    F W^{m}_{r,s}(x) &= f_{\frac{s-1}{2},m} W^{m-1}_{r,s}(x)
\end{aligned}
\end{equation}
where $e_{\ell,m}$ and $f_{\ell,m}$ are defined in \eqref{correct_fe}.
From this we can simply obtain the quantum group action on $A_m$ and $B_m$ by using the definition \eqref{eq:ABdef} and taking the $\mu \to 3$ limit.

We obtain immediately that the action of $E$ and $F$ on $B_m$ is the usual one
\begin{equation}
\begin{aligned}
    E B^{m}(x) &= e_{1,m} B^{m+1}(x)\,,\\
    F B^{m}(x) &= f_{1,m} B^{m-1}(x)\,,
\end{aligned}
\end{equation}
and the $B_m$'s form a genuine spin 1 subrepresentation of the $(\mathbb{V}_2,\mathbb{V}_1)$ representation.
Again, things become trickier when looking at $A_m$. For example, using \eqref{eq:ABdef}, we see
\begin{equation}
E A^m =
\frac{(-1)^m\alpha_1}{\sqrt{\epsilon}} e_{2,m} W_{1,5}^{m+1}+\frac{\alpha_2}{\epsilon} e_{1,m}Y^{m+1} +\ldots=  -e_{2,m} A^m + \frac{\alpha_2}{\alpha_3} \frac{e_{1,m}+e_{2,m}}{\epsilon} B^{m+1}+\ldots
\end{equation}
It might look like we have a $1/\epsilon$ divergence, but we are saved by the fact that, at $\mu =3 $, $e_{2,m} = -e_{1,m}$ for $m=-1,0,1$, so the term $\frac{e_{1,m}+e_{2,m}}{\epsilon}$ is actually of order one.
Similar cancellations also appear for the action of $F$ on these operators.

Plugging in our choices for $\alpha_j$ and taking the $\epsilon \to 0$ limit, we get
\begin{equation}
    \begin{aligned}
        E A^m &= e_{1,m} \left( A^{m+1} +\frac \pi2 B^{m+1}\right) \qquad m = -1,0\\
        F A^m &= f_{1,m} \left( A^{m-1} +\frac \pi2 B^{m-1}\right) \qquad m  =0,1
    \end{aligned}
\end{equation}
We consider separately the case $E A^1$ and $F A^{-1}$. Even though $e_{2,1}$ and $f_{2,-1}$ both vanish for $q \to q_I$, inverse powers of $\epsilon$ in \eqref{eq:ABdef} make the result finite
so that at $\mu = 3$, we have \begin{equation}
\begin{aligned}
    E A^{1} &= -\frac{i \sqrt{\pi}}{2^{1/4}} W^{2}_{1,5}\\
    F A^{-1} &= \frac{e^{i\pi 3/4} \sqrt{\pi}}{2^{1/4}} W^{-2}_{1,5}
\end{aligned}
\end{equation}

\begin{figure}
    \centering
    \includegraphics[width = .9\textwidth]{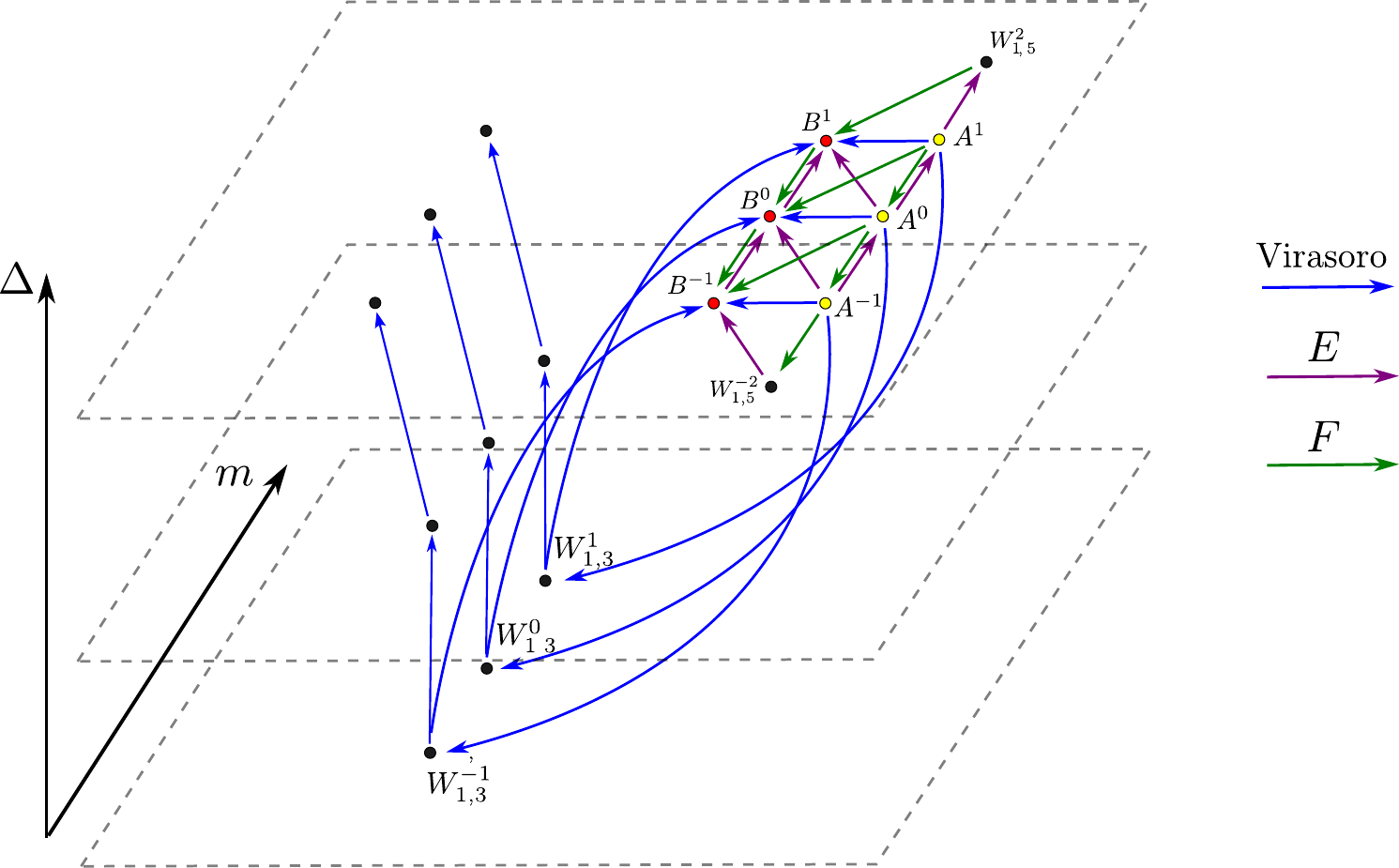}
    \caption{The action of Virasoro and the quantum group generators $E$ and $F$ on the operators transforming in the spin $1$ and $2$ representations of the quantum group in the Ising limit. For simplicity, we only indicate the action of $E$ and $F$ on the indecomposable representation formed by $A^m$, $B^m$ and $W_{1,5}^{\pm 2}$, and we include only a few descendant operators, without denoting them.}
    \label{fig:indecomposable_both}
\end{figure}

Therefore we see the same type of indecomposable but reducible action in the quantum group that we saw earlier under Virasoro, sketched in Figure \ref{fig:indecomposable_each}. Reducibility follows from the fact that the operators $B^m$ form a genuine spin one subrepresentation of the larger quantum group representation.
Putting both together, we can sum up the fate of the operators $W_{1,3}^m$ and $W_{1,5}^m$ in the $\mu \to 3$ limit as in Figure \ref{fig:indecomposable_both}.
While it might not be immediately manifest, an interested reader can check that the actions of Virasoro and the quantum group still commute.

\subsection{Identifying the Ising subsector}

Given the relation between the Ising spin chain \eqref{eq:HIsing} and the XXZ$_q$ spin chain, by invoking the state operator correspondence, we expect to find a subsector of operators of the CFT that corresponds to operators in the fermionic Ising CFT.

We will find strong evidence that this is indeed the case, and we find operators with the right dimensions, whose correlation functions are the same as those in the twisted Ising CFT up to the order we were able to check, namely, up to $6$ point functions. If the relation holds at all orders, these operators would form a closed subsector of the theory, which would mean that, for example, when looking at their correlation functions we would only see the exchange of operators belonging to this subsector. The fact that we reproduce the Ising model, a unitary theory, also means that this subsector is unitary.  We point out that both XXZ$_q$ for $q=q_I$ and Ising CFT have $c=\frac{1}{2}$. Nevertheless XXZ$_q$ CFT clearly contains additional operators, and lives in a larger Hilbert space. It would be interesting to understand how this happens.

We now make several extensive checks of this scenario. In order to identify this subsector, we are guided by the fact that while the full XXZ$_q$ theory is logarithmic, the unitary subsector is not; we will have to choose operators carefully so that logarithms drop out of correlation functions.
The situation will turn out to be subtle, with non-Ising operators appearing in the OPE of some Ising operators, but dropping out from Ising observables. This is possible only if the correlation functions of these non-Ising operators vanish. We will soon verify this for two and three-point functions, and provide hints pointing towards the same for the four-point function as well. To perform these computations, we will need not only the result \eqref{eq:C_as_MM_general} about OPE coefficients, but also to determine the sign ambiguities. We do this explicitly, by solving the BPZ equation for correlation functions and imposing crossing symmetry. The explicit computation is not too involved because the operators we are interested in throughout this section are $W_{1,3}$, $W_{1,5}$ and $W_{2,3}$, so the BPZ equation is at most of degree 6. 

\subsubsection{\texorpdfstring{$\varepsilon$}{}}
We now aim to identify which XXZ$_q$ operators correspond to the Ising operators. Remember that the spectrum of XXZ$_q$ can be read from \eqref{part_fn}, see also \eqref{isingOps} and Figure \ref{plot_delta_of_mu}. As a warm-up, we can immediately identify the energy operator $\vareps$ in the $\mu \to 3$ limit. The operator $W_{2,1}$ is a singlet of $U_q(sl_2)$ and has dimensions $(h_{2,1},h_{2,1})$, which becomes $(\frac 12,\frac 12)$ in the $\mu \to 3$ limit. This limit does not involve any logarithmic multiplets. Therefore, this operator behaves in the same way as $\phi_{2,1}$ for (generalized) minimal models; given that $W_{2,1}$ is a singlet of the quantum group the crossing equations are the same as those of $\phi_{2,1}$, meaning that the correlation functions match.
Therefore, in the $\mu \to 3$ limit we can make the identification
\begin{equation} \label{eq:eps_def}
    \vareps \equiv \lim_{\mu \to 3} W_{2,1}
\end{equation}
Note that while for general $\mu$ the $W_{2,1} \times W_{2,1}$ OPE has both the identity and $W_{3,1}$, the OPE coefficient for the latter vanishes for $\mu=3$; this can be checked by using the explicit form of OPE coefficients \eqref{eq:C_as_MM_general}. By using the BPZ differential equations of $W_{2,1}$ and taking the $\mu \to 3$ limit, we recover the expected Ising correlation function
\begin{equation}
    \braket{\vareps(0)\vareps(z,\bz)\vareps(1)\vareps(\infty)} = \left|\frac{z}{1-z}+\frac{1}{z}\right|^2
\end{equation}

\subsubsection{\texorpdfstring{$\psi$}{}} \label{sec:psi}
Given the twisting of the Ising model, see Figure \ref{fig:ZIsing}, one of the operators we are expecting to see is the fermion $\psi$, of dimension $(\frac{1}{2},0)$, arising from the $\mathbb{Z}_2$ twisted sector. This is indeed the dimension of the multiplet $W_{1,3}$ in the $\mu \to 3$ limit, and we make the identification.
\begin{equation}\label{eq:psi_def}
    \psi \equiv \lim_{\mu \to 3} \frac{W^1_{1,3} +W^{-1}_{1,3}}{2^\frac{1}{4}}
\end{equation}
As we have shown in Section \ref{sec:log_mult}, the four point function of $W_{1,3}^m$ generically contains $\log z$ terms, but we will see that with this specific choice they drop out from every $n$-point functions of $\psi$ we were able to check.
Notice that we are summing operators with different $U(1)$ charges in \eqref{eq:psi_def}, meaning that we will break the $U(1)$ symmetry of XXZ$_q$ and, therefore, the quantum group symmetry. This is a necessary step to reproduce the Ising CFT, which has only $\mathbb{Z}_2$ symmetry.

Using the explicit values of quantum Clebsch Gordan coefficients \eqref{qcg_formula}, and taking the $\mu \to 3$ limit, we can find that the two-point function is canonically normalized,
\begin{equation}
    \braket{\psi(x) \psi(0)} = \frac 1x
\end{equation}
the three-point function vanishes because of the $U(1)$ symmetry of the XXZ$_q$ theory
\begin{equation}
    \braket{\psi(x_1) \psi(x_2) \psi(x_3)} = 0\,.
\end{equation}
In order to study the four point function, we study the BPZ equation for $W_{1,3}$ together with the result for OPE coefficients \eqref{eq:C_as_MM_chiral}, and find the four point function explicitly to be\footnote{The BPZ equation for $W_{1,3}$ is of order three, and does not allow for closed solutions for generic $\mu$. In the $\mu \to 3$ limit, only the identity block is exchanged in the $\psi$ four point function, and this block takes the relatively simple form \eqref{eq:psi_4pf}. Notice that at no point we use the fact that $h_{1,3} = h_{2,1}$ at $\mu = 3$, because our theory is defined as a limit from $\mu \neq 3$, where this relation does not hold. However, the identity block we recover is the same that appears in the holomorphic part of the $\vareps$ four point function.}
\begin{equation} \label{eq:psi_4pf}
     \braket{\psi(0)\psi(z)\psi(1)\psi(\infty)} = \frac{z}{1-z}+\frac{1}{z}\,,
\end{equation}
which is exactly the four point function of the fermion operator in Ising. 

\subsubsection{\texorpdfstring{$\bar \psi$}{}}
The only primary we still need to identify from the Ising subsector is the operator $\bar \psi$. Notice that the XXZ$_q$ partition function is not symmetric under swapping $q \leftrightarrow \bar q$, and the operator that will give rise to $\bar \psi$ is unrelated to the operator giving rise to $\psi$. 

To correctly identify $\bar \psi$, rather than looking into the precise structure of these logarithmic multiplets, we can use the knowledge of operators $\vareps$ and $\psi$ we just obtained. We can identify $\bar \psi$ by using the OPE of these operators: we know from the Ising model that \begin{equation}
    \psi(0) \varepsilon(z) = \frac{\bar{\psi}(0)}{z} +\ldots
\end{equation}
From \eqref{eq:eps_def} and \eqref{eq:psi_def} we see that the relevant OPE that we should consider is
\begin{equation}
    W_{1,3}^{m}(0) W_{2,1}(z) = C_{(1,3),(2,1),(2,3)}\QCG{1,m,0,0,1,m,q} \frac{W_{2,3}^{m}}{z^{h_{13}+h_{21}-h_{23}}} +\ldots 
\end{equation}
Indeed, the operator $W_{2,3}$ has the correct dimensions in order to reproduce $\bar{\psi}$ in the $\mu \to 3$ limit.
It's easy to see that the Clebsch Gordan coefficient is trivially one, while
using the explicit forms of OPE from Section \ref{sec:general_OPE_coeffs}, together with an explicit computation of its sign, we see that the $\mu \to 3$ limit of this expression becomes
\begin{equation}
    \left. W_{1,3}^{m}(0) W_{2,1}(z)\right|_{\mu = 3} =\frac{W_{2,3}^m(0)}{z}+\ldots
\end{equation}
Therefore we can make the identification
\begin{equation}\label{eq:psib_def}
    \bar{\psi} \equiv \lim_{\mu \to 3} \frac{W^1_{2,3} +W^{-1}_{2,3}}{2^\frac{1}{4}}\ .
\end{equation}

Since the formation of logarithmic multiplet only has minor differences compared to that explained in \ref{sec:log_mult}, we will be pretty impressionistic with our explanation. At $\mu = 3$, $W_{2,3}$ develops a zero-norm descendant at level 1, as can be seen from the vanishing of the two point function of $L_{-1} W_{2,3}$ in this limit. At the same time, the dimension of operator $W_{2,5}$ becomes the same as that of $L_{-1} W_{2,3}$ at the Ising point, and therefore we expect logarithmic multiplets appearing in the form of  a level one staggered module.

We study the correlation function of the operator $W_{2,3}$ explicitly, computing the blocks from the BPZ differential equation and take the $\mu \to 3$ limit to find the expected Ising four point function
\begin{equation} \label{eq:psib_4pf}
     \braket{\bar{\psi}(0)\bar{\psi}(\bz)\bar{\psi}(1)\bar{\psi}(\infty)} = \frac{\bz}{1-\bz}+\frac{1}{\bz}\,.
\end{equation}

\subsubsection{Higher correlation functions}
For now the evidence for a unitary subsector of the theory corresponding to a twisting of the Ising model amounts to the matching of several correlation functions. Things become a bit more subtle when looking at the OPE of these operators.
If we look at the OPE of two fields in the Ising subsector, we might expect only the appearance of fields in the same subsector. We will see that the situation is not so simple.

For example, let's start with the $\psi \times \psi$ OPE, where a direct computation in the $\mu \to 3$ limit gives
\begin{equation}
    \psi(0)\psi(x) = \frac{\mathbb{1}}{x} + C_{\psi \psi \Lambda} x^\frac{3}{2} \Lambda(0)+\ldots
\end{equation}
where we neglect contributions of descendants. An explicit computation shows that the operator $\Lambda$ is given by
\begin{equation} \label{eq:lambda_def}
    \Lambda \equiv W_{1,5}^2 + W_{1,5}^{-2} - e^{i \pi/4} \sqrt{2\pi} B^0
\end{equation}
with the OPE coefficient given by
\begin{equation}
    C_{\psi \psi \Lambda} = \frac{1}{\sqrt2}C_{(1,3),(1,3),(1,5)}\big|_{\mu = 3}
\end{equation}
with
\begin{equation}
   C_{(1,3),(1,3),(1,5)}\big|_{\mu = 3} = \frac{e^{\frac{i \pi }{4}}}{\sqrt{55}}  \left( \frac{\Gamma \left(-\frac{1}{4}\right)^3 \Gamma \left(\frac{15}{4}\right)}{\Gamma \left(-\frac{11}{4}\right)  \Gamma \left(\frac{5}{4}\right)^{3}}  \right)^\frac14 \,.
\end{equation}
This might appear puzzling: how can we reproduce the Ising correlation functions if the OPE of $\psi$ with itself contains operators that do not belong to the Ising model? When we take a correlation function of several $\psi$, the operator $\Lambda$ should be exchanged. Already at the level of the four point function this operator, which does not belong to the Ising subsector, should be exchanged: how can we then recover the Ising result \eqref{eq:psi_4pf}? Something subtle must be going on.

It is easily checked that operator $\Lambda$ has a vanishing two-point function, given by the fact that
$B^0$ and $W^2_{1,5} +W^{-2}_{1,5}$ themselves have vanishing two-point functions at $\mu = 3$
\begin{equation}
    \braket{B^0B^0} = \braket{(W^2_{1,5} +W^{-2}_{1,5})(W^2_{1,5} +W^{-2}_{1,5} )} = 0\ .
\end{equation}
Therefore, the $\Lambda$ contribution to the four point function of $\psi$ vanishes, and the four point function of $\psi$ \eqref{eq:psi_4pf} matches the one of the Ising model.

What happens when we take higher point functions of $\psi$? Then higher point functions of $\Lambda$ might contribute. For example, we expect the six-point function of $\psi$ to have a contribution coming from the three-point function of $\Lambda$, the eight-point function of $\psi$ to have one from the four-point function of $\Lambda$, and so on. 

Using results from Appendix \ref{app:IsingPsiDetail}, we find the non-trivial result
\begin{equation}
   \braket{\Lambda \Lambda \Lambda} = 0
\end{equation}
which tells us that even in six point functions of the operator $\psi$, the operator $\Lambda$ does not contribute. This strengthens the conjecture that the XXZ$_q$ CFT has a subsector that reproduces all correlation function of the fermionic Ising CFT.

At this point, one would be interested in studying higher point functions of $\Lambda$ to check whether they also vanish which would mean that we indeed reproduce all correlation functions of $\psi$ in the Ising model. This task is more complicated, and would require studying either higher correlation functions of $\Lambda$, which becomes a much harder job, or studying the $\Lambda \times \Lambda$ OPE and then understanding which operators appearing in it also could potentially have zero correlation function. For the moment, we leave it as a conjecture that the operator $\Lambda$ does not contribute to any of the correlation functions of operators $\psi$. 
We will stop at the first non trivial check concerning the four point function of $\Lambda$. If the operator $A^0$ was exchanged, then we would have logarithms that could not be cancelled by anything else, given that $A^0$ is the unique logarithmic operator with dimension $h=5/2$. Therefore, the only possibility consistent with $\braket{\Lambda \Lambda \Lambda \Lambda} = 0$ is that $A$ is not exchanged. Since $L_{2}A$ is proportional to $W_{1,3}$, assuming a nonzero coefficient for $A$ would imply a nonzero coefficient for $W_{1,3}^0$ by the action of $L_2$ on both sides of the equation,
\begin{equation}
\label{eq:Aisthere?}
    C_{\Lambda\Lambda A} \neq 0 \quad\Rightarrow \quad C_{\Lambda\Lambda W_{1,3}^0} \neq 0 \ .
\end{equation}
We verify that $C_{\Lambda\Lambda W_{13}^0} = 0$, and hence $C_{\Lambda\Lambda A}=0$. This special occurrence makes us believe that the four point function of $\Lambda$ is indeed zero. 

A similar story applies to the $\bar{\psi}\times \bar{\psi}\,$ OPE. An explicit computation gives us that the combination of operators with dimension $\Delta = 5/2$ appearing in this OPE is precisely 
\begin{equation}
    \bar{\psi}(0)\bar{\psi}({x}) \supset \bar{x}^{-1} x^{5/2} C_{\bar \psi \bar \psi \Lambda}\Lambda(0)
\end{equation}
with
\begin{equation}
    C_{\bar \psi \bar \psi \Lambda} = \frac{1}{14} C_{\psi \psi \Lambda}
\end{equation}
Therefore the combination of operators appearing in this OPE is exactly the combination of operators \eqref{eq:lambda_def} which we expect to have zero $n$-point functions. This is a non-trivial result that strengthens our belief that a subsector of our CFT reproducing the Ising CFT exists.

The final OPE we should consider is of $\psi$ and $\bar{\psi}$,
\begin{equation}
    \psi(x)\bar{\psi}(0) = \varepsilon(0) + C_{\psi \bar{\psi} \tilde{\Lambda}} x^\frac{1}{2}\tilde{\Lambda}(0) +\ldots
\end{equation}
where a new operator $\tilde \Lambda$ appears. In Appendix \ref{app:other_multiplet}, we check explicitly that its two-point function vanishes as expected. Throughout this section we have provided strong evidence, based on the matching of several correlation functions, for the existence of an Ising subsector to the XXZ$_q$ CFT, as expected from the lattice model. 

Let us note that the mechanism for the decoupling of the unitary subsector is quite different from the one that occurs at integer $n$ and $Q$ in two-dimensional $O(n)$ and $Q$-state Potts models \cite{Gorbenko:2020xya}. In those models logarithmic multiplets present at generic $n$ ($Q$) decompose into regular multiplets which allows for the unitary theory to form a closed subalgebra of operators. Here, instead, logarithmic multiplets form at the special value of a parameter, and the unitary subsector does not form a subalgebra under OPE.

\section{Conclusions}

\subsection{Summary of results}

In this work we set out to answer the question of whether a QFT can have a quantum group as a global symmetry. The answer to this question is that it is certainly possible, but operators in nontrivial representations of the quantum group are typically defect ending operators rather than local operators. The lines these operators are attached to are topological, and can be moved freely as long as they do not cross other operators; they can be braided using the $\calR$-matrix of the quantum group \eqref{QFT:locality}.

We focused on the simplest quantum group, $U_q(sl_2)$, and showed how such a theory should look like. Correlation functions of operators are strongly constrained by the Ward identities of $U_q(sl_2)$ \eqref{QFT:wardid}. We showed that, in a $U_q(sl_2)$ symmetric QFT, the Ward identities together with the $\calR$-matrix imply that the spacetime spin and $U_q(sl_2)$ spin of operators have to satisfy a constraint \eqref{spin:option1}; this relation implies that, at generic $q$, the spacetime spin cannot be an integer for operators in nontrivial representations. Focusing on the case of CFTs, we gave an expression for the OPE of operators \eqref{CFT:OPE} and used this to build two \eqref{eq:2pf}, three \eqref{CFT:3pt1} and four point functions \eqref{eq:4pf_s}. The condition of crossing symmetry for four point function involves the $6j$-symbols of the quantum group \eqref{crossing:explicit1}.

For concreteness, we considered a specific example of a CFT with $U_q(sl_2)$ symmetry. Namely, we considered a spin chain, which we refer to as XXZ$_q$, governed by a $U_q(sl_2)$ symmetric Hamiltonian with non-local interactions \eqref{hamilt_XXZ_quantum_closed}. This model is not unitary, but its energy spectrum is still real -- a fact which is partially explained by its PT symmetry. The model is critical and is described at long distances by a $c\le 1$ CFT, where the central charge $c$ is related to the $q$ parameter of the quantum group. The XXZ$_q$ torus partition function, and therefore the spectrum of operators, is known \eqref{part_fn}.

Using the properties of $U_q(sl_2)$ symmetric theories, we then set off to fix the OPE coefficients of the XXZ$_q$ CFT. Exploiting the fact that all operators in our theory are degenerate, i.e. their holomorphic and antiholomorphic weights fall in the Kac table, we were able to fix the OPE coefficients modulo a sign. OPE coefficients turn out to be closely related to those of generalized minimal models, by the simple relation  \eqref{eq:C_as_MM_general}. An important insight in general 2d CFTs, without quantum group symmetry, arose from this analysis. It is well known that $U_q(sl_2)$ $6j$-symbols appear in the fusion kernel of Virasoro blocks of degenerate operators. 
This condition can be understood as a consequence of the existence and  crossing symmetry of  quantum-group symmetric CFTs, see \eqref{eq:crossing_6j} and discussion in the end of section \ref{sec:general_OPE_coeffs}.

Finally, we discussed a further connection of XXZ$_q$ with local and unitary CFTs. In the case where the theory has $c=\frac 12$, we have shown a connection with the Ising model. The theory contains a subsector similar to  the fermionic description of the Ising model. We have studied the XXZ$_q$ theory in this limit, and observed that we obtain peculiar representations, which are reducible but indecomposable, for both Virasoro and $U_q(sl_2)$. This means that the CFT is logarithmic; however, we identified which XXZ$_q$ operators form the Ising subsector. As must be, this identification gets rid of the logarithmic structure of the CFT, and also breaks the quantum group. The explicit computation of several correlation functions was given as evidence for the precise matching of the XXZ$_q$ subsector to Ising.

In a companion paper \cite{paper2}, we develop a Coulomb gas approach to study the XXZ$_q$ CFTs. This alternative method has the advantage of allowing a more explicit construction of the defect lines to which the operators of the theory are attached, as well as of some of the $U_q(sl_2)$ generators, written as integrals of screening charges. Using Coulomb gas we are able to fix completely (including the sign) the OPE coefficients of XXZ$_q$. This Coulomb gas approach can be used to study other theories as well, and, in particular, to explain the appearance of two quantum group $6j$-symbols in the crossing kernel of Virasoro blocks for operators of general weight $h_{r,s}$ in \eqref{eq:crossing_F}.

\subsection{Open questions}
Our work points to several interesting future directions to explore, both within the XXZ$_q$ model, as well as generalizations to different quantum-group symmetric systems. We list here a few of each:

\begin{itemize}
    \item It is important to understand the microscopic formulation of the topological lines.
    For example, one might want to measure a $n$-point function directly at the level of the spin chain. The state inserted on the cylinder in the past and in the future correspond to two operator insertions via the operator-state correspondence. There are two logically distinct admissible scenarios for the prescription needed for the other $n-2$ operator insertions. Namely, they can either be mapped to local operator or to defect-ending operator insertions on the spin-chain. In the first scenario it is the non-local Hamiltonian which dynamically gives rise to the defect line appearing in the continuum formulation, while in the second the defect needs to be inserted by hand.
    \item Historically, quantum groups have been discussed in physics mostly in the context of integrability. Indeed, both the XXZ$_q$ Hamiltonian \eqref{hamilt_XXZ_quantum_closed} from \cite{grosse1994quantum} as well as the closely related system with only local interactions \cite{Pasquier:1989kd} are integrable. It would be interesting to consider quantum group symmetry beyond integrable theories, for example by constructing a deformation of the spin chain which commutes with the quantum group, but breaks integrability.
    \item  Global symmetries can sometimes be gauged. Does this apply to our case as well? Concerning this point, the spin chain is invariant under an operator $G$ which implements translations when acting on the Hecke algebra elements. It is important to understand more about this operator and its anomalies. Previously, Yang-Mills theory based on gauged quantum group were discussed in \cite{Buffenoir:1994fh,Boulatov:1996bj,Aganagic:2004js}.
    \item Given the recent progress on generalized symmetries, and the topological lines implementing them, it is important to understand the connection, if any, between them and our work. For example, what are the fusion rules of the topological lines attached to the operators transforming under $U_q(sl_2)$? The lines that we constructed in this paper do not satisfy all the axioms usually assumed for topological defect lines. In particular, this is related to the fact that our lines are algebra-like elements, as opposed to group-like. It could be possible to construct in the CFT lines corresponding to  group-like elements of $U_q(sl_2)$. It would be interesting to see if quantum group lines form any kind of fusion category. We mention that, in the context of tensor networks, the work \cite{Couvreur:2022wyn}, with its construction of group-like elements as matrix product operators, is a step in this direction.
    \item We have provided a lot of evidence, thanks to the study of several of correlation functions, of the existence of a Ising subsector for the $c=\frac 12$ case. Can this be proven in general? Besides, a subsector is expected to exist for all integer values of $\mu\ge 3$. To what twisting of the respective minimal model does this sector correspond to? An analogous question is to relate the defect lines associated to the quantum group to some known defect lines of minimal models.
    \item Our work has focused mostly on CFTs, but the XXZ$_q$ CFT has a relevant deformation which does not break $U_q(sl_2)$, given by the operator $W_{2,1}$. This RG flow is integrable, and this could be used to study the S-matrix of a $U_q(sl_2)$ symmetric massive QFT. It would be interesting to see if the S-matrices appearing in the quantum group deformation of holography can also have quantum group as a global symmetry \cite{Thompson:2019ipl}. An example of a gapped spin chain with QG symmetry was studied in \cite{PhysRevB.102.081120}.

    \item The XXZ$_q$ model we studied is non-unitary. In principle, we do not see a reason why a $U_q(sl_2)$ theory should be non-unitary, which raises the question of whether a unitary theory with $U_q(sl_2)$ symmetry can exist.
    \item On the other hand, we showed that a necessary ingredient for $U_q(sl_2)$ symmetric QFTs is that operators have to be mutually non-local. Can this be relaxed in some way by modifying our assumptions? For example, by introducing more than one quantum group. 
    \item We see no obstruction in generalizing the construction shown in this work to lattice systems with more complicated quantum groups, such as $U_q(sl_{N\ge 3})$, see e.g. \cite{Karowski:1993nw,Lafay:2021scv}. Here the Virasoro algebra should be replaced by the $\mathcal{W}_N$ algebra. Similarly, the result could be generalized to $\widehat{su(N)}_k$ WZW models with continuous $k$.  
    \item An interesting question is whether theories that admit a Lagrangian description can have a quantum group symmetry. An interesting  example of a theory that could have such properties was considered, at a classical level, in \cite{Delduc:2016ihq}.
    \item It is possible that QFTs with quantum group as a global symmetry exist also in higher dimensions. This is a very exciting possibility. Of course, the lack of permutation invariance should again imply that most of the operators charged under the quantum groups should be non-local in some sense. For example in three dimensions one can imagine either point-like operators attached to dimension two topological defect surfaces, or line operators on which surface defects end.
\end{itemize}
Finally, let us come back to our original motivation of understanding general symmetries of QFTs, and in particular to the unknown symmetries of the two-dimensional $O(n)$ model that should explain degeneracies in its spectrum for a continuous range of $n$. While the symmetry structure of $O(n)$ model is related to quantum groups \cite{Fendley:1993wq,read2007enlarged} in some way, our present discussion shows that it is unlikely that a quantum group could act as global symmetry there since all operators appearing in the $O(n)$ torus partition function have an integer spin and are local. We may thus hope that there exists a structure, similar to a quantum group, which can transform local operators into themselves.

\section*{Acknowledgements}
We would like to thank Nils Carqueville, Michele Del Zotto, Matthias Gaberdiel, Jesper Jacobsen, Shota Komatsu, Sylvain Lacroix, Andreas Läuchli, João Penedones, Sylvain Ribault, Slava Rychkov, Volker Schomerus, Sahand Seifnashri, Didina Serban, Shu-Heng Shao and Sasha Trufanov for useful discussions. We are also grateful to the organizers and participants of the workshops “Bootstrap 2024” at Universidad Complutense in Madrid and “28$^\text{e}$ rencontre Itzykson: Analytic Results in Conformal Field Theory” at Institut de Physique Théorique in Saclay for their fruitful exchanges. The work of BZ has been partially supported by STFC consolidated grants ST/T000694/1 and ST/X000664/1, as well as David Tong's Simons Investigator Award. The work of VG and JQ is supported by Simons Foundation grant 994310 (Simons Collaboration on Confinement and QCD Strings).

\appendix

\section{Some properties of quantum Clebsch-Gordan coefficients}
The quantum Clebsch-Gordan coefficients enjoy several symmetry properties under manipulation of their indices. Here we cite those that we used in this work, and refer to \cite{QCG} for more details
\begin{subequations}
\label{eq:QCG_symmetries}
\begin{align}
\begin{split}
\QCG{\ell_1,m_1,\ell_2,m_2,\ell_3,m_3,q}&=\QCG{\ell_2,-m_2,\ell_1,-m_1,\ell_3,-m_3,q}
\end{split}\\[2ex]
\begin{split}
\QCG{\ell_1,m_1,\ell_2,m_2,\ell_3,m_3,q}&=(-1)^{\ell_3-\ell_1-m_2} q^{-m_2} \sqrt{\frac{[2\ell_3+1]_q}{[2\ell_1+1]_q}} \QCG{\ell_3,m_3,\ell_2,-m_2,\ell_1,m_1,q}
\end{split}\end{align}
\end{subequations}

The quantum Clebsch-Gordan coefficients with $q$ are related to the ones with $q^{-1}$ by the following identity
\begin{equation}
    \QCG{\ell_1,m_1,\ell_2,m_2,\ell_3,m_3,q}=(-1)^{\ell_1+\ell_2-\ell_3}\QCG{\ell_1,-m_1,\ell_2,-m_2,\ell_3,-m_3,q^{-1}}.
\end{equation}
This can be seen by substituting $r=\ell_1+\ell_2-\ell_3-s$ and summing over $s$ in \eqref{qcg_formula}.

The action of the $R$-matrix on the quantum Clebsch-Gordan coefficients give
\begin{equation} \label{eq:Rmat_QCG}
    \sum_{m_1',m_2'} [R_{\ell_2,\ell_1}]_{m_2,m_1}^{m_2',m_1'}\QCG{\ell_1,m_1',\ell_2,m_2',\ell_3,m_3,q} = (-1)^{\ell_1+\ell_2-\ell_3} q^{\ell_3(\ell_3+1)-\ell_1(\ell_1+1)-\ell_2(\ell_2+1)} \QCG{\ell_2,m_2,\ell_1,m_1,\ell_3,m_3,q}
\end{equation}
For the other choice of $R$-matrix $\tilde{R}$ we similarly get
\begin{equation}\label{eq:Rmattilde_QCG}
    \sum_{m_1',m_2'} [\widetilde{R}_{\ell_2,\ell_1}]_{m_2,m_1}^{m_2',m_1'} \QCG{\ell_1,m_1',\ell_2,m_2',\ell_3,m_3,q} = (-1)^{\ell_1+\ell_2-\ell_3} q^{-\ell_3(\ell_3+1)+\ell_1(\ell_1+1)+\ell_2(\ell_2+1)} \QCG{\ell_2,m_2,\ell_1,m_1,\ell_3,m_3,q}
\end{equation}

\section{Relation between fusion kernel and OPE coefficients in diagonal minimal models} \label{app:fusion_kernel_OPE}
The fusion kernel for the Virasoro blocks can be found explicitly in \cite{Furlan:1989ra}. The formulas involved are very unwieldy, so rather than inspecting them directly we would like to show the relation between the functions appearing in the fusion kernel and the OPE coefficients in diagonal minimal models.

We consider the correlation function $\braket{\phi_{r_1,s_1}(0)\phi_{r_2,s_2}(z,\bz)\phi_{r_3,s_3}(1)\phi_{r_4,s_4}(\infty)} $ in the generalized minimal model with charge $c=1-\frac{6}{\mu(\mu+1)}$;  the Virasoro blocks in the $s$ and $t$-channel are related as
\begin{equation}
    \calF_{r',s'}^{(t)}(z) = \sum_{r,s} F_{(r,s),(r',s')}\mathcal{F}^{(s)}_{r,s}(z)
\end{equation}
with the crossing kernel \cite{Teschner:1995ga,Cremmer:1994kc}
\begin{multline}
   F_{(r,s),(r',s')} =\\
   = \frac{g_{(r_1,s_1),(r_2,s_2)}^{(r,s)}g_{(r,s),(r_3,s_3)}^{(r_4,s_4)}}{g_{(r_2,s_2),(r_3,s_3)}^{(r',s')}g_{(r_1,s_1),(r',s')}^{(r_4,s_4)}}(-1)^{f(r_i,s_i,r,s,r',s')} \sixj{\frac{r_1-1}{2},\frac{r_2-1}{2},\frac{r-1}{2},\frac{r_3-1}{2},\frac{r_4-1}{2},\frac{r'-1}{2}}_{\tilde{q}} \sixj{\frac{s_1-1}{2},\frac{s_2-1}{2},\frac{s-1}{2},\frac{s_3-1}{2},\frac{s_4-1}{2},\frac{s'-1}{2}}_{q} \label{eq:fusion_kern_g}
\end{multline}
where for notational purposes we do not write the explicit dependence on the external operators $(r_i,s_i)$. While $q=e^{i \pi \frac{\mu}{\mu+1}}$, $\tilde q$ is defined as $\tilde q=e^{i \pi \frac{\mu+1}{\mu}}$. The sum is over all $(r,s)$ allowed by the fusion rules and we used the shorthand notation
\begin{equation}
    f(r_i,s_i,r,s,r',s') = \frac 12 (s_2-1)(r+r'-r_2-r_4) + \frac 12 (r_2-1)(s+s'-s_2-s_4)
\end{equation}
and $g$ has the symmetry property $g_{(r_1,s_1),(r_2,s_2)}^{(r,s)}=g_{(r_2,s_2),(r_1,s_1)}^{(r,s)}$.

We do not care about the explicit formula of $g$, but this can be deduced from the explicit formula for the fusion kernel of \cite{Furlan:1989ra}. Now we will show how the $g$'s are related to the diagonal minimal model OPE coefficients. For this, it will be enough to consider the case where $(r_3,s_3)=(r_1,s_1)$ and $(r_4,s_4)=(r_2,s_2)$, for which crossing symmetry amounts to
\begin{equation}
\sum_{r,s} (C^{\rm{MM}}_{(r_1,s_1),(r_2,s_2),(r,s)})^2\mathcal{F}^{(s)}_{r,s}(z)\mathcal{F}^{(s)}_{r,s}(\bz) = \sum_{r',s'} (C^{\rm{MM}}_{(r_1,s_1),(r_2,s_2),(r',s')})^2\mathcal{F}^{(t)}_{r',s'}(z)\mathcal{F}^{(t)}_{r',s'}(\bz) 
\end{equation}
and the minimal models OPE coefficients are symmetric under permutation of any of the operators. Using \eqref{eq:fusion_kern_g} for both the holomorphic and the antiholomorphic blocks, we get
\begin{equation}
\begin{aligned}
    \sum_{r,s} &(C^{\rm{MM}}_{(r_1,s_1),(r_2,s_2),(r,s)})^2\mathcal{F}^{(s)}_{r,s}(z)\mathcal{F}^{(s)}_{r,s}(\bz) =\\ 
    & \qquad =\sum_{r',s'} (C^{\rm{MM}}_{(r_1,s_1),(r_2,s_2),(r',s')})^2 \sum_{r_L,s_L}\sum_{r_R,s_R} F_{(r_L,s_L),(r',s')} F_{(r_R,s_R),(r',s')} \calF^{(s)}_{r_L,s_L}(z)\calF^{(s)}_{r_R,s_R}(\bz)
\end{aligned}
\end{equation}
The r.h.s. has to add to zero if $(r_L,s_L) \neq (r_R,s_R)$. Taking $(r_L,s_L) = (r_R,s_R)$, and matching the coefficients of the Virasoro blocks we get
\begin{equation}
    (C^{\rm{MM}}_{(r_1,s_1),(r_2,s_2),(r,s)})^2 = \sum_{r',s'} (C^{\rm{MM}}_{(r_1,s_1),(r_2,s_2),(r',s')})^2 F_{(r,s),(r',s')}^2
\end{equation}
which, using \eqref{eq:fusion_kern_g}, can be rearranged to
\begin{equation}
\begin{aligned}
    \left(\frac{C^{\rm{MM}}_{(r_1,s_1),(r_2,s_2),(r,s)}}{g_{(r_1,s_1),(r_2,s_2)}^{(r,s)}g_{(r_1,s_1),(r,s)}^{(r_2,s_2)}} \right)^2= \sum_{r',s'}  \sixj{\frac{r_1-1}{2},\frac{r_2-1}{2},\frac{r-1}{2},\frac{r_1-1}{2},\frac{r_2-1}{2},\frac{r'-1}{2}}_{\tilde{q}}^2 & \sixj{\frac{s_1-1}{2},\frac{s_2-1}{2},\frac{s-1}{2},\frac{s_1-1}{2},\frac{s_2-1}{2},\frac{s'-1}{2}}_{q}^2 \\ 
    & \times \left(\frac{C^{\rm{MM}}_{(r_1,s_1),(r_2,s_2),(r',s')}}{g_{(r_1,s_1),(r_2,s_2)}^{(r',s')}g_{(r_1,s_1),(r',s')}^{(r_2,s_2)}} \right)^2
    \end{aligned}
\end{equation}
and it can be checked, using the orthogonality relations of 6$j$ symbols \eqref{eq:6j_orthogonality}, that this equation is solved by
\begin{equation}
    \left(C^{\rm{MM}}_{(r_1,s_1),(r_2,s_2),(r,s)}\right)^2 = \beta_{12} \left(g_{(r_1,s_1),(r_2,s_2)}^{(r,s)}g_{(r_1,s_1),(r,s)}^{(r_2,s_2)} \right)^2\,
\end{equation}
with $\beta_{12}$ an arbitrary constant which depends only on the external operators, and not on $(r,s)$.

\section{Details about the Ising limit} \label{app:IsingPsiDetail}
\subsection{Vanishing of the \texorpdfstring{$\Lambda$}{} three-point function}
Using the definitions \eqref{eq:ABdef} and the explicit results for the OPE coefficients \eqref{eq:C_as_MM_general}, where we determined the sign by an explicit calculation, we can obtain
\begin{equation}
    \begin{aligned}
        \braket{B_0 B_0 B_0} & = \frac{117 i \sqrt{\frac{15}{14 \pi }}}{128 \, } \frac{ \Gamma (\frac{3}{4})}{\Gamma (\frac{1}{4})^\frac14 \Gamma(\frac{5}{4})^\frac34} \frac{1}{x_{12}^{5/2} x_{13}^{5/2} x_{23}^{5/2}}\\
        \braket{B_0 W_{1,3}^2   W_{1,3}^{-2}} &=\braket{W_{1,3}^2 W_{1,3}^{-2} B_0 } = -\braket{W_{1,3}^{-2} B_0 W_{1,3}^2 } = \sqrt{2} \pi e^{-i \pi /4} \braket{B_0 B_0 B_0}\\
        \braket{B_0 W_{1,3}^{-2}   W_{1,3}^{2}} &=\braket{W_{1,3}^{-2} W_{1,3}^{2} B_0 } = -\braket{W_{1,3}^{2} B_0 W_{1,3}^{-2} } = -\sqrt{2} \pi e^{i \pi /4} \braket{B_0 B_0 B_0}
    \end{aligned}
\end{equation}
where we use the shorthand notation $\braket{A B C} = \braket{A(x_1) B(x_2) C(x_3)}$. These are all of the ingredients we need to evaluate,
\begin{equation}
    \braket{B_0 \Lambda \Lambda} = \braket{B_0 W_{1,5}^2 W_{1,5}^{-2}} + \braket{B_0 W_{1,5}^{-2} W_{1,5}^{2}} +2 \pi e^{i\pi/2}\braket{B_0 B_0 B_0} = 0\ ,
\end{equation}
and
\begin{equation}
    \braket{W_{1,5}^{+ 2} \Lambda \Lambda} +\braket{W_{1,5}^{- 2} \Lambda \Lambda} = 0\ .
\end{equation}
We conclude that $\braket{\Lambda \Lambda \Lambda} = 0$.

\subsection{Another logarithmic multiplet and vanishing of \texorpdfstring{$\braket{\tilde \Lambda \tilde \Lambda}$}{}} \label{app:other_multiplet}
In the $\mu \to 3$ limit, a logarithmic multiplet also forms between descendants of $W_{2,3}$ and $W_{2,5}$. Given that the operator $\bar \psi$ is identified as a combination of the components of $W_{2,3}$, this logarithmic multiplet plays a role in the Ising limit.

Let's consider operator $W_{2,3}$ first. When $\mu=3$ it develops a level one zero-norm descendant. We define,
\begin{equation}
    \tilde{Y}^m = L_{-1} W_{2,3}^m\ . 
\end{equation}
This operator will have a vanishing two-point function (but finite higher point functions), and just as earlier, there will be divergences for other correlators. The correct procedure is again to organize things into a log multiplet,
\begin{equation}
\begin{aligned}
    \tilde{A}^m &= \lim_{\epsilon \to 0}   \frac{(-1)^m\tilde{\alpha}_1}{\sqrt{\epsilon}} W_{2,5}^m+ \frac{\tilde{\alpha}_2}{\epsilon}\tilde{Y}^m\\
    \tilde{B}^m &=\lim_{\epsilon \to 0}   \tilde{\alpha}_3 \tilde{Y}^m \label{eq:ABtildedef}
\end{aligned}
\end{equation}
where we make the choice 

\begin{equation}
    \tilde{\alpha}_1 = \sqrt{2}e^{-i \pi/4}\,, \qquad  \tilde{\alpha}_2 = 2 i \sqrt{6}+i \sqrt{\frac{3}{8}}\pi\epsilon \,, \qquad \tilde{\alpha}_3 = -i\sqrt{\frac{3}{2}}\ .
\end{equation}
This gives
\begin{align}
    L_0 \begin{pmatrix}
        \tilde{A}^m \\ \tilde{B}^m 
    \end{pmatrix} = \begin{pmatrix}
        1 & 1 \\ 0 &1
    \end{pmatrix}\begin{pmatrix}
         \tilde{A}^m \\ \tilde{B}^m
    \end{pmatrix}\ ,
\end{align}
and the finite two-point functions
\begin{equation}
\begin{aligned}
    \braket{\tilde A^m (0) \tilde A^{-m}(x)} &= \frac{\log x}{x^2 \bar x} \QCG{1,m,1,-m,0,0,q_I}\ ,\\
    \braket{\tilde A^m (0) \tilde B^{-m}(x)} &= -\frac{1}{2 x^2 \bar x} \QCG{1,m,1,-m,0,0,q_I}\ ,\\
    \braket{\tilde B^m (0) \tilde B^{-m}(x)} &= 0\ .
\end{aligned}
\end{equation}
Notice that, contrary to $L_0$, the action of $\bar L_0$ on this multiplet is diagonal.

As mentioned in the main text, the OPE of $\psi$ and $\bar{\psi}$ contains another non-Ising operator $\tilde \Lambda$,
\begin{equation}
    \psi(x)\bar{\psi}(0) \sim \varepsilon + C_{\psi \bar{\psi} \tilde{\Lambda}} x^\frac{1}{2}\tilde{\Lambda} +\ldots
\end{equation}
where
\begin{equation}
    C_{\psi \bar{\psi} \tilde{\Lambda}} = \sqrt{\frac{10}{7}} C_{\psi {\psi} {\Lambda}}
\end{equation} 
and
\begin{equation}
    \tilde{\Lambda} = W_{2,5}^2 + W_{2,5}^{-2} - e^{i \pi/4}\sqrt{2\pi} \tilde{B}^0\
\end{equation}

Then, as argued earlier for $\Lambda$, it better be that the correlation functions of several $\tilde \Lambda$-s vanish. As a starting point, it is easy to check that the two-point function of $\tilde{\Lambda}$ vanishes. We also underline the fact that the operator $\tilde A$ is not exchanged in the OPE of $\psi$ and $\bar \psi$, in agreement with the fact that correlation functions of these operators do not have logarithms.

\bibliography{biblio_genQG}
\bibliographystyle{utphys}
\end{document}